\documentclass[twoside,11pt]{article}
\usepackage{jmlr2e}
\usepackage{color}
\usepackage{algorithm}
\usepackage{algpseudocode}
\usepackage{amsmath}
\usepackage{amssymb}
\usepackage{amsfonts}
\usepackage{amsbsy}
\usepackage{subfigure}
\usepackage{url}
\usepackage{siunitx}
\usepackage{fullpage}
\usepackage{natbib}
\usepackage{cancel}
\usepackage{epstopdf}
\usepackage{float}
\usepackage{tabto}
\usepackage{mathtools, nccmath}

\def\cbk{\color{black}~}
\providecommand\given{}
\newcommand\SetSymbol[1][]{%
\nonscript\:#1\vert
\allowbreak
\nonscript\:
\mathopen{}}
\DeclarePairedDelimiterX\Set[1]\{\}{%
\renewcommand\given{\SetSymbol[\delimsize]}
#1
}

\begin{document}

\title{High Dimensional Cluster Analysis Using Path Lengths}
\author{\name Kevin L. McIlhany  \email mcilhany@usna.edu \\
        \addr Physics Department \\
              U. S. Naval Academy, Stop 9c \\
              572c Holloway RD\\
              Annapolis, MD 21402-5002, USA~~~~~~~~~~~~~~~~~~~~~~~~~~~~~~~~~~~~~~~~~~~~~~~~~~~~~~~~~~~~~~~~~~\today
        \AND
        \name Stephen Wiggins \email s.wiggins@bristol.ac.uk \\
        \addr School of Mathematics\\
              University of Bristol\\
              University Walk \\
              Bristol, BS8 1TW, \\
              United Kingdom }

\editor{}

\maketitle

\begin{abstract}
A hierarchical scheme for clustering data is presented which applies to spaces with a high number of dimension ($N_{_{D}}>3$).  The data set is first reduced to a smaller set of partitions (multi-dimensional bins). Multiple clustering techniques are used, including spectral clustering, however, new techniques are also introduced based on the path length between partitions that are connected to one another.  A Line-Of-Sight algorithm is also developed for clustering.  A test bank of 12 data sets with varying properties is used to expose the strengths and weaknesses of each technique.  Finally, a robust clustering technique is discussed based on reaching a consensus among the multiple approaches, overcoming the weaknesses found individually.
\end{abstract}


\section{Introduction}
“Clustering” is a fundamental technique and methodology in data analysis and machine learning. The explosion of the field of data science has, consequently, led to an expansion in how this notion is applied. In this respect, it would be more appropriate to refer to “clustering” as “data organization”, which would encompass the ideas of 1) data reduction, 2) data identification, 3) data clustering, and 4) data grouping.

Data reduction is the process of converting raw data into a form that is more amenable for the application of a specific analytical and/or computational methodology. Data identification is the process of analysing trends or distributions within the data. Data clustering is the process of associating data through proximity, similarity, or dissimilarity. Data grouping refers to breaking down data into groups according to a criterion that is appropriate for the specific application under consideration.

The literature on clustering is extensive and it is beyond the scope of this paper to provide an adequate review of this topic. However, the following papers \cite{jain1999data,ng2002spectral,barbakh2009review,jain2010data}, for background on the clustering  methods in this paper and the book \cite{kaufman2009} provides a broad overview of clustering methodologies, as well as their numerical implementation.

There is no single algorithm that realizes all four of these aspects of data organization. The approach to this problem pursued in this paper is to develop a hierarchical scheme leading to a cluster analysis that encompasses the issues raised above and is adapted to high dimensional spaces.

The data analysis scheme presented in this paper uses a blend of traditional data analysis via a multi-variate histogram along with standard clustering techniques, such as k-means, k-medoids and spectral clustering. By binning the data onto a multi-dimensional grid, data is partitioned into regions on the grid which may be connected or separated depending on the character of the data set. Data reduction is realized by only retaining bins that have a population above a user selected threshold. The resulting multidimensional bins are referred to as partitions. The passage to partitions is the data reduction step.

Data indentification is the process of assigning known data distributions (parent) to an entangled set of data.  Typical examples are found in the literature of Bayesian analysis \cite{binder1978,kass1995}, however, this pursuit dates farther back to earlier attempts to understand how to distinguish data from two or more distributions with overlapping tails.  In more difficult scenarios, several distributions might overlap within the peak regions, changing the problem to the identification of subdomains of the mixed versus non-mixed distributions.

Data clustering traditionally refers to assigning data to subsets based on the proximity of data to one another.  The goals of the field of data clustering have expanded from this definition, taking on some of the other roles identified here.  For the purposes of this study, the term clustering will refer to both the overall techniques applied as well as the specific property a set has when its members are close to on another when appropriate.   In the broadest sense, a cluster is simply a label given to data to identify common features.

Data grouping is the process of assigning labels to data, without regard for proximity or parent distributions.  An example might be to segregate a class of thirty $2^{nd}$ grade children into five subgroups before entering a museum for a tour.  How the larger group is broken apart is unimportant, merely that the larger group is distributed into smaller groups.

In this study, standard clustering techniques are applied such as kmeans, kmedoids and spectral clustering, along with new path-based approaches.  After data reduction, data within partitions may be connected in regions where a path length can be calculated along the grid of partitions between any two data.  Several new clustering algorithms have been developed using the path length. Further, if two partitions are visible to each other by a Line-Of-Sight criteria, the relationship between them is given additional significance. These ideas are used, in conjunction with standard clustering techniques, to construct 26 different clustering algorithms.

An analysis configuration is the set of all 26 clustering techniques along with the choices made for which variables represent the data manifold as well as how the data space is partitioned.  The choice of variables used to represent the data defines the number of dimensions as well as the character of the data manifold formed.  In some cases, two or more variables may provide redundant information, while other choices may add noise.  Changes to the resolution of how the data space is partitioned may lead to changes in a datum's cluster assignment.   For each choice of clustering technique, variables used (dimensions) and resolution (partitioning), each datum is assigned to a cluster.  When data consistently cluster in one arrangement across multiple analysis configurations, the data is assigned robustly to its cluster.  To determine a {\em robust} clustering assignment, a polling technique is used to arrive at a consensus amongst the clustering algorithms.  While any one technique has faults, the consensus of techniques overcomes any one failure mode, giving the best all-round identification \cite{strehl2003}.

This paper is organized as follows, section \ref{sec:data-reduce} defines the basic components used in this study.  Section \ref{sec:calcs-1} shows the calculations of several values used throughout the analysis.  Section \ref{sec:strategy-1} outlines the strategy taken for this study and it lists the comprehensive set of arrays calculated that are needed for the suite of algorithms.  This section also introduces a test-bank of data sets used for clustering.  Section  \ref{sec:algorithms-1} presents each algorithm, with details left for the appendix.  Section \ref{sec:results-1} shows the results for each clustering algorithm, discussing the strengths and weaknesses of each approach.  Section \ref{sec:robust-1} introduces the approach to robust clustering, employing multiple techniques and how a consensus is reached.  Section \ref{sec:conclusions} concludes with suggestions for extending this suite of clustering techniques.  In the appendix, Sec. \ref{sec:algorithms-general-1} discusses algorithms used by more than one clustering technique.  The next appendices show the details of the clustering algorithms as applied in this study for:  clustering to maxima globally and using path length, Sec. \ref{sec:local-max-1}, as well as clustering based on Line-Of-Sight, Sec. \ref{sec:los-3}.


\section{Terminology, Definitions, Notation and Data Reduction}
\label{sec:data-reduce}
This study has five basic components that discussed in detail;  data, partitions, clusters, clustering algorithms and configurations of analysis.  The definitions for data, partitions and clc usters are given in this section, while the clustering algorithms and configurations of analysis given in Sec. \ref{sec:algorithms-1} and Sec. \ref{sec:robust-1} respectively.  \cbk Quantities denoted by a tilde refer to the original data set while quantities without a tilde symbol refer to partitions (a reduced data set defined in Sec.\ref{sec:partitions}).\cbk

\subsection{Data}
\label{sec:defs-1}
The term data used in this study will refer to a set of measurements taken of a system.   The values of the measurements may vary in nature, from logical values, lists of characters, images, vector or scalar values, either real or imaginary.  In the case of complex values or vectors, each component will be taken separately as a real valued scalar.  For parts of the data which are non-numerical, the values of the measurements will be mapped to a numerical value.  A {\em data vector} is formed whose components are the set of numerical measurements.  The collection of these data vectors form the data set.   For each component, the range is determined either by the extrema taken from the data set or by the limits based on the functional form of the measurement.   The {\em data space} is the space spanned by the set of data vectors.

The total number of data vectors in the set is denoted by ``N''.  Each component of the data vector represents a dimension of the data space, with the total number of components, $N_D$.  The data vector formed will be denoted by a capital ``{\bf X}'' and the components of the vector will be denoted in lower case, ``x'', together given as ${\bf \tilde{X}}(\tilde{x})$.

\begin{definition}[{\bf Data, Data Vector, Data Space}]~\\
    \label{def:data-1}
    $\hspace{-0.1cm}${\bf Data:} $\hspace{1.3cm}$ Collection of events, $\mathcal{D}$, forming a set, described by a set of variables.  The variables are often mapped to numerical values. \\
    {\bf Data Vector:}           $\hspace{-0.1cm}$ Vector of data values, ${\bf \tilde{X}} = [\tilde{x}_{1},\tilde{x}_{2},...\tilde{x}_{N_D}]$. \\
    {\bf Data Space:}            $\hspace{0.0cm}$ Space which contains the set of data vectors; however, the data space may extend further to the natural limits for each variable.
\end{definition}
\begin{definition}[{\bf Numbering}]~\\
    $N              \hspace{0.70cm}\leftarrow$ the number of data within $\mathcal{D}$. \\
    $N_D            \hspace{0.49cm}\leftarrow$ the number of dimensions (data vector components). \\
    $N_{B,i}        \hspace{0.32cm}\leftarrow$ the number of bins per component,  indexed by i, defined in Sec. \ref{sec:defs-2}\\
    $N_{B,tot}      \hspace{0.09cm}\leftarrow$ the highest possible bin index value, $N_{B,tot} = \Pi_{i=1}^{N_D} N_{B,i}$ \\
    $N_P            \hspace{0.50cm}\leftarrow$ the number of partitions, defined in Sec. \ref{sec:partitions}. \\
    $N_{CL}         \hspace{0.31cm}\leftarrow$ the number of clustering techniques, defined in Sec. \ref{sec:clusters-1}\\
    $N_{C,m}        \hspace{0.20cm}\leftarrow$ the number of clusters for a technique, with techniques indexed by $m$.\\
    $N_{sought}     \hspace{0.10cm}\leftarrow$ the number of clusters sought using k-means and k-medoids algorithms.\\
    $N_{spec}       \hspace{0.10cm}\leftarrow$ the number of spectral clusters sought after.\\
\label{def:nums1}
\end{definition}

\subsection{Multi-variate Histogram, Bin Addresses and Data Reduction}
\label{sec:defs-2}

\begin{figure*}[t!p]
    \label{fig:partid4x}
	\centering
	\subfigure[] {
    	\label{fig:grid2d7}
		\includegraphics[width=0.45\linewidth]{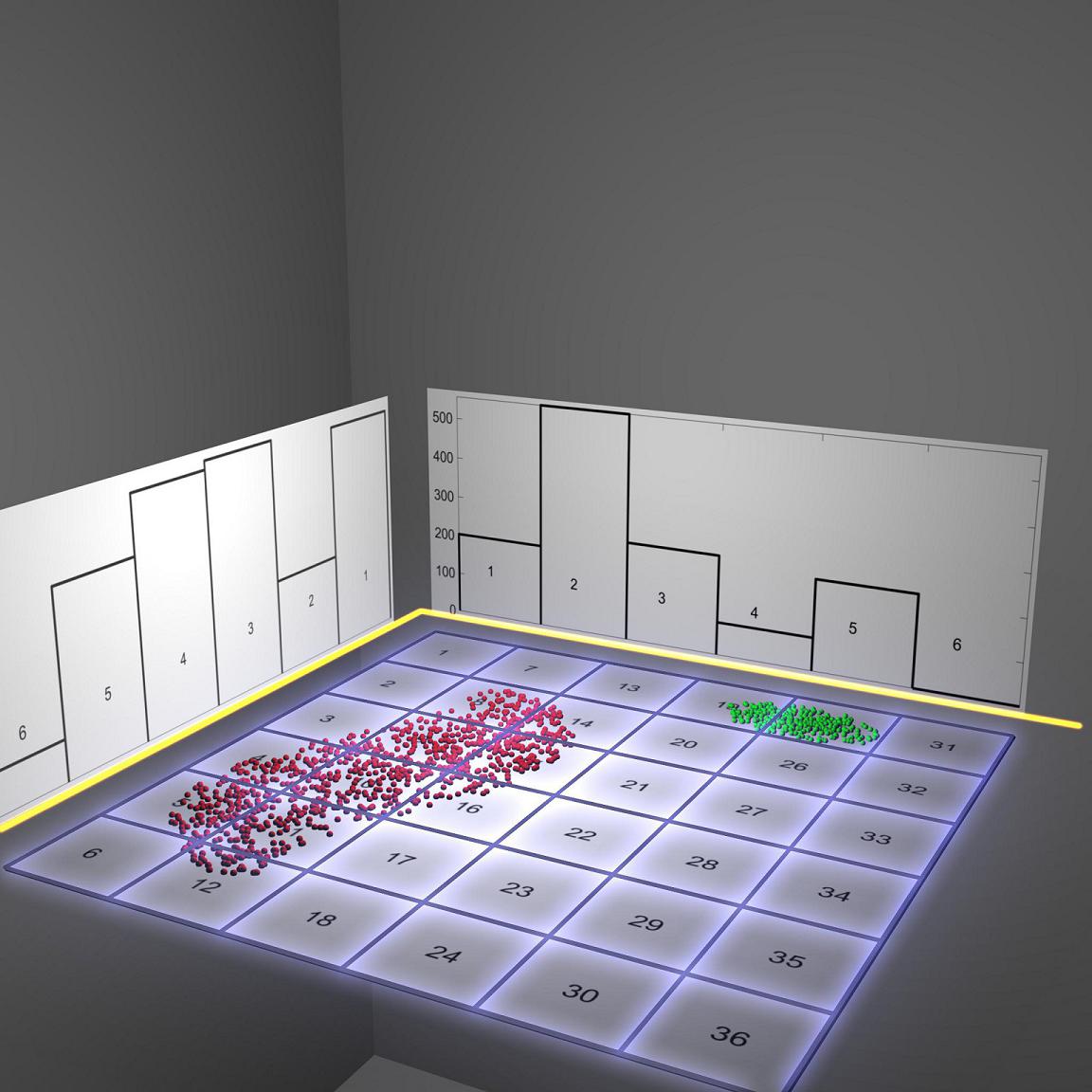} }
\qquad\hskip -0.3cm
	\subfigure[] {
    	\label{fig:grid3d8}
		\includegraphics[width=0.45\linewidth]{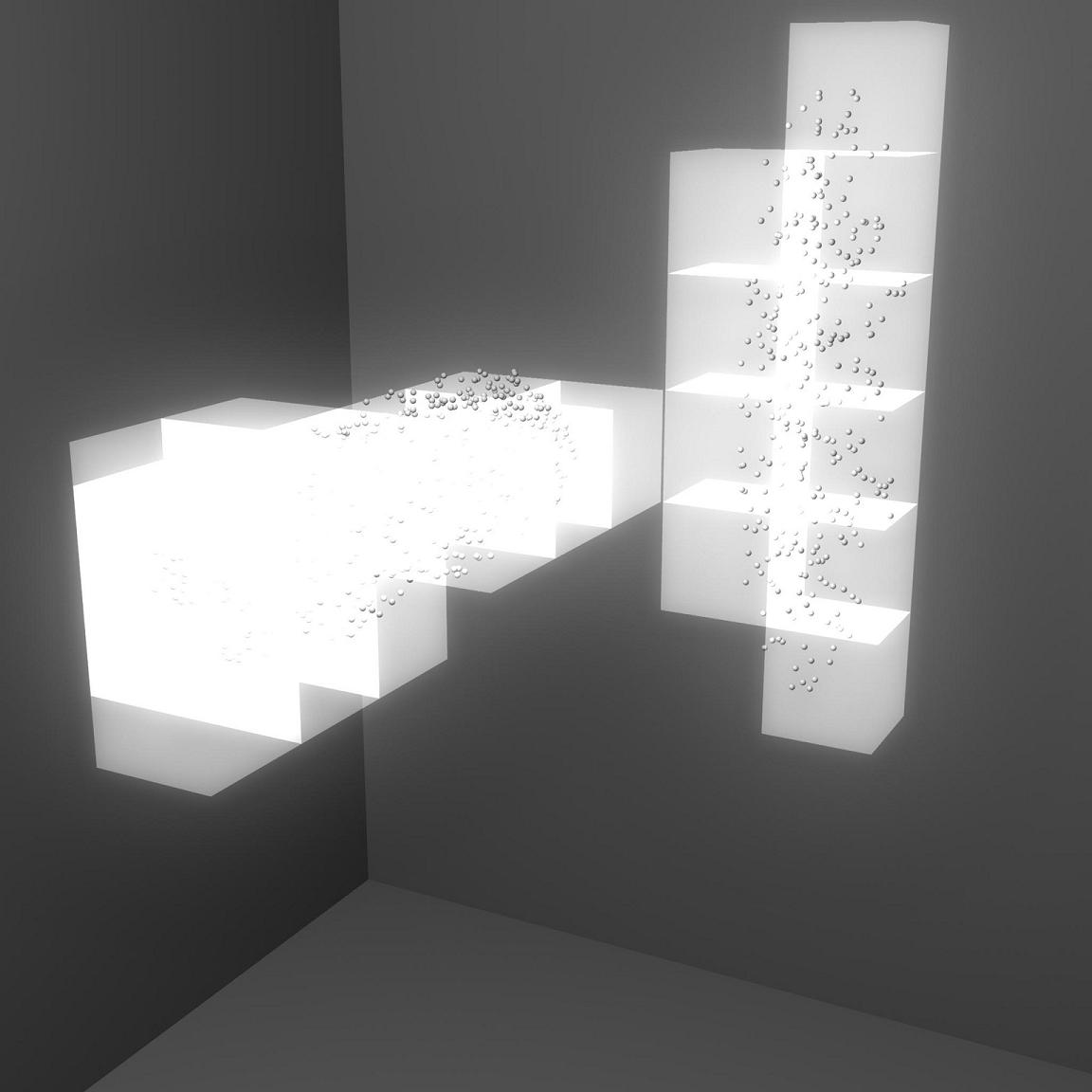} }
    \caption{Panel \ref{fig:grid2d7} illustrates the idea of partitioning the data into bins.  Each axis is binned with an index from 1 to 6, leading to the combined binning from 1 to 36, with data residing in bins: 4,5,8,9,10,11,12,14,15,16,19 and 25.  Panel \ref{fig:grid3d8} illustrates the same data set, where a third dimension has been added, leading to 216 possible bins, filling 16 of them.  The data represents two groups of data, shown in red and green, yet the task of separating the data is the goal of the clustering algorithm.  As such, data initially is presented without knowledge of what distributions it breaks into. }
\end{figure*}

The primary mechanism to reduce the total data set to a smaller size is to employ a single multi-dimensional histogram.   The details of forming the bins of the multi-variate histogram are given below as they relate to how neighboring bins are determined.   A histogram is a frequency distribution along one or more dimensions.  For a histogram along one component, each datum is assigned a bin number based on the minimum and maximum value, [$x_{min}, x_{max}$], and number of bins given along each component, ($N_{B}$).  The bin value for a datum along one dimension is then given by:

\begin{definition}[{\bf Bin Index, Bin Address}, $\tilde{b},{\bf \tilde{B}}$]
    \label{def:binaddr1}
    Vector of bin indices, ${\bf \tilde{B}} = [\tilde{b}_{1},\tilde{b}_{2}...\tilde{b}_{i},...\tilde{b}_{N_D}]$.
    \begin{equation}
    \tilde{b} = {\rm ceil} \left[ \frac{ \tilde{x} - x_{min} } { x_{max} - x_{min} }   N_{B}  \right]
    \end{equation}
\end{definition}
The bin indices ($\tilde{b}$) form a vector, the ``bin address'', ${\bf \tilde{B}}(\tilde{b})$, assigning each datum to a multi-dimensional bin.  A serial bin address index, $\tilde{k}$, is then formed from the bin address vector.

\begin{definition}[{\bf Serial Bin Address Index - Datum}, $\tilde{k}$]
    \label{def:binaddr2}
    Serialized bin index mapping the bin address vector to a single value, $\tilde{k}$.
    \begin{equation}
        \tilde{k} = \sum_{i=1}^{N_D} \left[\left({\rm {\bf \tilde{B}}}(\tilde{b}_i)-1\right) \ast \prod_{q=0}^{i-1} N_{B,q}\right] ~+~1
    \end{equation}
\end{definition}
with each component indexed by $i$ or $q$ and $N_{B,0}=1$.

Data are grouped into multi-dimensional bins in this manner.   Figure 1 illustrates the labelling of bins in 2D and 3D for a sample data set.  Each multi-dimensional bin will contain a subset of the data, whose population will be used as a weighting factor in later analyses.  The range of values for the bin address index can be quite large, given a high number of dimensions.  The possible range of values for this index is: ${1... N_{B,tot}}$,  where the maximal value is simply the product of the number of bins used along each dimension.  Further, this index will have large gaps in regions representing empty space.

Once a serial bin address index is assigned to each datum, the population of each bin is found by summing over the data with the same bin address index, $\tilde{k}$.   This can be shown as a 1D histogram using the serial bin address index as the bins and setting the number of bins equal to the largest value of $\tilde{k}$.  An example of this  histogram is shown in Fig. 2(d), where the vertical axis of this histogram is the population of each bin.

This process reduces the data set from $N$ data down to a smaller set of bins containing the data, where each bin is identified by two numbers, the serial bin address index and the weight.  Because each datum within a bin has the same address index, {\em all} data within the bin is simply defined by two values $[\tilde{k}, \tilde{w_k}]$.   The number of multi-dimensional bins should be significantly less than the number of data, provided the data are not homogeneously distributed across the space.

\cbk
As a possible means to reduce the data set further, only bins with a higher population will be considered for further analysis.  Bins not considered are found  either by selecting bins with specific low populations (bins with one datum, two data, etc...),  or by finding the cumulative set of all bins whose summed population is some small percentage of the total data size.  By defining the parameter, $\Theta_{low}$, as a percentage of the total data set size, only consider bins above this threshold to define a new data set, $\mathcal{D'}$.  The complementary
dataset, $\mathcal{D''}=\mathcal{D}-\mathcal{D'}$, either represents noise in the data set or low population data bins compared to the larger data set. Figure 2(e) shows an example histogram and thresholds, where the sum of all bins below threshold is less than $\Theta_{low}~N$.  \cbk The data set formed above this lower threshold, $\mathcal{D'}$, is defined as:

\begin{equation}
    \mathcal{D'} = \Set*{ \tilde{x},\tilde{w} \in \mathcal{D} \given \mathcal{F}(\tilde{k})>\Theta_{low}\mathcal{F}(N_{B,tot})}, \\
    \label{eqn:data2}
\end{equation}
\begin{equation}
    {\rm where~~}\mathcal{F}(\tilde{k}) ~=~ \sum_{\tilde{k'}=1}^{\tilde{k}} \tilde{w}(\tilde{k}') \;~~~~~~{\rm for~~} \tilde{k} = \{1...N_{B,tot}\}.
\end{equation}

\subsection{Partitions}
\label{sec:partitions}

\begin{figure*}[t!]
    \label{fig:partid1x}
	\centering
	\includegraphics[width=0.85\linewidth]{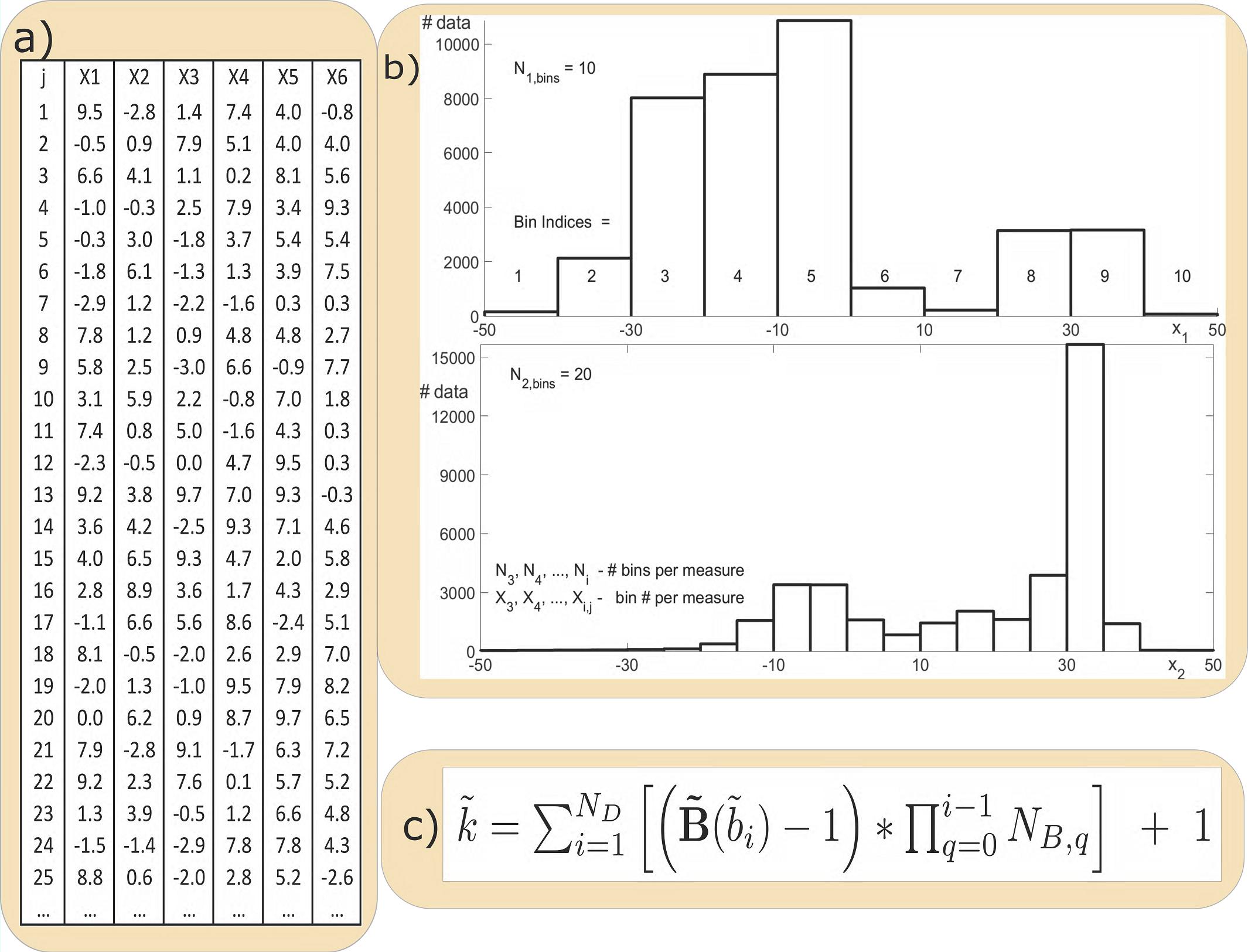}
    \caption{Data reduction process shown starting from a data set (a) to the individual histograms (b), where the bin indices are collected, leading to the calculation (c) of the unique bin address, $k_j$, assigned to each datum.
    \vspace{0.0cm}}
\end{figure*}
\setcounter{figure}{1}
\begin{figure*}[t!]
    \label{fig:partid1xx}
	\centering
	\includegraphics[width=0.85\linewidth]{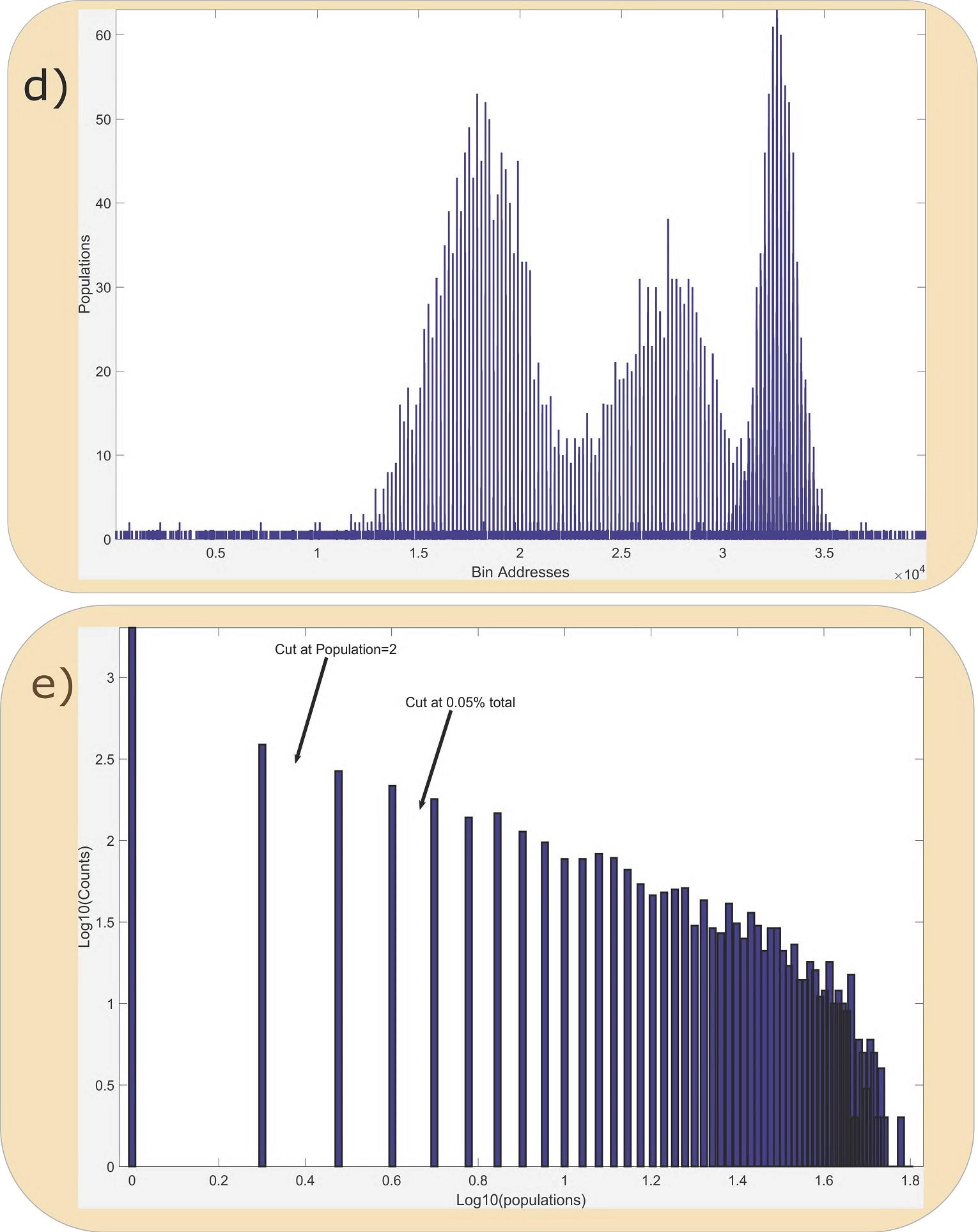}
    \caption{Continuation of the data reduction process shown.  The last two panels show  the histogram of the bin addresses (d) which leads to the last histogram of the bin address frequencies, where a cut is placed to eliminate the cumulative group of partitions below the lower distribution threshold, $\Theta_{low}=0.05$ {\em or} a direct cut on the population of the bins at two or less.
    \vspace{0.0cm}}
\end{figure*}
\cbk
The set of multi-dimensional bins above threshold, $\mathcal{D'}$, contains a majority of the data, defined by their bin index and population, $[\tilde{k}, {\tilde{w_k}}]$.  Within each bin, the data all share a common bin index, $\tilde{k}$.  As such, the bins of $\mathcal{D'}$ will be referred to as ``partitions'' for the remainder of the analysis.  The tilde is dropped to indicate the new data set formed from the bins themselves,  $\mathcal{P}({\bf k}, {\bf w})$.  The index, $k$, is the mapping from the serial bin index per datum, $\tilde{k}$ to a sequential bin index, where any gaps in the values of $\tilde{k}$ are removed, giving:  $k \in \{1,2,...,N_P\}$, with $N_P$ the number of bins above threshold (equal to the number of partitions).\cbk    Each partition is defined by two values, the partition index, $k$, and the population of that partition, $w_k$.  A partition address vector is obtained by mapping the partition index back to the original multi-dimensional bin in the data space, giving the binning of each partition, ${\bf B}(b)$ and its location within the space, ${\bf X}(x)$.  The set of partitions, $\mathcal{P}$, will be defined on the bin index space, where each partitions' width has unit length.

\begin{definition}[{\bf Partition}]
    \label{def:part1}
    Set of multi-dimensional bins, $\mathcal{P}$, populated with high density data, whose population is the number of data located within a bin, effectively weighting the bin.
\end{definition}

\begin{definition}[{\bf Sequential Bin Index - Partition}, $k$]
    \label{def:binaddr2}
    Sequential bin index, ${k} \in \{1...N_P\}$, obtained from the serial bin index, $\tilde{k}$ removing any gaps and renumbering sequentially.
\end{definition}
\begin{definition}[{\bf Partition Bin Address, ${\rm {\bf B}}$}]
    \label{def:binaddr3}
    Vector of bin indices for a partition, mapping from $k \rightarrow \tilde{k} \rightarrow {\rm b_i}$, the individual bin indices per component, forming ${\rm {\bf B}} = [{\rm b_{1},b_{2}...b}_{i},...{\rm b}_{N_D}]$.
\end{definition}

\begin{definition}[{\bf Partition Population, $w$}]
    \label{def:pop1}
    Number of data located within a partition.
\end{definition}

\begin{definition}[{\bf Partition space}]
    \label{def:pop1}
    Space containing the set of partitions with axes defined by bin indices, ${\rm {\bf B}}$, where each partition is defined as a unit hypercube. The partitions are arranged to form a multidimensional grid, with most of the grid empty (no partitions).
\end{definition}
\cbk During this analysis, values computed between any two partitions form a matrix with indices given by the current partition being investigated, $k$, and another partitions, labeled, $\ell$.  Each calculation is then represented by either a vector of values with $k$ as the index or as a matrix of values with $[k,\ell]$ as the indices.  Unless required for clarification, these array values will be shown without indices. \cbk

\subsection{Clusters}
\label{sec:clusters-1}
The idea of a cluster is that data can be grouped based on common features found within subsets.  When data is described by several variables, one possible definition is that data near one another within the data space belong to a {\em cluster}.  Clustering can also be defined as a simple grouping of the data, which could be based alphabetically, by income, or some property that is difficult to map numerically such as an objects shape.  Proximity of data to one another is a common feature of data clustering.  Proximity alone can fail to cluster data that has a direction with respect to its relation to other data, such as a directed graph.  By altering the definition of ``proximity'' to include distance measures such as path length, clustering can still be viewed as a local grouping.  Broadly, clustering refers to any choice of grouping of data into subsets.  This paper explores multiple clustering algorithms to later sort the clustering assignments into groupings reached by consensus.
\begin{definition}[{\bf Cluster}]
    \label{def:cluster-1}
    A subset of data defined by a common feature.
\end{definition}


\section{Intermediary Calculations - Euclidean Distance, Nearest Neighbors, Path Length}
\label{sec:calcs-1}

Several calculations are common to multiple techniques which require only the partition bin address vector.  These low level calculations define geometrical features of how the partitions are related to one another. These calculations are represented by a matrix, where each row represents a partition and the columns represent all other partitions.  Specific algorithms for each calculation can be found in the appendix.

The Euclidean distance is calculated between all partitions within the partition space.  First, the difference between two partitions' bin address are calculated for one component, ${\bf \Delta b_i}$.  The Euclidean distance is then calculated from the sum over ${\bf \Delta b}_i$:
\begin{eqnarray}
  {\bf \Delta b}_{i} & =  &  b_{i,k} - b_{i,\ell} \\
  {\bf \Delta R}     & =  & \sqrt{ {\Sigma_{i=1}^{N_D}} {\bf \Delta b}_{i}^2}
\end{eqnarray}
with indices $(k,\ell)$ representing the two partitions.  As each partition is a unit hypercube, the distances range from $\{1...\sqrt{N_D}\}$.

The First Nearest Neighbor matrix, ${\bf NN1}$, defines the Euclidean distance between any two bins that are in contact with one another.  Two partitions are in contact with one another if there exists no bin address component difference greater than one in magnitude, leading to the interpretation that they share a common geometric feature; a point, line, area, etc...  The elements of the matrix given by:

\begin{eqnarray}
{\bf \mathcal{I}} & \equiv & \left\{
    \begin{array}{ll}
        0 & \quad \mid {\bf \Delta b}_{i} \mid    > 1,~~~~{\rm for~any~} i \\
        1 & \quad \mid {\bf \Delta b}_{i} \mid \leq 1,~~~~{\rm for~all~} i \\
    \end{array}
\right. \\
   {\bf NN1}  & = & {\bf \mathcal{I}} \circ {\bf \Delta R}.
\end{eqnarray}
The ${\bf NN1}$ matrix is the adjacency matrix weighted by Euclidean distance, ${\bf \Delta R}$.

In this analysis, the usage of the Path Length is an integral part of many of the clustering algorithms.   Because the data has been reduced to a set of partitions defined by a grid of integer bin address, the Path Length is the distance between any two partitions taken by stepping from one partition to another through $NN1$ steps, summing the Euclidean distance (L2-norm) on this grid {\it between} each step along the path:
\begin{equation}
{\bf \Delta L} ~=~ \sum_{j=k}^{\ell} {\bf NN1}_{j,j+1}
\end{equation}
where the initial partition is $k$, interim partitions, $j$, up to the final partition, $\ell$.  \cbk Partitions are {\em connected} when a path is found, and for partitions that have no connecting path, the path length is set to $\infty$.  The number of steps taken between any two partitions is the Path Count, ${\bf \Delta L}_C \equiv P_C$,\cbk where $P_C$ is used when a fixed pair of partitions has been established for a calculation.

\begin{definition}[{\bf Path}]
    \label{def:path-1}
 A set of partitions where each member has a minimum of one shared $NN1$ partition in common with another partition in the set.
\end{definition}
\begin{definition}[{\bf Path Length}]
    \label{def:pathl-1}
    The sum of $NN1$ steps from one partition to another, with $\infty$ assigned when two partitions have no path between them.
\end{definition}
\begin{definition}[{\bf Connected}]
    \label{def:conn-1}
    Set of all partitions having a non-infinite path length between them.
\end{definition}
\begin{definition}[{\bf Line-Of-Sight: LOS}]
    \label{def:los-1}
    State of a path between two partitions having the minimal L2 path length, being closest to the straight line path, which does not intersect any empty multi-dimensional bins.
\end{definition}
\begin{definition}[{\bf Visibility}]
    \label{def:vis-1}
    The number of partitions that are Line-Of-Sight to a partition.
\end{definition}
\cbk

Between any two partitions, many different paths may be taken to connect them.  In order to find if two partitions have a Line-Of-Sight (LOS) to one another, only paths that fall within a rectangular convex hull are considered.  Placing one partition as the origin with the second partition as the distant corner, the convex hull is the set of all partitions whose bin address indices are equal to or fall between the two partitions along each component.

Six values are calculated for specific paths unique to the LOS criteria.   The true path length is the distance between any two partitions taken by stepping from one partition to another through the {\em least} number of $NN1$ steps and the minimal sum of Euclidean distances from step to step along the path:
\begin{equation}
{\bf \Delta L2}_{T} ~=~ {\rm min} \left[\sum_{j=k}^{\ell} {\bf NN1}_{j,j+1} \right]
\label{eqn:truepathl}
\end{equation}
with indices for the initial partition, $k$, final partition, $\ell$, and the interim partitions up to the final partition, $j$.  For partitions that have no connecting path, the true path length is set to $\infty$.  \cbk

The Summed L1-norm, ${\bf \Delta SL1}$, is calculated between any two partitions, using the initial partition as the origin to the convex hull.  Starting from the origin, the summed L1 distance from origin to each path step is calculated:
\begin{eqnarray}
\label{eqn:l1norm-01}
 {\bf \Delta L1} & = & \sum_{i=1}^{N_D} \mid {\bf \Delta b}_{i} \mid \\
{\bf \Delta SL1} & = & \sum_{j=k}^{\ell} {\bf \Delta L1}_{kj}
\end{eqnarray}

Similar to the case with the Path Length, a Minimal L1-norm Path Length as well as the True L1-norm Path Length are needed to establish the LOS criteria.   The Minimal Summed L1-norm is the length of the path whose summed L1-norms add together giving the least value between the two partitions, while the True Summed L1-norm is the sum of L1-norm values taken along the True Path established in Eqn.\ref{eqn:truepathl}.   Finally, the squared difference along a path of the summed L1-norm to the True Summed L1-norm is used to find the true path.  Details of these calculations will be given in Appendix \ref{sec:pathl01}.
\begin{eqnarray}
\label{eqn:l1norm-01}
{\bf \Delta SL1}_{min} & = & {\rm min} \left[ \sum_{j=k}^{\ell} {\bf \Delta L1}_{kj} \right] \\
{\bf \Delta SL1}_{T}   & = & \sum_{j=k}^{\ell} {\bf \Delta L1}_{kj,true}\\
{\bf SL1VAR}           & = & \sum_{j=k}^{\ell} \mid {\bf \Delta SL1}_{kj} - {\bf \Delta SL1}_{T,kj} \mid^2
\end{eqnarray}

By finding these six values from all paths connecting any pair of partitions, the Line-Of-Sight criteria is formed using the True Path Length, Path Count, Summed L1-norm, True Summed L1-norm, Minimum Summed L1-norm and the Summed L1-norm Variance, where the Line-Of-Sight condition is discussed in section \ref{sec:los-1} and appendix \ref{sec:pathl01}.


\section{Strategy}
\label{sec:strategy-1}

The clustering strategy presented in this paper uses a blend of traditional data analysis via a multi-variate histogram along with standard clustering techniques such as k-means, k-medoids and spectral clustering.  By binning the data onto a multi-dimensional grid, data is partitioned into regions on the grid which may be connected or separated depending on the character of the data set.  A novel approach is taken by calculating the path length between any two populated partitions provided they are connected.  Further, if two partitions are visible to each other by a Line-Of-Sight criteria, LOS, the relationship between them is given additional significance.

\cbk
The initial data set, $\mathcal{D}$, is reduced to a smaller set of partitions based on the number of variables chosen to represent the data, the number of bins for each variable and a threshold placed on each bin to ensure the population of the bins is above a minimal value.  Taking only the set of data with higher population bins to analyze further, $\mathcal{D'}$, the data has effectively been reduced to a set of {\em partitions}, $\mathcal{P}$, whose weight represents the population of data in the partition.  Each partition is given a unique address identifying its location within the data space.

Matrices used in this analysis each have rows and columns representing the partition set, with off-diagonal entries representing values which depend on two partitions.  Based on any given partitions proximity to another partition, a first nearest-neighbor, ${\bf NN1}$, matrix is created where the neighborhood is defined by fixing one partition (a row) as the center of the neighborhood and defining partitions that share a common geometrical border with the center partition as non-zero, with the entries along the columns representing the Euclidean distance of the $NN1$ to the center partition. \cbk

From the set of variables defined in Sec. \ref{sec:defs-3},\ref{sec:calcs-1}, twenty-six differing clustering techniques are employed to determine any given partitions' overall cluster identity.  To {\em robustly} determine a final clustering assignment, a polling technique is used to arrive at a consensus amongst the clustering algorithms.  While any one technique has faults, the consensus of techniques overcomes any one failure mode, giving the best all-round identification.

The group of clustering techniques applied to the analysis is called an {\em analysis configuration}.  Analysis configurations include the 26 clustering techniques along with variations in binning choice and variable choice.  As the binning is changed, resolution of the partitions changes, leading to possible changes in cluster assignment.  Also, the choice of variables used to represent the data defines the number of dimensions as well as character of the data manifold.   When agreement is reached across multiple techniques, differing bin resolutions as well as data manifolds, if data consistently cluster  in one manner, then the data is assigned {\em robustly} to its cluster.  This study presents 26 different clustering algorithms, multiplied by each choice in binning and variables, leading to new variations which will likely lead to new cluster assignments to individual data.  Each choice of (variables, binning, clustering algorithms) creates a configuration for an analysis.  For each configuration, the consensus approach then determines clusters based on the best agreement between the assignments.    The best choice for clustering can then be determined by the analyst of the data as their needs might favor one configuration over another.

\subsection{Arrays: Indices, Scalars, Vectors and Matrices}
\label{sec:defs-3}
Adding to the definitions already given, a comprehensive list of several arrays integral to the clustering algorithms are given here:
\begin{definition}[{\bf Indices}]~\\
    $i              $\tabto{2.20cm}$\leftarrow$ component index (vectors), dimension index (space)\\
    $k,\ell,j       $\tabto{2.20cm}$\leftarrow$ partition indices \\
    $m              $\tabto{2.20cm}$\leftarrow$ clustering algorithm index
    \label{def:indices2}
\end{definition}

\begin{definition}[{\bf Scalars}]~\\
    $\tilde{x}_i    $\tabto{2.20cm}$\leftarrow$ values of the data for each component \\
    $\tilde{b}_i    $\tabto{2.20cm}$\leftarrow$ bin index value (data) for component $i$\\
    $\tilde{k}      $\tabto{2.20cm}$\leftarrow$ serialized bin index value for each datum\\
    $b_i            $\tabto{2.20cm}$\leftarrow$ bin index value (partition) along component $i$\\
    $k              $\tabto{2.20cm}$\leftarrow$ sequential bin index value for a partition
    \label{def:scalars2}
\end{definition}

\begin{definition}[{\bf Vectors}]~\\
    ${\bf k}      $\tabto{2.20cm}$\leftarrow$ array of indices for partitions\\
    ${\bf w}      $\tabto{2.20cm}$\leftarrow$ array of weights (populations) for partitions\\
    ${\bf k'}     $\tabto{2.20cm}$\leftarrow$ array of partition indices for maximal weights
    \label{def:arrays2}
\end{definition}

\begin{definition}[{\bf Matrices}]~\\
    ${\bf \tilde{X}}        $\tabto{2.20cm}$\leftarrow$ Data values, size: $(N,N_D)$, requires: $\tilde{x}_i$\\
    ${\bf \tilde{B}}        $\tabto{2.20cm}$\leftarrow$ Bin indices, size: $(N,N_D)$, requires: $\tilde{b}_i$\\
    ${\bf X}                $\tabto{2.20cm}$\leftarrow$ Partition values, size: $(N_P,N_D)$, requires: $x_i$ \\
    ${\bf B}                $\tabto{2.20cm}$\leftarrow$ Partition bin indices, size: $(N_P,N_D)$, requires: $b_i$\\
    ${\bf \Delta b}_i       $\tabto{2.20cm}$\leftarrow$ Partition bin index differences, size: $(N_P,N_P)$, requires: $b_i$\\
    ${\bf \Delta R}         $\tabto{2.20cm}$\leftarrow$ Bin based Euclidean distance between two partitions, size: $(N_P,N_P)$, requires: ${\bf \Delta b_i}$\\
    ${\bf NN1}              $\tabto{2.20cm}$\leftarrow$ First Nearest Neighbor, size: $(N_P,N_P)$, requires: ${\bf \Delta b_i}$\\
    ${\bf \Delta L}         $\tabto{2.20cm}$\leftarrow$ Path Length - distance between two partitions, size: $(N_P,N_P)$, requires: ${\bf NN1}$\\
    ${\bf \Delta L2}_T      $\tabto{2.20cm}$\leftarrow$ True Path Length - minimal distance two partitions, size: $(N_P,N_P)$, req: ${\bf NN1}$\\
    ${\bf \Delta L}_C \equiv {\bf P_C}       $\tabto{2.20cm}$\leftarrow$ Path Length count - \# of steps between two partitions, size: $(N_P,N_P)$, req: ${\bf NN1}$\\
    ${\bf \Delta L1}        $\tabto{2.20cm}$\leftarrow$ L1-norm between two partitions, size: $(N_P,N_P)$, requires: ${\bf \Delta b_i}$\\
    ${\bf \Delta SL1}       $\tabto{2.20cm}$\leftarrow$ Summed L1-norms for all steps along the path, size: $(N_P,N_P)$, requires: ${\bf \Delta L1}$\\
    ${\bf \Delta SL1_{min}} $\tabto{2.20cm}$\leftarrow$ Minimal Summed L1-norms steps along the path, size: $(N_P,N_P)$, requires: ${\bf \Delta L1}$\\
    ${\bf \Delta SL1}_T     $\tabto{2.20cm}$\leftarrow$ True Summed L1 - distance along L2 true path , size: $(N_P,N_P)$, req: ${\bf \Delta SL1, SL1_{min}}$\\
    ${\bf SL1VAR}           $\tabto{2.20cm}$\leftarrow$ Summed L1-norm variance path to true path, size: $(N_P,N_P)$, requires: ${\bf \Delta SL1, SL1_{T}}$\\
    ${\bf LOS}              $\tabto{2.20cm}$\leftarrow$ Line-Of-Sight condition  of two partitions, size: $(N_P,N_P)$, requires: ${\bf \Delta L, SL1}$\\
    ${\bf LAPNN1}           $\tabto{2.20cm}$\leftarrow$ Numerical Laplacian using $NN1$, size: $(N_P,N_P)$, requires: ${\bf NN1}$\\
    ${\bf LAPLOS}           $\tabto{2.20cm}$\leftarrow$ Numerical Laplacian using LOS, size: $(N_P,N_P)$, requires: ${\bf LOS}$\\
    ${\bf LAPGAU}           $\tabto{2.20cm}$\leftarrow$ Numerical Laplacian of Gaussian, size: $(N_P,N_P)$, requires: ${\bf \Delta R, \sigma}$\\
    ${\bf CONN}             $\tabto{2.20cm}$\leftarrow$ Connection between two partitions, size: $(N_P,N_P)$, requires: ${\bf NN1}$\\
    ${\bf CLUS_m}           $\tabto{2.20cm}$\leftarrow$ Cluster for technique $m$ with members for each row per cluster, size: $(N_{CL},N_{C,m})$
    \label{def:matrices2}
\end{definition}
For matrices used to represent partition calculations $(N_P,N_P)$, each row represents one partition and each column represents all other partitions.

\begin{figure*}[t!]
	\centering
	\subfigure[] {
    	\label{fig:l-1}
		\includegraphics[width=0.20\linewidth]{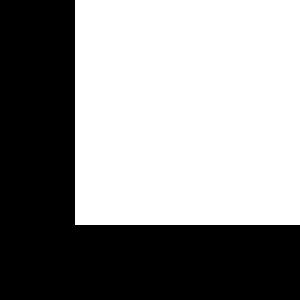} }
	\subfigure[] {
    	\label{fig:plus-1}
		\includegraphics[width=0.20\linewidth]{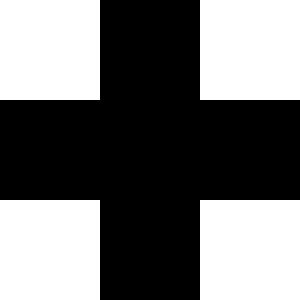} }
	\subfigure[] {
    	\label{fig:plus-2}
		\includegraphics[width=0.20\linewidth]{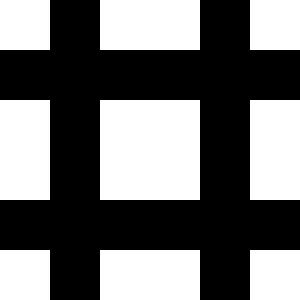} } \\
	\subfigure[] {
    	\label{fig:concentric-1}
		\includegraphics[width=0.20\linewidth]{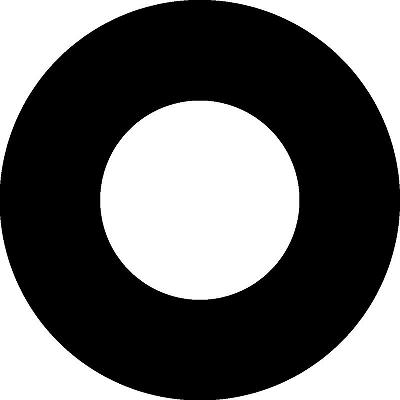} }
	\subfigure[] {
    	\label{fig:concentric-2}
		\includegraphics[width=0.20\linewidth]{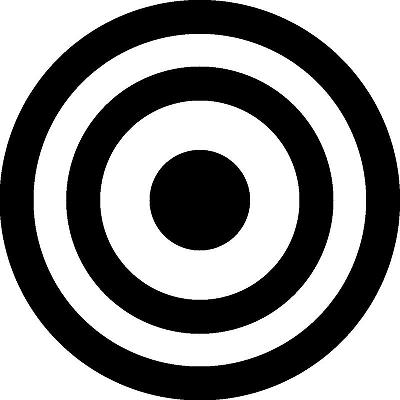} }
	\subfigure[] {
    	\label{fig:flame-1}
		\includegraphics[width=0.20\linewidth]{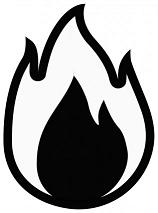} }\\
	\subfigure[] {
    	\label{fig:flame-2}
		\includegraphics[width=0.15\linewidth]{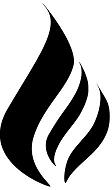} }
	\subfigure[] {
    	\label{fig:flame-3}
		\includegraphics[width=0.50\linewidth]{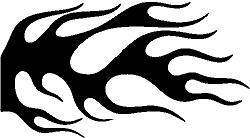} } \\
	\label{fig:testbank-1}
	\caption{Test bank of 8 shapes: L, Plus-1, Plus-2, Concentric-1, Concentric-2, Flame-1, Flame-2, Flame-3. }
\end{figure*}

\subsection{Data Test Cases - 12 Shapes}
\label{sec:test-bank-1}
\begin{figure*}[t!]
	\centering
	\subfigure[] {
    	\label{fig:data2d1-1}
		\includegraphics[width=0.37\linewidth]{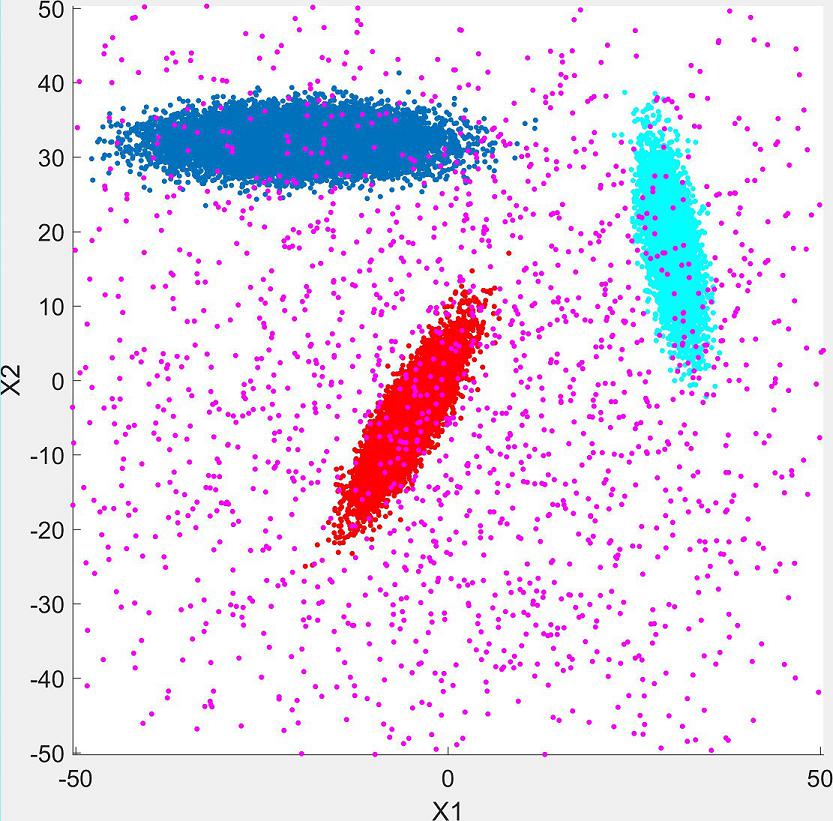} }
\qquad\hskip -0.3cm
	\subfigure[] {
    	\label{fig:data2d2-1}
		\includegraphics[width=0.37\linewidth]{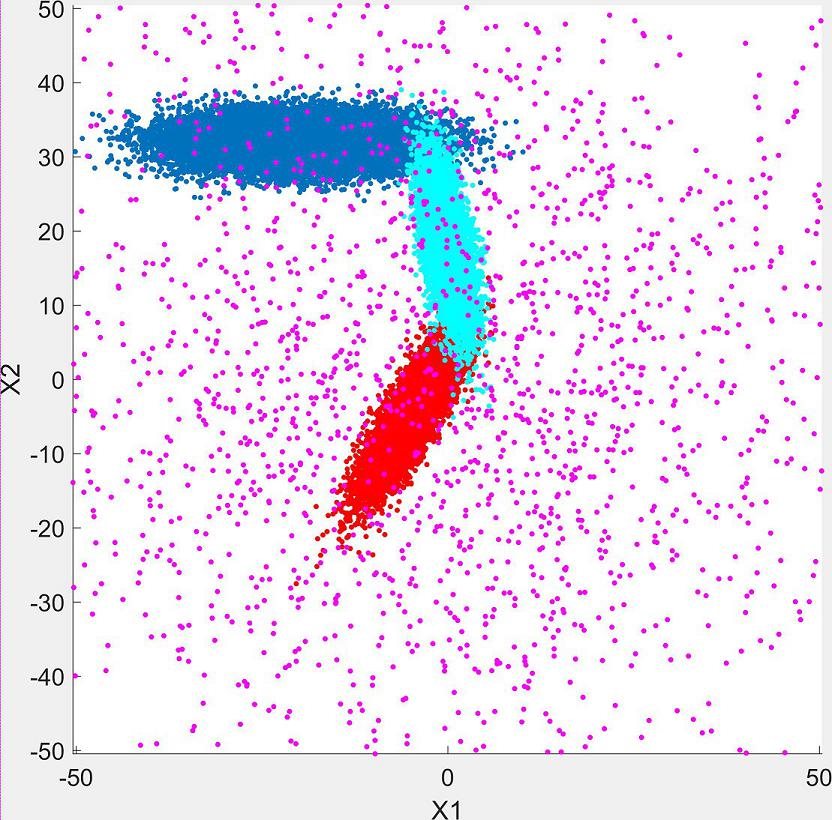} } \\
\vspace{-0.2cm}
	\subfigure[] {
    	\label{fig:data3d1-1}
		\includegraphics[width=0.37\linewidth]{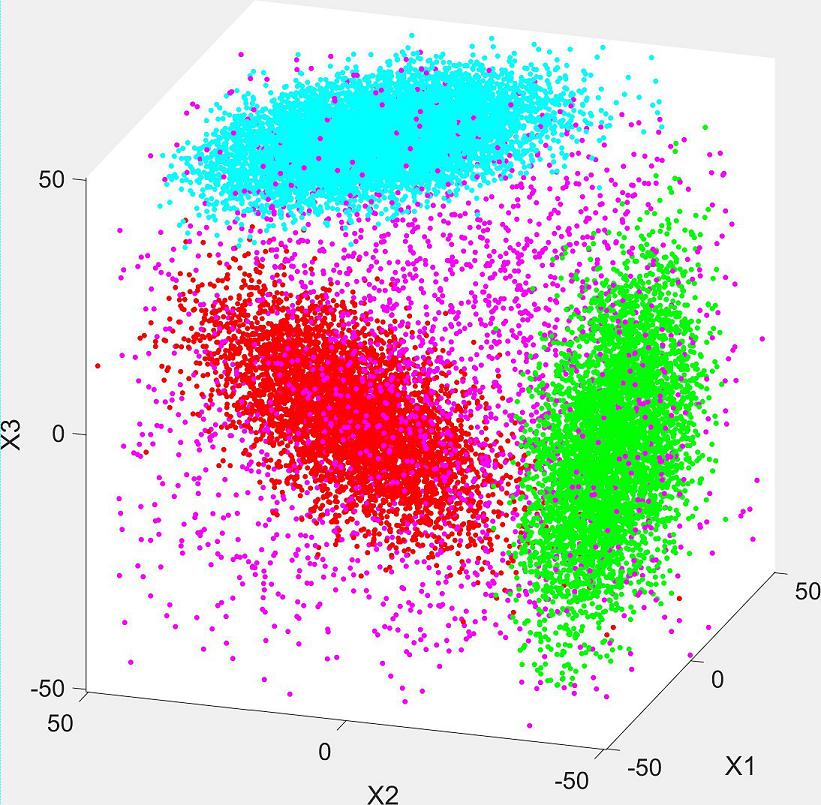} }
\qquad\hskip -0.3cm
	\subfigure[] {
    	\label{fig:data3d2-1}
		\includegraphics[width=0.37\linewidth]{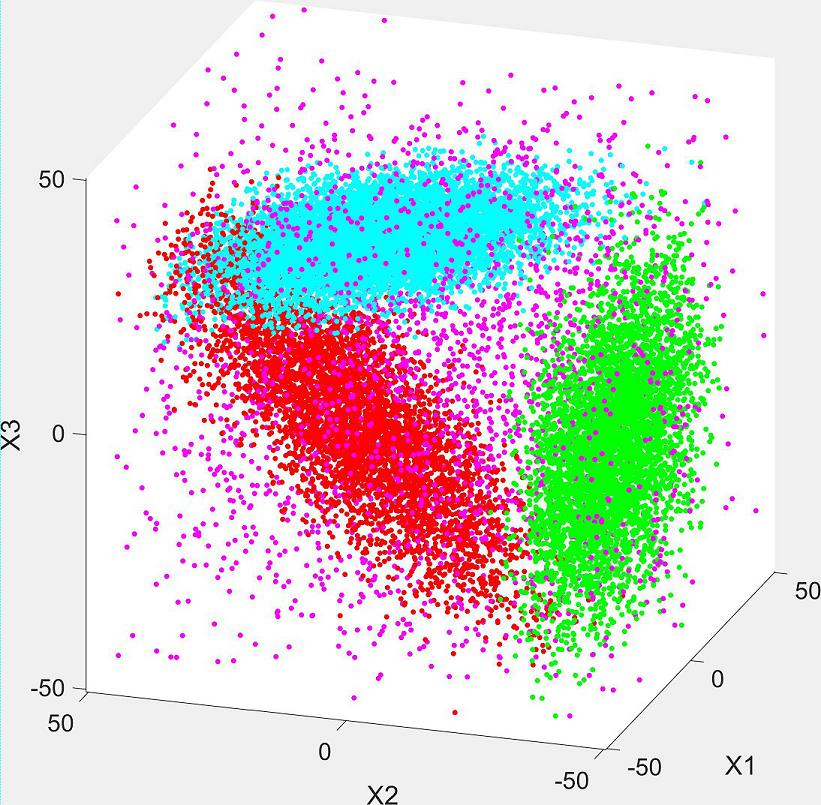} }
\vspace{-0.2cm}
	\label{fig:testbank-2}
	\caption{Test bank of 4 distributions: Data2D-1 (a), Data2D-2 (b), Data3D-1 (c) and Data3D-2 (d). }
\end{figure*}

\begin{table}[htpb]
\label{tab:testbank}
\centering
{\small
\begin{tabular}{|l|r|cccccc|}\hline
                &   & \multicolumn{6}{c|}{Test Bank Data Sets} \\ \hline
  Labels        & \# & dim & size (pixels/pts) &   connected & symmetry &  plateau  & filamentary  \\ \hline \hline
 L              &  1 &  2D &     1200x1200     &    $\surd$  &    X     &  $\surd$  &      X       \\
 Plus1          &  2 &  2D &     1200x1200     &    $\surd$  &  $\surd$ &  $\surd$  &      X       \\
 Plus2          &  3 &  2D &     1200x1200     &    $\surd$  &  $\surd$ &  $\surd$  &      X       \\
 Concentric1    &  4 &  2D &     1200x1200     &    $\surd$  &  $\surd$ &  $\surd$  &      X       \\
 Concentric2    &  5 &  2D &     1200x1200     &      X      &  $\surd$ &  $\surd$  &      X       \\
 Flame1         &  6 &  2D &     1200x1200     &    $\surd$  &    X     &  $\surd$  &      X       \\
 Flame2         &  7 &  2D &     1200x1200     &      X      &    X     &  $\surd$  &      X       \\
 Flame3         &  8 &  2D &     1200x1200     &    $\surd$  &    X     &  $\surd$  &  $\surd$     \\
 Data2D-1       &  9 &  2D &      200,000      &      X      &    X     &    X      &  $\surd$     \\
 Data2D-2       & 10 &  2D &      200,000      &  $\surd$/X  &    X     &    X      &  $\surd$     \\
 Data3D-1       & 11 &  3D &      200,000      &      X      &    X     &    X      &  $\surd$     \\
 Data3D-2       & 12 &  3D &      200,000      &  $\surd$/X  &    X     &    X      &  $\surd$     \\ \hline \hline
\end{tabular}
}
\caption{Test bank data sets.}
\end{table}
A test bank of data sets was used to develop the algorithms for this study comprised of various shapes, both connected and disconnected as well as a point cloud in both 2D and 3D.  In each of the point clouds, four gaussian distributions were placed near one another, with three densely populated regions and a fourth low density gaussian which spans the domain.  The point clouds were further varied in 2D by creating two differing sets, one where the three dense set of points are clearly separated from each other and another set where the three dense populations have overlapping regions.   Similarly in 3D, two point clouds were made where the first has the three high density regions fully separated and the second has two of the three overlapping with a lone third set.  Figures 3,4 illustrate the test bank used.  For each test, a differing feature was sought to examine.  Table 1 lists the testbank set as well as the features sought to examine in each case.  The first test is the simple {\em L} as discussed in section \ref{sec:simple-L}.  The {\em Plus1} and {\em Plus2} cases are extensions to the {\em
L} case where the symmetry of the algorithms can be understood as well as how the routines respond to void regions in the data, {\em Plus2} case.  {\em Concentric1} and {\em Concentric2} test how the routines respond to curved domains with symmetry and whether the domain is connected or not.  {\em Flame1, Flame2} and {\em Flame3} test how asymmetry is dealt with as well as connected versus disconnected regions.  {\em Flame3} also tests how well ``tendrils'' or filamentary data is handled.  As a test of a 2D point cloud, {\em Data2D-1} and {\em Data2D-2} test how well four gaussian point clouds can be clustered for the case of three separated clusters,{\em Data2D-1}, and three close-by clusters, {\em Data2D-2}, where the fourth gaussian is evenly distributed across the domain as noise.   {\em Data3D-1} and {\em Data3D-2} show the point cloud in 3D of four gaussian distributions similar in definition to the 2D cases, where the first are three disconnected elliptical distributions with a fourth acting as a background of noise, while the second shows the same three elliptical distributions moved closer to one another such that two of the tails overlap.  For all cases other than the point clouds, the data is derived from an image, where a binary set of points is established for all 8-bit grey-scale values above 100 (1) or below (0).  The image sizes when possible are 1200x1200, unless the aspect ratio prevented that exact size.  The point clouds are based on four distributions with a summed value of 200,000 points.  Figures \ref{fig:l-1}-\ref{fig:data3d2-1}  show the test bank in this order:  {\em L, Plus1, Plus2, Concentric1, Concentric2, Flame1, Flame2, Flame3, Data2D-1, Data2D-2, Data3D-1, Data3D-2}.


\section{Clustering Algorithms}
\label{sec:algorithms-1}

This section discusses the clustering algorithms used in this approach.  Some techniques are standard approaches, but several are variations on existing techniques with new approaches.  The new approaches involve changing the distance metric used from a traditional L2-norm to a path length along a grid of partitions.  Along with investigating path length based clustering, a Line-Of-Sight, LOS, criteria is also developed.  An alternative approach to using spectral clustering is also used, utilizing a different set of eigenvectors to establish clusters, as well as an alternative to the traditional Laplacian operator.  Once all twenty-six clustering techniques are used to assign a cluster identity, an overall cluster identity is given to each data based on the consensus of the set of techniques, similar to how ensemble modeling reduces systemic errors for simulation.  Table 2 lists the twenty-six techniques used in this study.

\subsection{K-Means and K-Medoids Clustering - KMEANS, KMEDOIDS}
\label{sec:kmeanskmedoids-1}

K-means is a well established clustering technique, seeking from a data set, the lowest possible distance to a set of mean positions based on the overall positions of the data.  As an input, the user is required to provide a number of means to seek $(k\equiv N_{sought})$, at which point, the algorithm proceeds to find exactly that number of mean positions, regardless of whether the data actually cluster into as many groups.  As a result, k-means suffers from an inability to ``stop early'' in its search for clusters.  Silhouette plots help determine the appropriate number of means to seek, making this process computationally expensive as it requires several passes to find reasonable clustering.  Finally, data is not always meaningful as a continuous valued set, making the concept of a ``mean'' of the data invalid.  Examples include data sets based on finite categorizations or possible mixed data sets of discreet and continuous data.  Finally, the mean location may be located outside of the data set, making the position of the mean difficult to interpret in terms of the data axes, examples include a concentric distribution, where two differing data sets are interweaved, yet share similar mean locations in the space.  Several variants on k-means exist which address many of its problems, however, its interpretation remains problematic.

K-medoids is similar in its approach to clustering as k-means, yet assigns positions based on locations of data points {\em within the set} and not mean locations (medoids).    Further, k-medoids minimizes the dissimilarity for each medoid compared to other data points within a proposed cluster compared to other clusters.  As a result, k-medoids tend to provide more meaningful cluster definitions which are less sensitive to noise \citet{kaufman2009}.   Initially, a number of medoids is sought after $(k\equiv N_{sought})$, however, once k-medoids has found the cluster definitions that have the least dissimilarity, the search ends, allowing the algorithm to stop early.

By shifting the analysis from individual datum to partitions with weights, the K-means and K-medoids algorithms are adjusted to accommodate the weighted bins.  All calculations for distance between two partitions are multiplied by the weight of each partition and any centroid calculation must be treated as a weighted value.

\subsection{Maxima Clustering - Global And Path Length}
\label{sec:maxima-clustering}
Section \ref{sec:partitions} discussed how the initial data set is reduced to a smaller set of multi-dimensional bins referred to as ``partitions''.  Each partition has a unique location in the bin address space formed from the bin index given along each component.  Further, each partition has a ``weight'' which is equivalent to the population of data located in that particular bin.

Clustering by proximity is a standard approach to finding which data are similar to one another based on how close the data are within the space.   This approach can be problematic in that data from one distribution can be near data from a separate distribution, yet might be clustered together due to their mutual closeness.  Two algorithms are applied that cluster partitions based on the weights (population) of the partitions and their proximity to a nearby local maxima of weights.   A cluster which is {\em global} in scope has a clustering relationship which does not require the partitions be in direct contact to one another, in contrast to the {\em path length} analysis which requires a cluster to be contiguous.   The global scheme can associate partitions together even if they are not connected to one another by using the Euclidean distance between partitions as the distance metric.  In contrast, the path length algorithm only associates partitions that are connected to one another via a path and uses the path length as the distance metric.   Partitions in both schemes are assigned to peaks in the weight distribution based on the distance from any given partition to a local maxima as well as the slope to the nearest peak. Local maxima, peaks and slopes are defined in the appendix \ref{sec:local-max-1}.  A peak and all of the partitions  associated  with it are then assigned a cluster identification number.

\begin{table}[t!]
\label{tab:clustering}
\centering
{\footnotesize
\begin{tabular}{|l|r|cccccccc|}\hline
                     &   & \multicolumn{8}{c|}{Clustering Algorithms} \\ \hline
  Labels             & \#  & connected & balanced &   LOS   & sens. noise &  fixed N &   ID   & Cluster & Group  \\ \hline \hline
 KMEANS              &  1  &     X     &    X     &    X    &       X     &  $\surd$ &    X   & $\surd$ &   X    \\
 KMEDOIDS            &  2  &     X     &    X     &    X    &       X     &  $\surd$ &    X   & $\surd$ &   X    \\
 MAXGLOB             &  3  &     X     &    X     &    X    &       X     &     X    &    X   & $\surd$ &   X    \\
 MAXPATHL            &  4  &  $\surd$  &    X     &    X    &       X     &     X    &    X   & $\surd$ &   X    \\
 CONN                &  5  &  $\surd$  &    X     &    X    &       X     &     X    &    X   &    X    &   X    \\
 LOS-MAXVIS          &  6  &  $\surd$  &    X     & $\surd$ &       X     &     X    & $\surd$&    X    &   X    \\
 LOS-MUTUAL          &  7  &  $\surd$  & $\surd$  & $\surd$ &       X     &     X    & $\surd$&    X    &   X    \\
 SPEC-NN1-12-2DHIST  &  8  &  $\surd$  &    X     &    X    &    $\surd$  &     X    &    X   &    X    &   X    \\
 SPEC-NN1-12-KMEANS  &  9  &  $\surd$  &    X     &    X    &    $\surd$  &  $\surd$ &    X   &    X    &   X    \\
 SPEC-NN1-12-KMEDS   &  10 &  $\surd$  &    X     &    X    &    $\surd$  &  $\surd$ &    X   &    X    &   X    \\
 SPEC-NN1-23-2DHIST  &  11 &  $\surd$  & $\surd$  &    X    &    $\surd$  &     X    &    X   &    X    & $\surd$\\
 SPEC-NN1-23-KMEANS  &  12 &  $\surd$  & $\surd$  &    X    &    $\surd$  &  $\surd$ &    X   &    X    & $\surd$\\
 SPEC-NN1-23-KMEDS   &  13 &  $\surd$  & $\surd$  &    X    &    $\surd$  &  $\surd$ &    X   &    X    & $\surd$\\
 SPEC-LOS-12-2DHIST  &  14 &  $\surd$  &    X     & $\surd$ &    $\surd$  &     X    & $\surd$&    X    &   X    \\
 SPEC-LOS-12-KMEANS  &  15 &  $\surd$  &    X     & $\surd$ &    $\surd$  &  $\surd$ & $\surd$&    X    &   X    \\
 SPEC-LOS-12-KMEDS   &  16 &  $\surd$  &    X     & $\surd$ &    $\surd$  &  $\surd$ & $\surd$&    X    &   X    \\
 SPEC-LOS-23-2DHIST  &  17 &  $\surd$  & $\surd$  & $\surd$ &    $\surd$  &     X    & $\surd$&    X    & $\surd$\\
 SPEC-LOS-23-KMEANS  &  18 &  $\surd$  & $\surd$  & $\surd$ &    $\surd$  &  $\surd$ & $\surd$&    X    & $\surd$\\
 SPEC-LOS-23-KMEDS   &  19 &  $\surd$  & $\surd$  & $\surd$ &    $\surd$  &  $\surd$ & $\surd$&    X    & $\surd$\\
 SPEC-GAU-12-2DHIST  &  20 &     X     &    X     &    X    &       X     &     X    &    X   & $\surd$ &   X    \\
 SPEC-GAU-12-KMEANS  &  21 &     X     &    X     &    X    &       X     &  $\surd$ &    X   & $\surd$ &   X    \\
 SPEC-GAU-12-KMEDS   &  22 &     X     &    X     &    X    &       X     &  $\surd$ &    X   & $\surd$ &   X    \\
 SPEC-GAU-23-2DHIST  &  23 &     X     & $\surd$  &    X    &       X     &     X    &    X   &    X    & $\surd$\\
 SPEC-GAU-23-KMEANS  &  24 &     X     & $\surd$  &    X    &       X     &  $\surd$ &    X   &    X    & $\surd$\\
 SPEC-GAU-23-KMEDS   &  25 &     X     & $\surd$  &    X    &       X     &  $\surd$ &    X   &    X    & $\surd$\\
 LMH-POS             &  26 &     X     &    X     &    X    &       X     &  $\surd$ &    X   & $\surd$ &   X    \\ \hline \hline
\end{tabular}
}
\caption{Clustering techniques for 26 algorithms highlighting requirements, pros and cons in each case.  Some algorithms required partitions to be connected in order to search for clustering within the connected region.  Clusters can be feature driven or can even the distribution of cluster assignments (balanced, group).  LOS is a criteria for some clustering, which in turn can help identify data distributions.  Finally, some algorithms treat isolated partitions on equal footing with larger connected regions, making the clustering sensitive to noise.}
\end{table}

\subsubsection{Global Maxima Clustering - MAXGLOB}
\label{sec:glob-1}
The MAXGLOB scheme seeks to form clusters based on the proximity of a partition to local maxima among the population of the partitions.  Appendix \ref{sec:local-max-1} details how peaks are determined relative to nearby partitions using two values, the Euclidean distance to a peak, ${\bf \Delta R}$, as well as the slope between the value of the current partitions weight and the weight of a nearby local maxima.  When several peaks are considered to be associated with a partition, a hierarchy exists among the peaks to properly associate any given partition to the correct peak.  The results of the MAXGLOB scheme is shown in Sec. \ref{sec:results-glob-pathl} and most closely resemble clustering done via k-means as well as k-medoids.  The advantage of using the MAXGLOB approach is that clustering is performed on partitions ($\mathcal{P}$) and not directly on the original dataset ($\mathcal{D'}$), giving it a significant computational advantage.

\subsubsection{Path Length Maxima Clustering  - MAXPATHL}
\label{sec:pathl-1}
The MAXPATHL scheme seeks to form clusters based on the closest distance from partitions to local maxima in the populations of partitions using the path length, ${\bf \Delta L}$, as the distance metric. \cbk The clustering relationship is based on proximity {\em within a connected} group of partitions, making the algorithm sensitive to the shape of the distribution by using the path length.  \cbk  Similar to the MAXGLOB algorithm, the path length and slope are used in conjunction to determine which peak is best associated with a partition where ${\bf
\Delta R}$ is simply replaced by ${\bf \Delta L}$.  Figure 7 in Sec. \ref{sec:results-glob-pathl} compare the results of MAXGLOB and MAXPATHL to KMEANS, KMEDOIDS.

\subsection{Clustering via Connection - CONN}
\label{sec:connection-1}
In cases where local clusters of partitions are sparsely found within the data space, a simple clustering algorithm is to determine which partitions are connected to one another using first nearest neighbor steps, ${\bf NN1}$.  All partitions connected to one another, ${\bf CONN}$,  are given a single cluster ID.  Any partitions that are alone, disconnected from all other partitions, are given their own cluster ID of zero.  By giving all isolated partitions a cluster ID of zero, allows for a simple cluster designation based on a logical value $({\bf CONN}>0)$ which then distinguishes between connected regions in the partition data space and isolated regions, which may be viewed as noise.  An isolated partition may contain a large amount of data, in which case, care should be taken to consider the weight of each partition such that a better method of looking for noise would be the condition $(({\bf CONN}=0)AND({\bf w}<\Theta_{noise}))$, where $\Theta_{noise}$ is a user defined threshold.

The simplest technique for determining whether a partition is connected to another within a set of partitions is to begin with the first nearest neighbor matrix, ${\bf NN1}$.  If this matrix is block diagonal, then all partitions within a block on the diagonal are connected to each other and are disconnected from any partitions in a separate block.  A ${\bf NN1}$ matrix is most likely not block diagonal initially, however, it can be readily made block diagonal using a Dulmage-Mendelsohn decomposition (\citeyear{dmperm}). \cbk  A ${\bf NN1}$ matrix is most likely not block diagonal initially, however, can be readily made block diagonal using the decomposition of \cite{dmperm}.   \cbk In this approach, for the symmetric ${\bf NN1}$ matrix, a series of row and column interchanges occurs until the matrix has become block diagonal, at which point, the blocks represent subsets of connected partitions.

\cbk Many problems in clustering are made difficult by having multiple clusters in close proximity to one another, yet not being contiguous.  \cbk  By using the connection criteria, as long as the data space has been resolved well enough by choosing appropriate bin sizes along each axis, clusters should be resolvable to within the resolution established across the space.

\subsection{Clustering by Line-Of-Sight - LOS}
\label{sec:los-1}
Clustering by Line-Of-Sight is motivated by the idea that data within a convex hull has a higher chance of being correlated than data separated by distance and visibility.   Although distributions of data may follow traditional gaussian shapes, it is also possible for a distribution to be bent within the data space.  Distributions of data can appear to follow a curve within the space which may simply reflect a functional dependence between one or more variables, yet it may also form when two or more distributions have means near one another and tails that overlap, leading to the appearance of a single bent distribution.  Clustering via {\bf CONN} will associate all data in the bent distribution, however, checking whether two data lie within a convex hull more closely associates those data with one another.  A Line-Of-Sight, LOS, criteria determines which partitions are convex to one another. As examples, figures \ref{fig:data2d1-1}-\ref{fig:data3d2-1} illustrate several distributions which have both convex regions as well as overlapping tails of distributions.  In this discussion, the term {\em visibility} refers to the number of partitions that are LOS to a specific partition.  A detailed discussion is given in appendix \ref{sec:los-3} of the algorithms used.

 The ${\bf LOS}$ matrix is formed where each row represents a partition and each column represents all other partitions where a logical value indicates whether the two are LOS.   The ${\bf LOS}$ matrix is symmetric, further, squaring the ${\bf LOS}$ matrix, ${\bf LOS^2}$, gives a matrix whose values along each row tally the number of partitions which are mutually LOS to one another.  The ${\bf LOS^2}$ matrix may contain non-zero values for partitions that share a mutual visibility with any given row, but are not LOS to the current partition.  An example will given next to expand on this idea.  In order to handle these entries in ${\bf LOS^2}$ that are not present in the ${\bf LOS}$ matrix, a Hadamard product is taken between ${\bf LOS}$ and ${\bf LOS^2}$ yielding a third matrix, ${\bf LLL}$.   The LOS criteria is detailed in the appendix \ref{sec:los-3}.

\subsection{Simple Example: L}
\label{sec:simple-L}
\begin{figure*}[t!]
\label{fig:simplel}
\begin{minipage}{0.35\linewidth}
	\centering
		\includegraphics[width=1.0\linewidth]{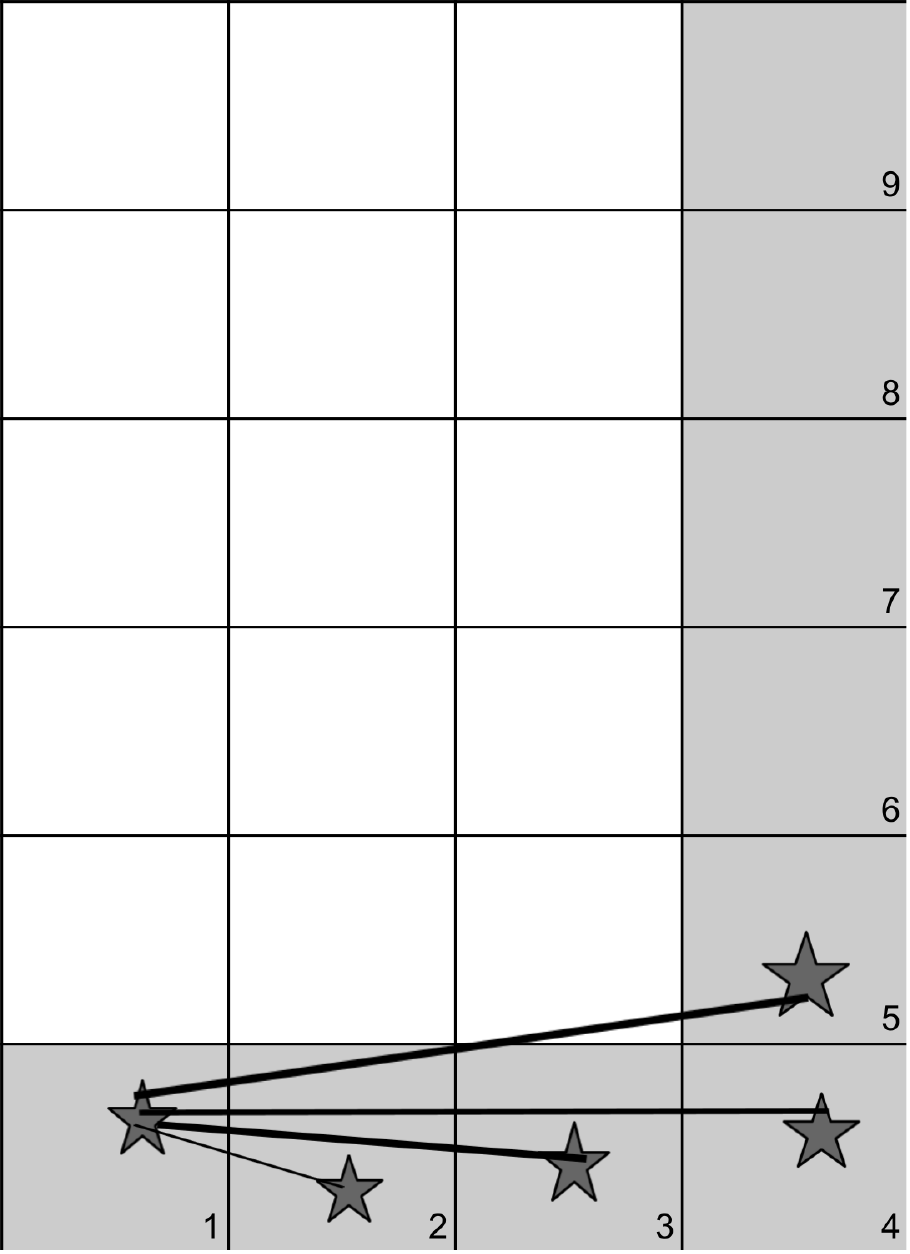}
	\label{fig:simple-domain-01}
\end{minipage}
\begin{minipage}{0.28\linewidth}
\centering
{\footnotesize
\[
\label{eqn:NN1-01}
\begin{pmatrix}
    1 &      1 &      0 &      0 &      0 &      0 &     0 &      0 &      0  \\
    1 &      1 &      1 &      0 &      0 &      0 &     0 &      0 &      0  \\
    0 &      1 &      1 &      1 &      1 &      0 &     0 &      0 &      0  \\
    0 &      0 &      1 &      1 &      1 &      0 &     0 &      0 &      0  \\
    0 &      0 &      1 &      1 &      1 &      1 &     0 &      0 &      0  \\
    0 &      0 &      0 &      0 &      1 &      1 &     1 &      0 &      0  \\
    0 &      0 &      0 &      0 &      0 &      1 &     1 &      1 &      0  \\
    0 &      0 &      0 &      0 &      0 &      0 &     1 &      1 &      1  \\
    0 &      0 &      0 &      0 &      0 &      0 &     0 &      1 &      1
\end{pmatrix}
\]
$NN1$ matrix
\[
\label{eqn:LOS-L2}
\begin{pmatrix}
    5 &      5 &      5 &      5 &      5 &      3 &     3 &      3 &     3     \\
    5 &      5 &      5 &      5 &      5 &      3 &     3 &      3 &     3     \\
    5 &      5 &      9 &      9 &      9 &      7 &     7 &      7 &     7     \\
    5 &      5 &      9 &      9 &      9 &      7 &     7 &      7 &     7     \\
    5 &      5 &      9 &      9 &      9 &      7 &     7 &      7 &     7     \\
    3 &      3 &      7 &      7 &      7 &      7 &     7 &      7 &     7     \\
    3 &      3 &      7 &      7 &      7 &      7 &     7 &      7 &     7     \\
    3 &      3 &      7 &      7 &      7 &      7 &     7 &      7 &     7     \\
    3 &      3 &      7 &      7 &      7 &      7 &     7 &      7 &     7
\end{pmatrix}
\]
$LOS^2$ matrix
}
\end{minipage}
\hspace{0.5cm}
\begin{minipage}{0.28\linewidth}
\centering
{\footnotesize
\[
\label{eqn:LOS-L}
\begin{pmatrix}
    1 &      1 &      1 &      1 &      1 &      0 &     0 &      0 &     0     \\
    1 &      1 &      1 &      1 &      1 &      0 &     0 &      0 &     0     \\
    1 &      1 &      1 &      1 &      1 &      1 &     1 &      1 &     1     \\
    1 &      1 &      1 &      1 &      1 &      1 &     1 &      1 &     1     \\
    1 &      1 &      1 &      1 &      1 &      1 &     1 &      1 &     1     \\
    0 &      0 &      1 &      1 &      1 &      1 &     1 &      1 &     1     \\
    0 &      0 &      1 &      1 &      1 &      1 &     1 &      1 &     1     \\
    0 &      0 &      1 &      1 &      1 &      1 &     1 &      1 &     1     \\
    0 &      0 &      1 &      1 &      1 &      1 &     1 &      1 &     1
\end{pmatrix}
\]
``L'' $LOS$ matrix
\centering
\[
\label{eqn:LOS-L3}
\begin{pmatrix}
    5 &      5 &      5 &      5 &      5 &      0 &     0 &      0 &     0     \\
    5 &      5 &      5 &      5 &      5 &      0 &     0 &      0 &     0     \\
    5 &      5 &      9 &      9 &      9 &      7 &     7 &      7 &     7     \\
    5 &      5 &      9 &      9 &      9 &      7 &     7 &      7 &     7     \\
    5 &      5 &      9 &      9 &      9 &      7 &     7 &      7 &     7     \\
    0 &      0 &      7 &      7 &      7 &      7 &     7 &      7 &     7     \\
    0 &      0 &      7 &      7 &      7 &      7 &     7 &      7 &     7     \\
    0 &      0 &      7 &      7 &      7 &      7 &     7 &      7 &     7     \\
    0 &      0 &      7 &      7 &      7 &      7 &     7 &      7 &     7
\end{pmatrix}
\]
$LOS^2 \cdot LOS \equiv LLL$ matrix
}
\end{minipage}
\caption{Simple example to illustrate the ideas behind LOS clustering.  This ``L'' shaped domain has nine populated bins.  Starting from the bottom left to right then and moving upwards,the bins are numbered initially by rasterizing the domain, then contracting the bin indices to simply number from $\#$1-9.   Beginning with bin $\#$1, the line-of-sight bins are indicated by the starred ($\bigstar$) bins.  Matrices calculated for the simple example, the ${\bf NN1}$ (upper left) ${\bf LOS}$ (upper right), ${\bf LOS^2}$(lower left), ${\bf LOS^2} \cdot {\bf LOS} \equiv {\bf LLL}$(lower right).}
\end{figure*}

A simple example can serve to realize the idea of these matrices and how they interact with one another.   Consider a simple distribution of partitions forming a 6x4 grid connected to each
other in an ``L'' configuration as  shown in Fig. 6.  In order to follow the serialization of partitions given in Sec. \ref{sec:defs-2}, it helps to invert the L from right to left ($\leftrightarrow$).  Numbering the partitions along the bottom row first (1-4), then along the vertical right side (5-9).   In this case, there are only nine partitions connected to each other, requiring a 9x9 matrix to represent the information.  As each partition is connected to all of the other partitions, the ${\bf CONN}$ matrix is full, with values of one.  The ${\bf NN1}$ matrix reflects which partitions share a common geometrical feature.  The ${\bf LOS}$, ${\bf LOS^2}$ and ${\bf LLL}$ matrices show which partitions can ``see'' another partition.  Note that partition five is visible to partition one, meaning that partitions can see the edges of one another.  From the matrices shown, partitions (3,4,5) form a cluster with the maximal visibility, followed by partitions (6,7,8,9) then (1,2) (LOS-MAXVIS).  Partitions (3,4,5,6,7,8,9) form a cluster with the highest mutual visibility followed by (1,2) with the lowest (LOS-MUTUAL).

In the language established in previous sections, the data points are pixels gathered from an image where two axes are used to describe the image, making the data space 2D.  The data are integers for the $(x_1,x_2)$ positions ranging from $(0...4],(0...6]$ respectively.  Each axis is binned horizontally (4 bins) and vertically (6 bins). The data bin addresses are:\\
\tabto{4cm}$[(1,1),(1,2),(1,3),(1,4),(2,4),(3,4),(4,4),(5,4),(6,4)]$\\
and the serial bin address ranges from $[1...24]$, of which only: $[1,2,3,4,8,12,16,20,24]$ are filled.  These filled bins become the partitions, labeled by the partition sequential bin addresses: $[1...9]$.  The partition space is also 2D with partition bin addresses having the same values as the data bin addresses.

\subsubsection{LOS Clustering With Maximal Visibility - LOS-MAXVIS}
\label{sec:los-maxvis}
The ${\bf LOS}$ matrix contains for each row the logical status of which partitions are LOS to the current partition. Further, the ${\bf LLL}$ matrix shows the number of mutually visible partitions within LOS of the current.  From the ${\bf LLL}$ matrix, two values can be used to determine clustering using LOS.  The highest value in the ${\bf LLL}$ matrix indicates which partitions are within LOS of the {\em most} other partitions.  These highest valued ${\bf LLL}$ partitions have the {\em maximal visibility}, LOS-MAXVIS, of the set of partitions that are LOS.  An example would be any partition that is located at an intersection of several distributions of partitions.  Consider the test cases:  {\em L} and {\em Plus1}, where the corner of the {\em L} and the center of the {\em Plus1} will have maximal visibility.

Clustering by LOS-MAXVIS is achieved by taking the diagonal from ${\bf LLL}$, which gives the total visibility of each partition.  A histogram of the visibility is shown in Fig.
7.  The horizonal axis indicates the number of partitions LOS to others.  Being a histogram, the vertical axis is the number of partitions sharing a common visibility
value.  Starting from the maximal value of the visibility, a cluster is formed by taking all partitions sharing the maximum or nearby, defined by including all bins in the histogram starting from the topmost until a minimum in the bins is reached.  In the case of the simple L, the most visible partitions are the corner partitions with a ${\bf LLL}$ value of nine.  Further clusters are then identified by taking partitions associated with the next highest visibility bin in the histogram, beginning where the last set left off, and including all partitions with successively lower visibilities until the next minimum in the bins is reached.  This process continues until the set of partitions is fully associated with clusters.

\cbk
\subsubsection{LOS Clustering With Mutual Visibility - LOS-MUTUAL}
\label{sec:los-mutual}
The ${\bf LLL}$ matrix can alternatively be used to cluster partitions with the {\em highest mutual visibility} (LOS-MUTUAL) by selecting clusters with the most common shared ${\bf LLL}$ value instead of the maximal value.  In this manner, clusters are formed around partitions that can mutually see each other the most.  Clustering by LOS-MUTUAL is achieved by first creating
a histogram of ${\bf LLL}$ values over all partitions.  Figure 6 shows the histogram of visibility values for the {\em Data3D2} test case.  The horizonal axis indicates the
 visibility of partitions.  The vertical axis is number of partitions sharing a common visibility value.  Starting from the bin with the most frequent visibility, a cluster is formed by seeking the minima on both sides of the peak in the histogram nearest the most populated bin.  Once the lower and upper bins are found, all partitions which have any visibility values in ${\bf LLL}$ within this range are clustered together.  After identification, these clusters are removed from future searches by setting the rows and columns of the ${\bf LLL}$ matrix to zero for these partitions.  The process is repeated until all partitions are identified.   LOS-MUTUAL clustering finds the largest sets of partitions that are LOS to each other first, then searches for the next largest set of partitions that do not include the first set and so on.

In the case of the simple L, the highest mutually visible partitions are the partitions forming the long arm of the L.  For the Chesapeake Bay, all partitions with an visibility between 1700 up to 3000 are included in the first cluster found.

\cbk
\begin{figure}[t!]
	\label{fig:gathering1vis}
	\centering
	\includegraphics[width=0.80\linewidth]{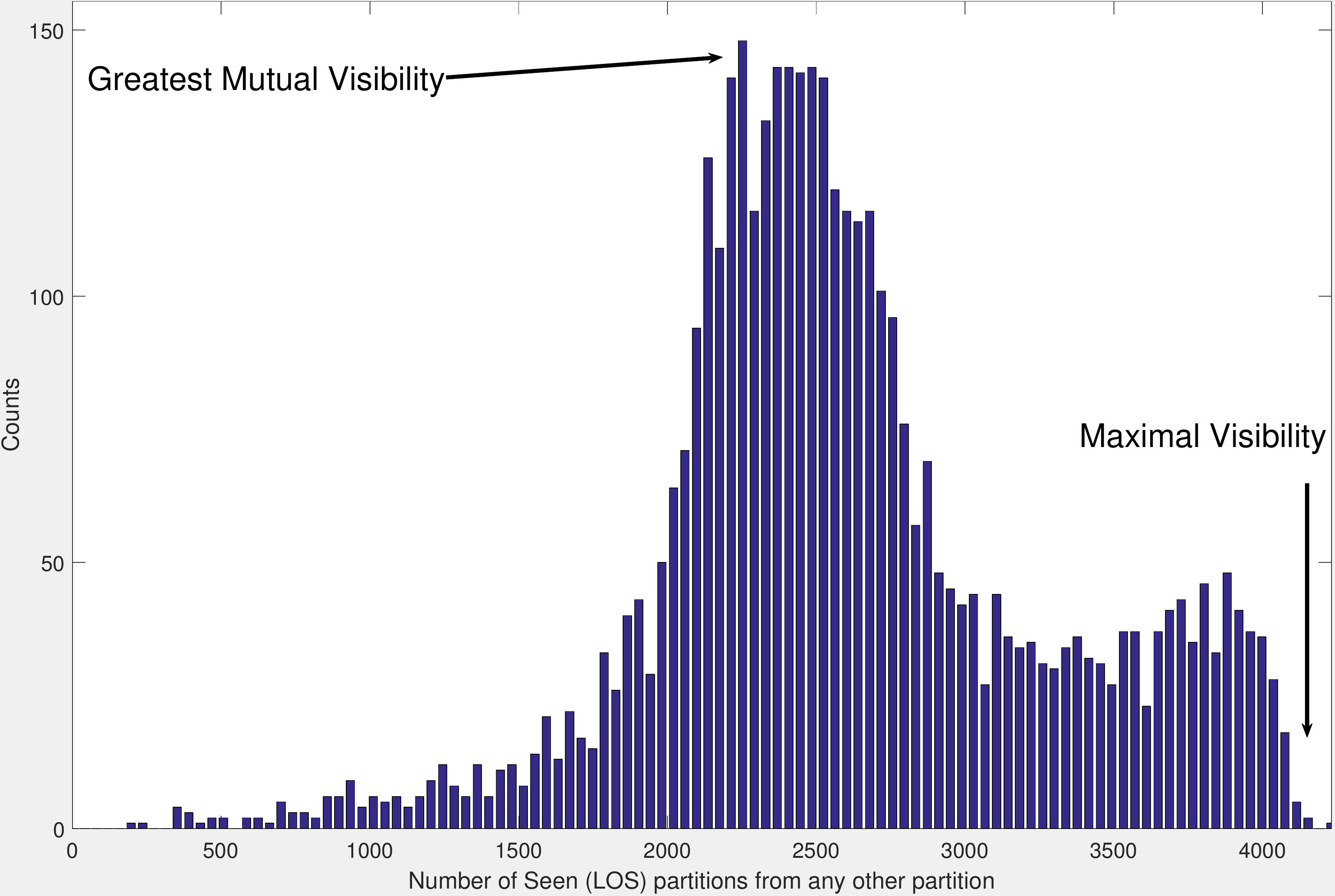}
	\caption{Histogram of ${\bf LLL}$ values, the mutual visibility from one partition to all others for the {\em Data 3D-2} case.  The horizonal axis indicates the number of mutually seen LOS to this partition.  The vertical axis is number of other partitions LOS to the current.  From the horizontal axis, this partition is visible to $\approx4100$ other partitions, giving it a MAXVIS of 4100.  Also, this partition is part of a group of partitions that maximally and mutually see each other with a visibility of $\approx2200$ whose set size is $\approx150$ partitions in membership, starting off the search for MAXMUT clusters.  In both cases, the starting point for the cluster search defines which other partitions are near to the goal, either maximal visibility or highest mutual visibility.  The clusters are formed by grouping LOS partitions to the current around the closest peak in the histogram, where peaks are separated by the basins.}
\end{figure}

\subsection{Spectral Clustering}
\label{sec:spec-1}
Spectral clustering [\cite{chung1997,vonLuxburg2007}] represents data as a graph, where data becomes vertices and relationships between data points are represented by edges and weights in the graph.  This analysis uses the ${\bf NN1}$ matrix, the ${\bf LOS}$ matrix as well as a Gaussian radial basis function to form a graph.  A Laplacian operator is formed by setting the degree of vertices along the diagonal of the Laplacian matrix with off diagonal elements set to a negative weight factor.  When using the ${\bf NN1}$ matrix, the Euclidean distance is given a binary value  whereas the ${\bf LOS}$ matrix uses the binary connections between partitions that are LOS to one another.  For the Gaussian radial function, the Euclidean distance between any two partitions is used as the basis to an exponential term with a constant allowing control over the degree of locality.   Radial spectral clustering differs from the other two methods in that ${\bf NN1}$ and ${\bf LOS}$ clustering require a connection to exist between two partitions, limiting the scope of how the clusters form.  Radial clustering is performed over all pairs of partitions regardless of connection.  Given two parameters, ($\sigma_{rad}, \Theta_{rad}$), the Gaussian radial approach can be limited by setting the falloff ($\sigma_{rad}$) as well as a distance cutoff ($\Theta_{rad}$ - measured in units of $\sigma_{rad}$), where all distances greater than the cutoff set the Laplacian of Gaussian term to zero.  In all cases, the analysis that follows is similar.  The eigenvectors are calculated for the Laplacian, where the lowest two eigenvectors are typically used to define a {\em new data space} using each eigenvector as a basis.  The partitions are then mapped to the eigenspace and clusters within the space are sought using novel 2D clustering techniques, either KMEANS, KMEDOIDS or a simple 2D histogram over the domain.

This analysis employs all three clustering techniques in the eigenspace as well as explores using two differing sets of eigenvectors, the lowest pair (1,2) as well as the next two lowest pair (2,3).  In the first case using eigenvectors (1,2), the first eigenmode accentuates a single large feature within the eigenspace, where the second eigenvector segments the space into two symmetric regions.   By using the next lowest pair of eigenvectors, surpassing the lowest eigenmode, the partitions are segregated differently, more evenly distributed, leading to a different interpretation of clustering.

\subsubsection{Spectral Cluster Gathering By K-Means, K-Medoids or 2D Histograms}
\label{sec:spec-3}
Once the eigenspace has been populated with the partitions, k-means, k-medoids as well as traditional 2D histograms can be used to collect the partitions and assign them to cluster IDs.   K-means and k-medoids have been discussed earlier in Sec. \ref{sec:kmeanskmedoids-1} as to their strengths and weaknesses. As an alternative approach to finding the clusters within the eigenspace, simply histogram the 2D eigenspace and assign to each non-zero bin a different cluster ID (2DHIST).  This approach has the advantage of simplicity and finds exactly the number of clusters that fill bins within the eigenspace, not requiring an initial guess as the number of possible clusters, as in the case of k-means or k-medoids, however a maximum possible count of clusters is set by the bin size of the eigenspace ($N_{spec},N_{spec}$).

\subsection{Clustering By Coarse Position (LMH-POS)}
\label{sec:lmh-pos-1}
The most obvious form of clustering is to associate a partition solely by its {\em position} (LMH-POS) using a coarse binning within the partition space.  By setting the number of bins along each dimension to three, the bins are interpreted as being {\em low, medium} or {\em high} for the values represented along each axis.  In this case, the {\em sequential partition bin index}, ${\rm {\bf k}}$, becomes the cluster ID, with the maximum number of possible clusters at $3^{N_D}$, for the three bins along each axis.   This approach is a coarse designation for clustering as it employs no complicated algorithms, and data with similar values are associated irrespective of all other factors.     When handling large data sets, this approach allows for a quick look at where the data reside within the larger space.

This approach suffers from many problems in that data in one bin will not be clustered with data from a neighboring bin no matter how close in proximity the two are to one another.   In the extreme case of using only three bins per axis, the data are characterized in the crudest sense with no refinement for the shape of a distribution or even the relative sizes of the distribution.

One advantage to this approach is that it is easy to understand, even while spanning multiple dimensions.  As an entry point for a discussion about the data, the LMH-POS approach eschews complication for simplicity and frames the discussion to evolve towards the nuances within the data set, its shape, dimension, span, relative size, etc...

\subsection{User Choice in Clustering - Variables, Thresholds, Binning}
\label{sec:user-1}
Throughout this study, several parameters have been defined which affect the outcome of the clustering algorithms.  Each of these user defined values reflects how knowledge of the data set can lead to appropriate choices for clustering.  A list of the user defined variables, thresholds and binning choices is given here:
\vspace{-0.1cm}
\begin{definition}[{\bf Variables}]~\\
    $x_i            $\tabto{2.20cm}$\leftarrow$ variables chosen forming the data space.  \\
    $N_D            $\tabto{2.20cm}$\leftarrow$ the number of dimensions. \\
\label{def:nums11}
\end{definition}
\vspace{-0.4cm}
\begin{definition}[{\bf Binning}]~\\
    $N_{B,i}        $\tabto{2.20cm}$\leftarrow$ the number of bins for component $i$. \\
    $N_{spec}       $\tabto{2.20cm}$\leftarrow$ the number of spectral clusters sought after.
\label{def:nums12}
\end{definition}
\vspace{-0.4cm}
\begin{definition}[{\bf Thresholds}]~\\
    $\Theta_{low}            $\tabto{2.20cm}$\leftarrow$ threshold for bins with low population, set either as an integer cutoff for the per bin population, or as a percentage cutoff for the sum of all low population bins compared to the total population. \\
    $\Theta_{pathmin}        $\tabto{2.20cm}$\leftarrow$ threshold similar to $\Theta_{low}$ to ignore partitions whose maximum pathcount is low, effectively removing isolated partitions. \\
    $\Theta_{overlap}         $\tabto{2.20cm}$\leftarrow$ threshold set for LOS maximum visibility gathering process requiring an percentage of overlap between smaller clusters and total visible set of partition from larger clusters, with typical values set at 90\% or above.\\
    $\sigma_{rad}            $\tabto{2.20cm}$\leftarrow$ sets distance scale used in spectral clustering with a radial basis (Gaussian), typically set at the longest distance scale in the domain, ${\bf \Delta R_{max}}$. \\
    $\Theta_{rad}            $\tabto{2.20cm}$\leftarrow$ threshold for the value for the Laplacian of the Gaussian used in spectral clustering with a radial basis.\\
    $\Theta_{cons}            $\tabto{2.20cm}$\leftarrow$ threshold for a consensus to be reached, where a simple majority is 50\%.
\label{def:nums13}
\end{definition}


\begin{figure*}[t!]
    \label{fig:comparison1}
    \centering
	\subfigure[] {
    	\label{fig:kmeans}
		\includegraphics[width=0.40\linewidth]{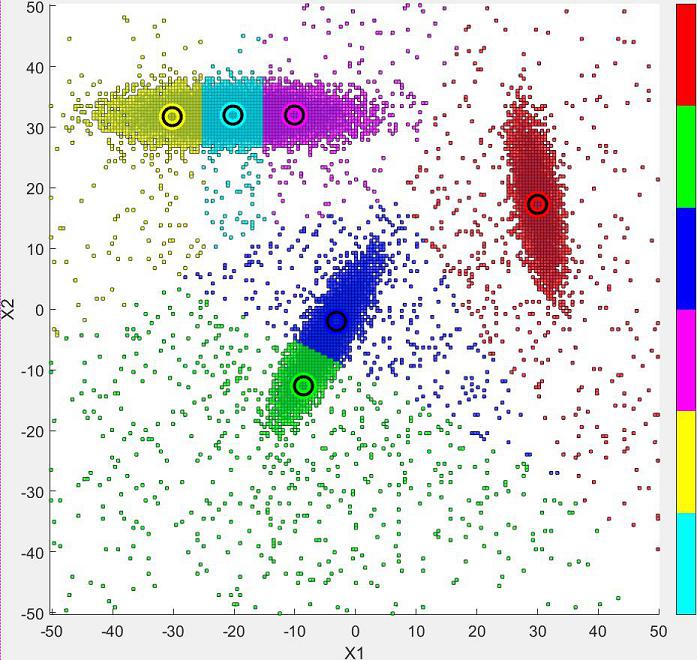} }
\qquad\hskip -0.3cm
	\subfigure[] {
    	\label{fig:kmedoids}
		\includegraphics[width=0.40\linewidth]{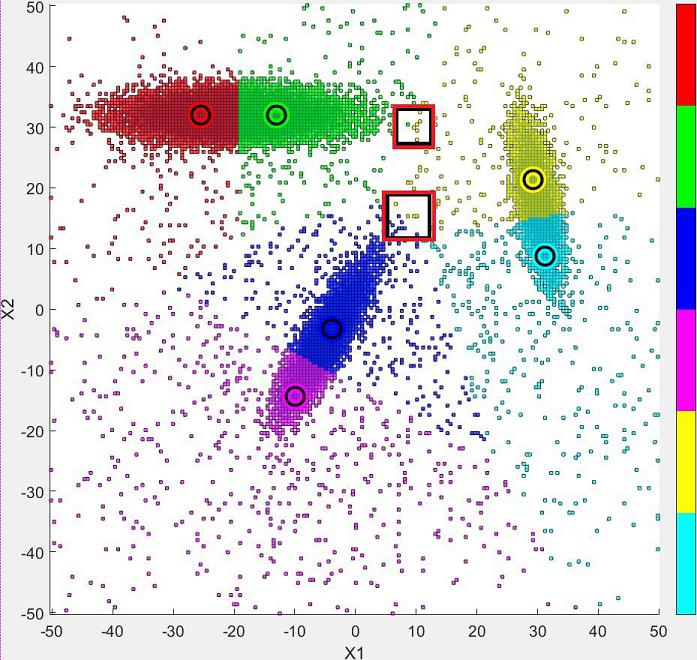} } \\
	\subfigure[] {
    	\label{fig:global}
		\includegraphics[width=0.40\linewidth]{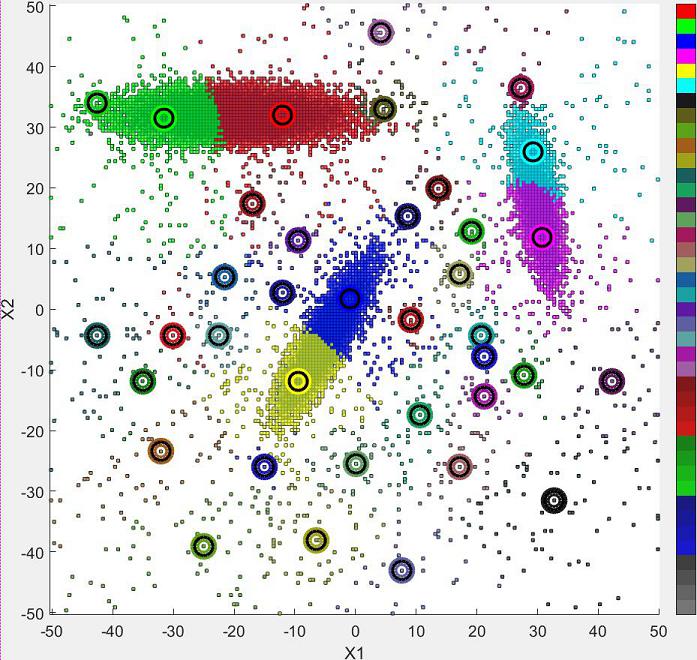} }
\qquad\hskip -0.3cm
	\subfigure[] {
    	\label{fig:pathlwgt}
		\includegraphics[width=0.40\linewidth]{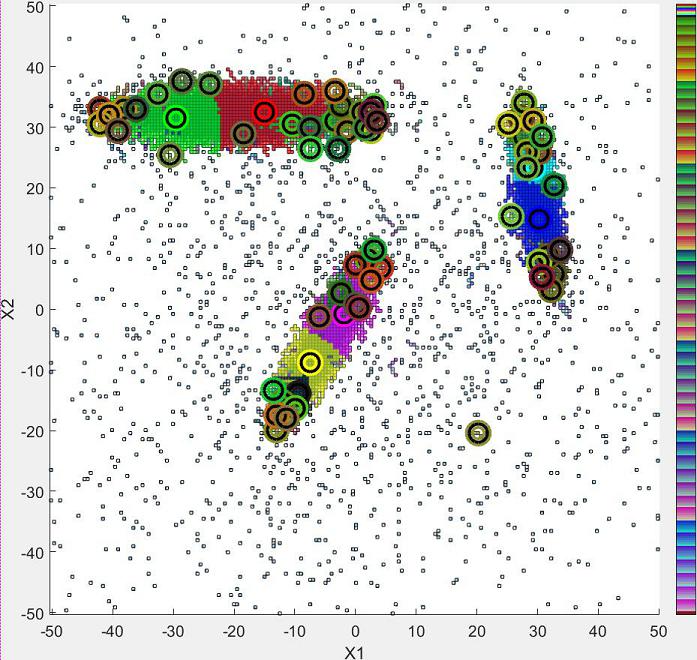} }
	\caption{Comparison of k-means \ref{fig:kmeans}, k-medoids \ref{fig:kmedoids}, MAXGLOB \ref{fig:global}, and MAXPATHL \ref{fig:pathlwgt}.  In the figure for k-medoids, two red/black squares have been added to illustrate a failure of the method, namely, elongated distributions can falsely associate with nearby data, when the data is transverse  and far from the short axis of an ellipsoidal distribution while being close to the longitudinal axis of another. }
	\label{fig:global02}
\end{figure*}

\section{Results}
\label{sec:results-1}

\subsection{Global and Path Length Maxima Clustering}
\label{sec:results-glob-pathl}
Results from MAXGLOB clustering are shown in Figs:\ref{fig:kmeans}-\ref{fig:pathlwgt}, for KMEANS, KMEDOIDS, MAXGLOB and MAXPATHL maxima clustering for the {\em Data2D-1} data set.  Global clustering performs similarly to KMEANS and KMEDOIDS, however, this approach does not require an initial guess at how many clusters may be present, which is a problem at times for k-means.  The MAXPATHL approach uses the same algorithm as MAXGLOB, however, uses the path length as the distance metric, allowing it to associate clusters following the envelope of the distribution, avoiding confusion between nearby, yet disconnected dense regions.

\begin{figure*}[t!]
    \label{fig:losmaxvismut1}
    \centering
	\subfigure[] {
    	\label{fig:maxvis1}
		\includegraphics[width=0.40\linewidth]{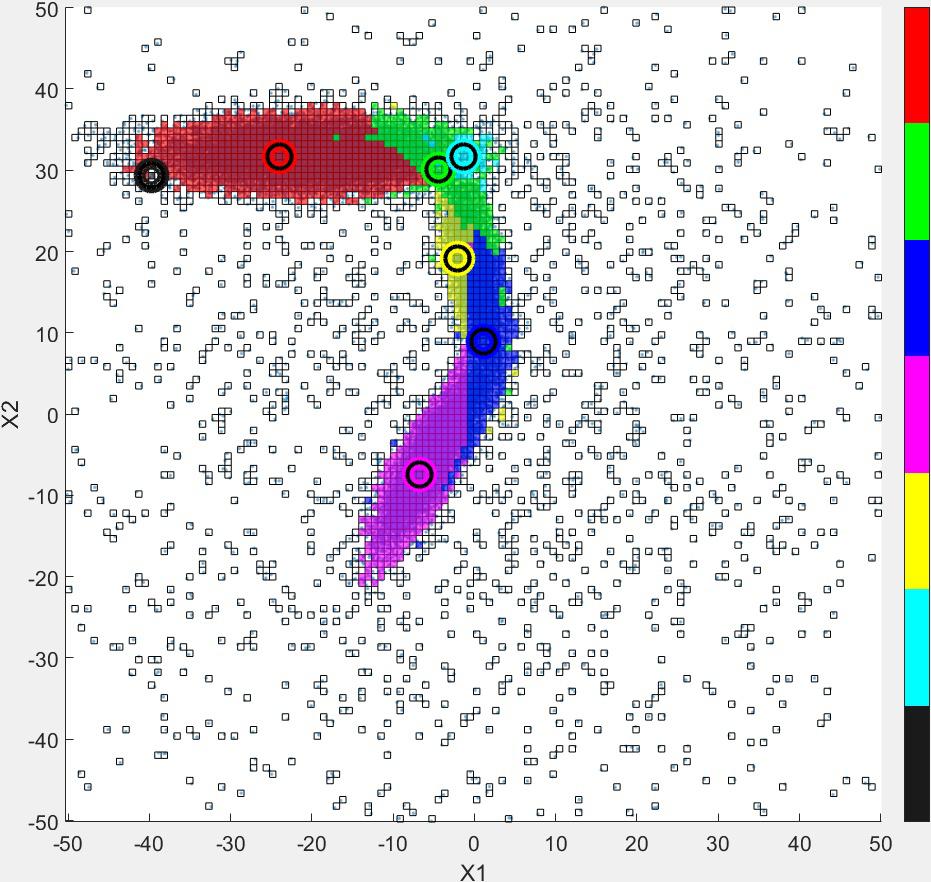} }
\qquad\hskip -0.3cm
	\subfigure[] {
    	\label{fig:mutual1}
		\includegraphics[width=0.40\linewidth]{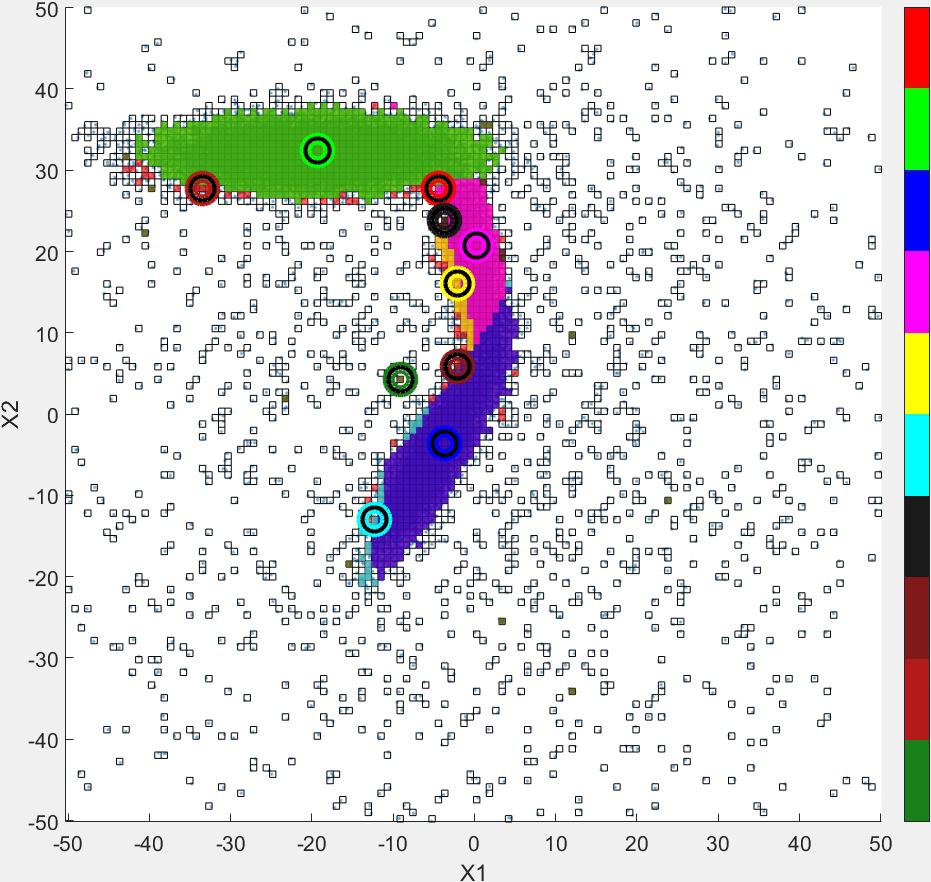} } \\
	\subfigure[] {
    	\label{fig:maxvis3}
		\includegraphics[width=0.40\linewidth]{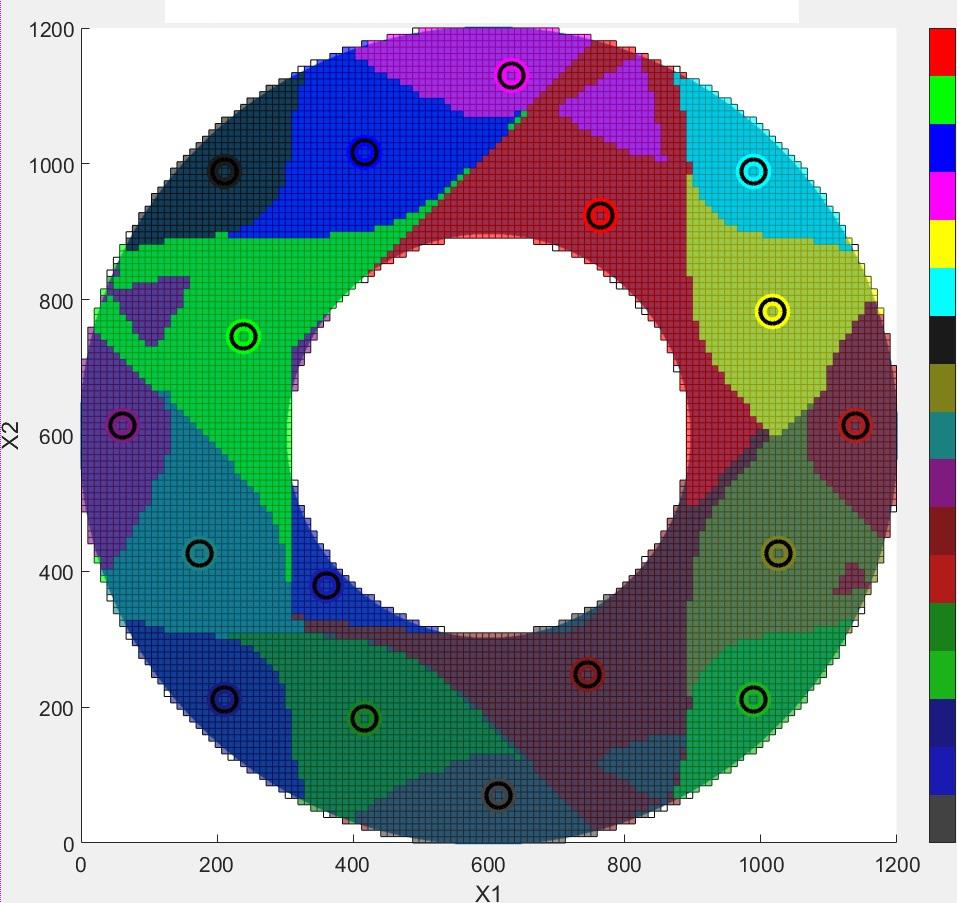} }
\qquad\hskip -0.3cm
	\subfigure[] {
    	\label{fig:mutual3}
		\includegraphics[width=0.40\linewidth]{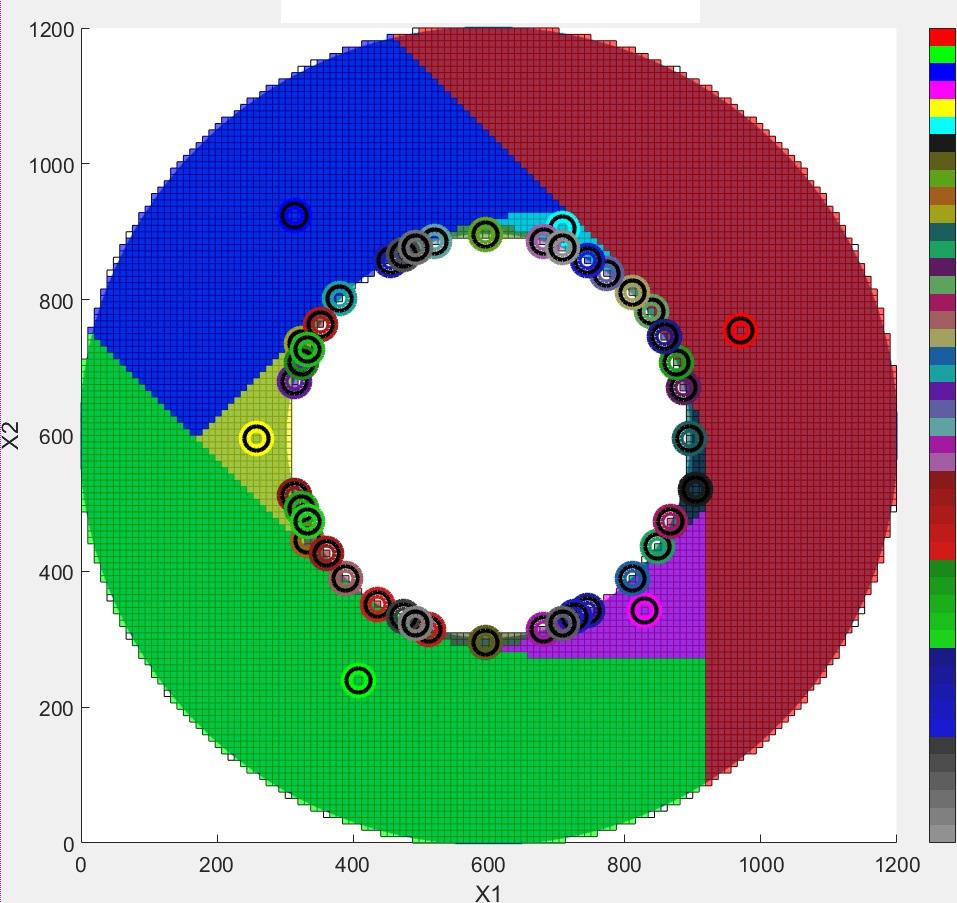} } \\
	\subfigure[] {
    	\label{fig:maxvis2}
		\includegraphics[width=0.45\linewidth]{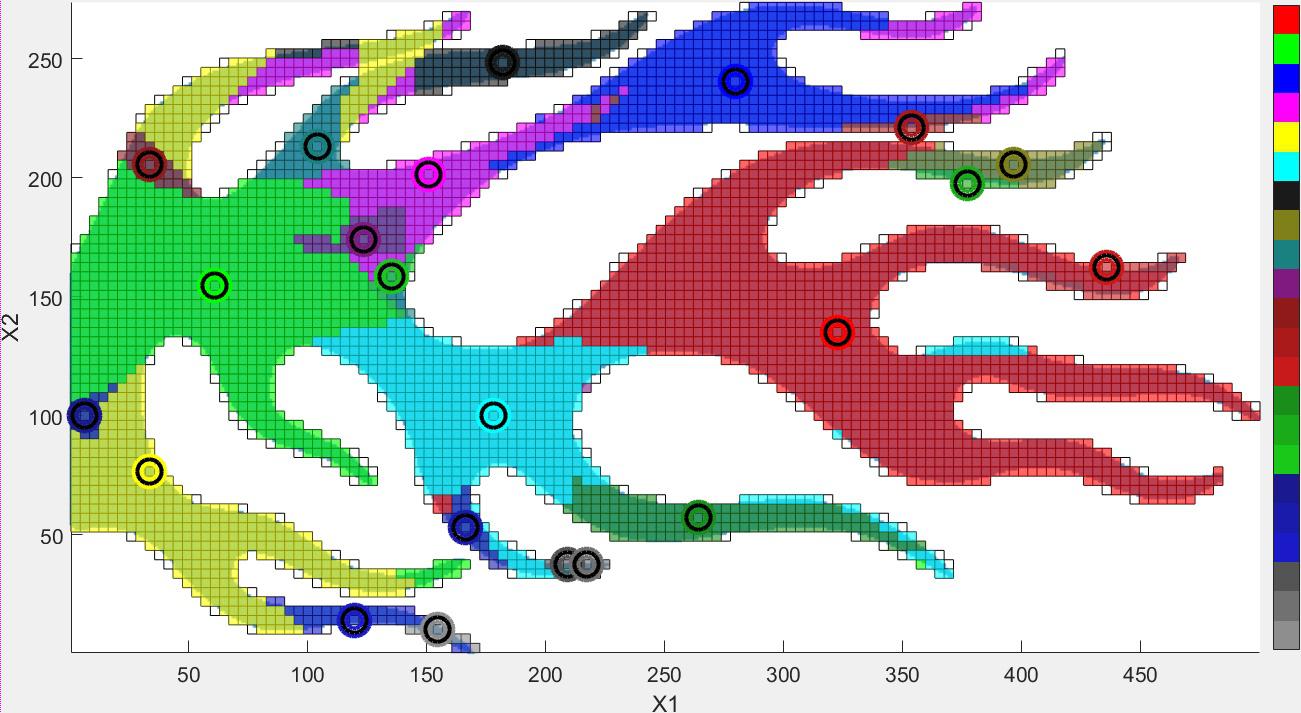} }
\qquad\hskip -0.3cm
	\subfigure[] {
    	\label{fig:mutual2}
		\includegraphics[width=0.45\linewidth]{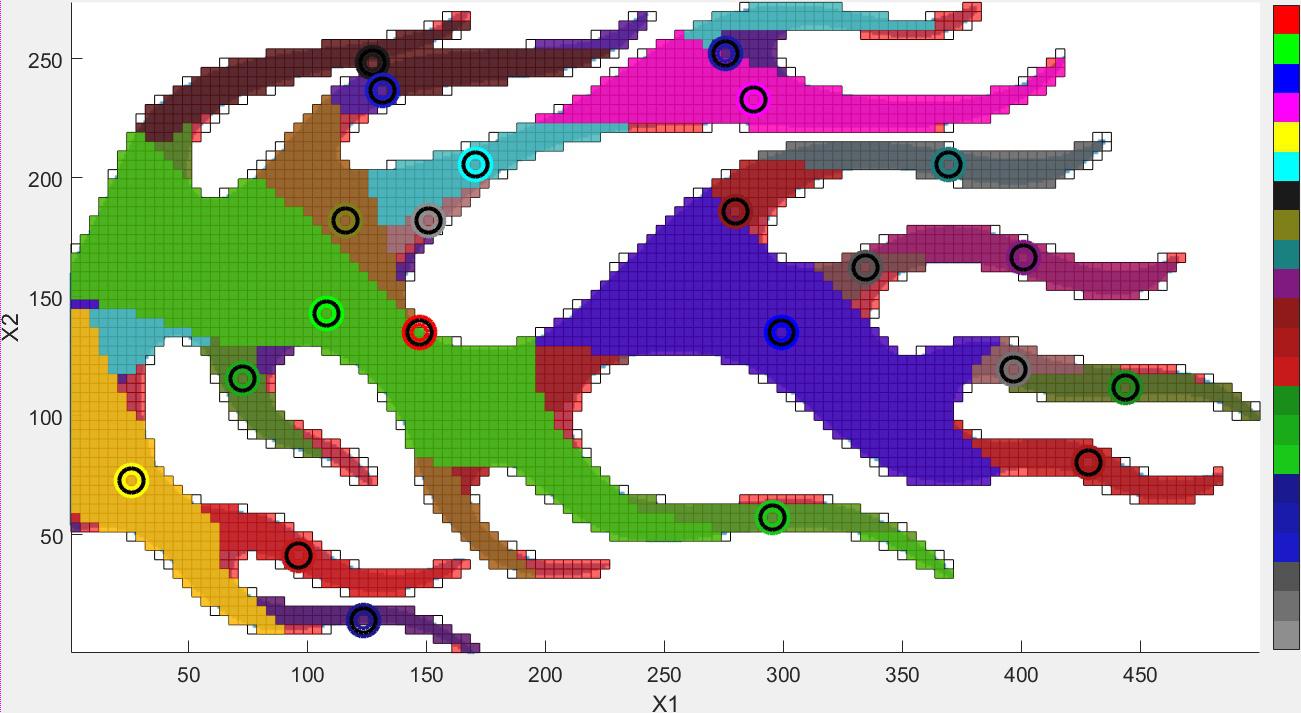} } \\
	\caption{Panels \ref{fig:maxvis1},\ref{fig:maxvis3},\ref{fig:maxvis2} illustrate LOS-MAXVIS clustering.   Panels \ref{fig:mutual1},\ref{fig:mutual3},\ref{fig:mutual2} show the LOS-MUTUAL clustering applied to the same test cases.  The test cases chosen illustrate how the LOS criteria affects ellipsoidal distributions, cases with a high degree of symmetry and filamentary cases.  }

\end{figure*}
\clearpage
\subsection{LOS Clustering - Maximal Visibility and Highest Mutual Visibility}
\label{sec:results-LOS-1}
Results from LOS-MAXVIS and LOS-MUTUAL clustering are shown in Figs:\ref{fig:maxvis1}-\ref{fig:mutual2}, for  clustering for the {\em Data2D-1} data set.  Maximum visibility finds clusters initially in regions with the highest coverage, establishes a cluster, then successively removes those partitions from further searches.  The process is repeated until all partitions are assigned to clusters.  This process can find a high number of clusters depending on how the partitions are arranged.  It is useful to then gather the large number of initial clusters found and require that the smaller clusters regroup with a larger cluster.  The condition for regrouping is that the smaller cluster share a large percentage of membership from its partitions to the larger clusters {\em visible} partition set, typically requiring more than 90\% overlap from the smaller cluster to the larger.  Figure \ref{fig:maxvis1} illustrates the property of LOS-MAXVIS to find the regions with the largest coverage, which tend to be the corners or elbows of a set of partitions.  The maximum visibility algorithm also finds symmetry in groups based on visibility as is shown in Fig. \ref{fig:maxvis3}.  LOS-MAXVIS is also useful in finding clusters along filamentary distributions as is shown in Fig. \ref{fig:maxvis2}.

LOS-MUTUAL shares many properties with LOS-MAXVIS, however, the order of grouping is performed differently.  The largest mutual visible group is clustered first, regardless of the degree of coverage.  When a partition has a visibility that is shared in common with the most amount of other partitions, a cluster is formed.  The effect is shown in Figs. \ref{fig:mutual1}-\ref{fig:mutual2}.  In this approach, the largest common group of visible partitions is assigned to a cluster first, then the next largest group and so on.  This has the effect of grabbing a single large group of partitions near one another, such as taking the long arm of the simple ``L'' example, then leaving the short arm to be taken as the second cluster, ending the search for clusters.  By comparison, the LOS-MAXVIS algorithm finds the corner region as a cluster first, then searches along the arms.  LOS-MUTUAL finds clusters along filamentary distributions, yet tends to break symmetry by taking the largest pieces first.

Figures \ref{fig:maxvis1},\ref{fig:maxvis3},\ref{fig:maxvis2} illustrate LOS-MAXVIS clustering.  In the first case ({\em Data2D2}), the maximum visibility identifies the corner regions  separately from the centers of the ellipsoids.  For the {\em Concentric1} case, maximal visibility first finds eight symmetric smaller regions hugging the outer radius, then finds interior regions extending to the inner radius, each with lesser and lesser visibility.  The last case {\em Flame3} shows how maximal visibility finds the central regions first, then finds successive filamentary regions.

Figures \ref{fig:mutual1},\ref{fig:mutual3},\ref{fig:mutual2} show the LOS-MUTUAL clustering applied to the same test cases.  In each case, the largest common visibility region is found first, then assigned a cluster ID.  Each subsequent search grabs the next largest regions to cluster until no further regions exist.  This approach identifies larger distributions then proceeds to track down the remaining ones, also finding clusters within filaments.

\subsection{Spectral Clustering}
\label{sec:results-spec}
Spectral clustering is a commonly used approach, generally using the first two eigenvectors formed from a numerical Laplacian based on a graphs first nearest neighbors.  In this study, the graph would be based on the lattice of partitions.  The definition of a neighbor can be simple, such as a geometric neighbor on a grid (${\bf NN1}$) or it may be more abstract, such as all partitions visible to one another ${\bf LOS}$.  Further, the neighborhood may be defined globally by taking the Laplacian of a Gaussian for the distance between two neighbors, using the L2-norm ${\bf \Delta R}$.

\begin{figure*}[htbp]
	\centering
	\subfigure[] {
    	\label{fig:spectral1}
		\includegraphics[width=0.35\textwidth]{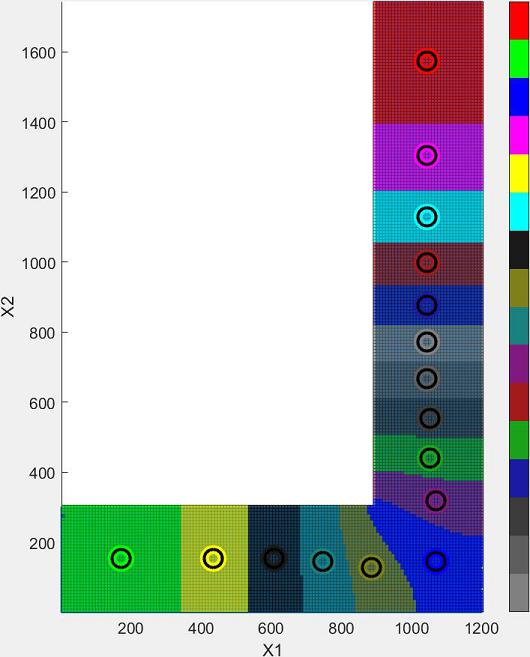} }
\qquad\hskip -0.4cm
	\subfigure[] {
    	\label{fig:spectral2}
		\includegraphics[width=0.35\textwidth]{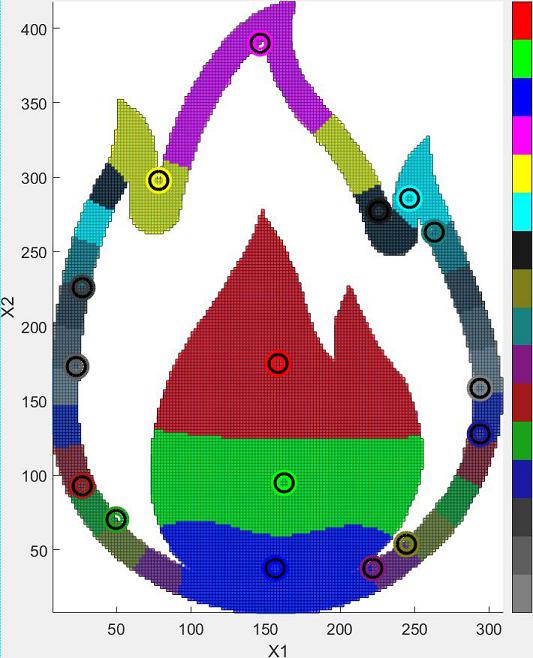} }
	\subfigure[] {
    	\label{fig:spectral3}
		\includegraphics[width=0.35\textwidth]{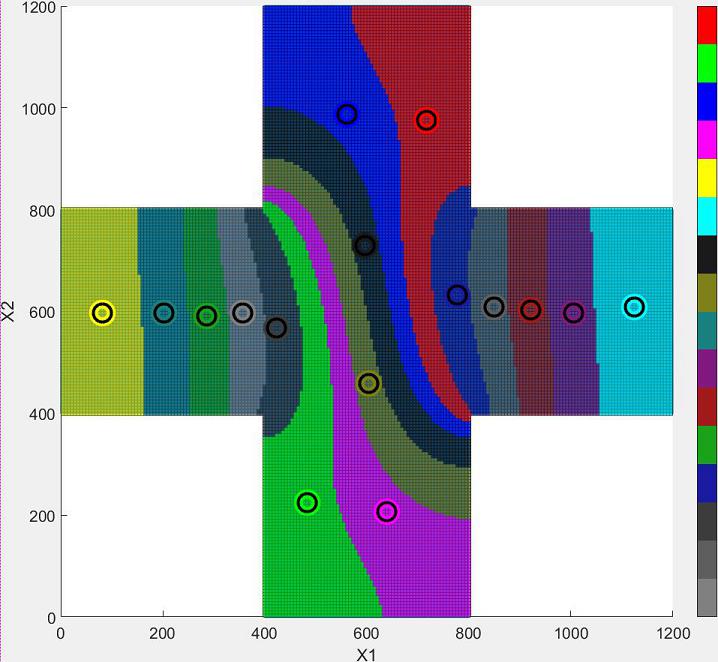} }
\qquad\hskip -0.4cm
	\subfigure[] {
    	\label{fig:spectral4}
		\includegraphics[width=0.35\textwidth]{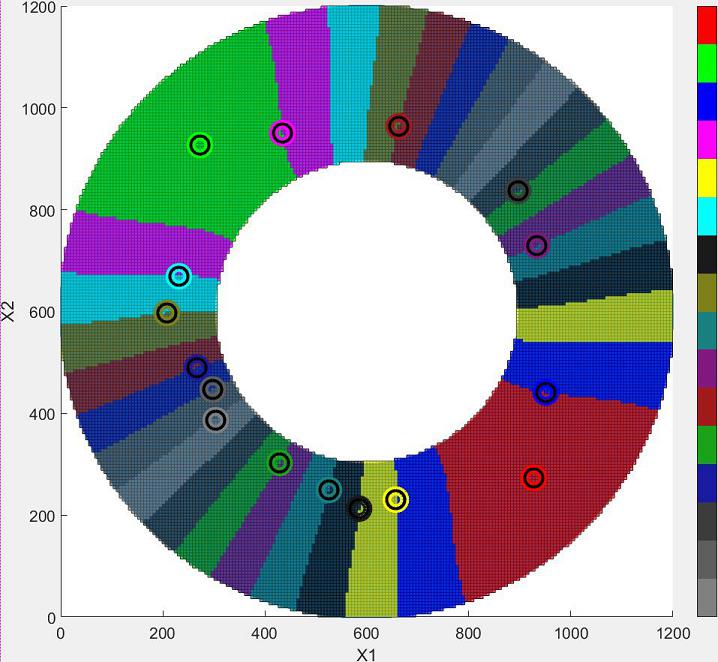} }
	\subfigure[] {
    	\label{fig:spectral5}
		\includegraphics[width=0.35\textwidth]{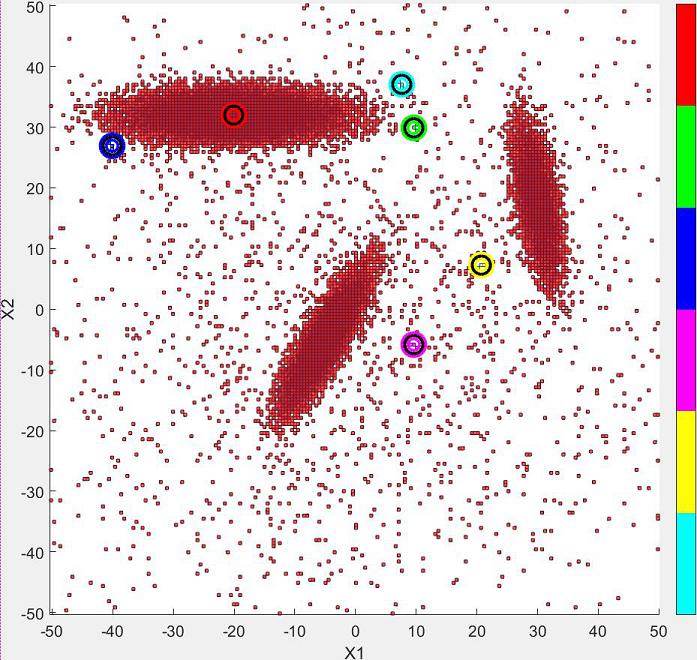} }
\qquad\hskip -0.4cm
	\subfigure[] {
    	\label{fig:spectral6}
		\includegraphics[width=0.35\textwidth]{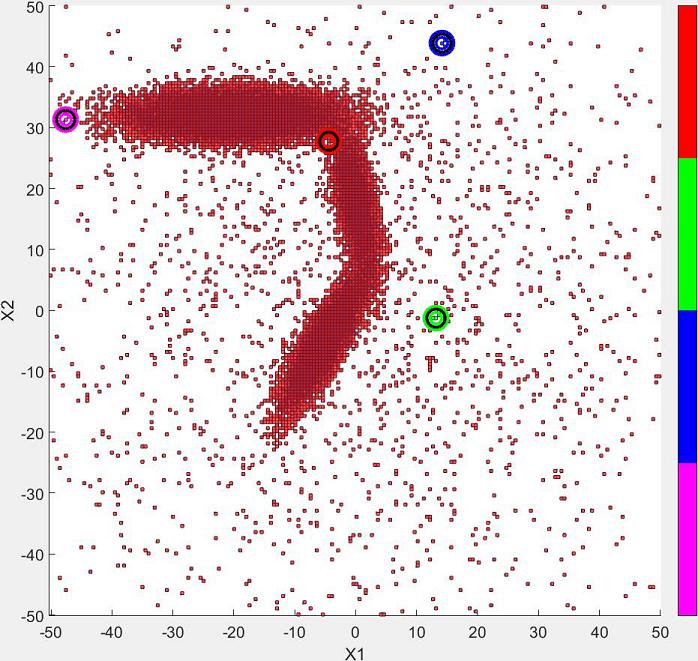} }
\vspace{-0.4cm}
	\caption{Test bank data sets for {\em L, Plus1, Concentric1, Data2D1, Data2D2} and {\em Flame1} are shown using spectral clustering with a Laplacian based on $NN1$ using eigenmodes one and two collected in the eigenspace using K-means seeking 16 possible clusters.}
	\label{fig:testbank3}
\end{figure*}

The choice of eigenvectors applied changes the interpretation of the results of clustering within the eigenspace formed from the two eigenvectors.  When using the lowest two eigenvectors to form the eigenspace, the lowest eigenvector has the longest feature scale in the eigenspace, where its values cluster near the largest feature seen in the domain of the partitions.  The result is that a large cluster is formed near the largest subregion in the partition space.  Further clusters are found at increasingly smaller feature sizes, as is shown in Figs. \ref{fig:spectral1}-\ref{fig:spectral6}.  Clustering values in the eigenspace correlates to identifying differing length scales in the position space, with oscillatory behavior along the length of any subdomains in the position space.

\begin{figure*}[htbp]
	\centering
	\subfigure[] {
    	\label{fig:spectral7}
		\includegraphics[width=0.35\textwidth]{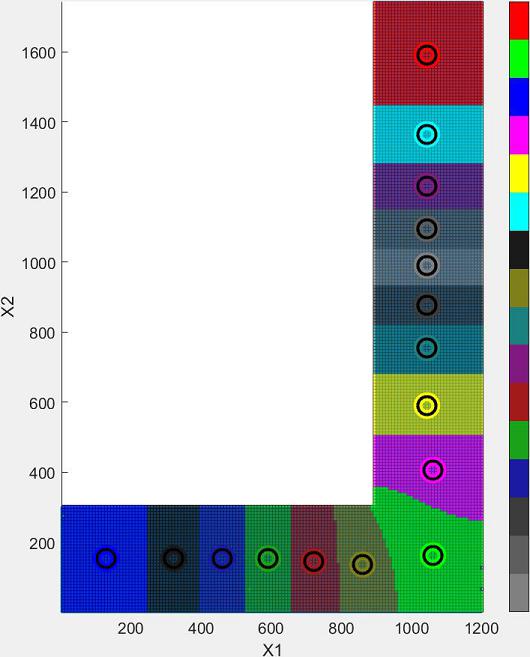} }
\qquad\hskip -0.4cm
	\subfigure[] {
    	\label{fig:spectral8}
		\includegraphics[width=0.35\textwidth]{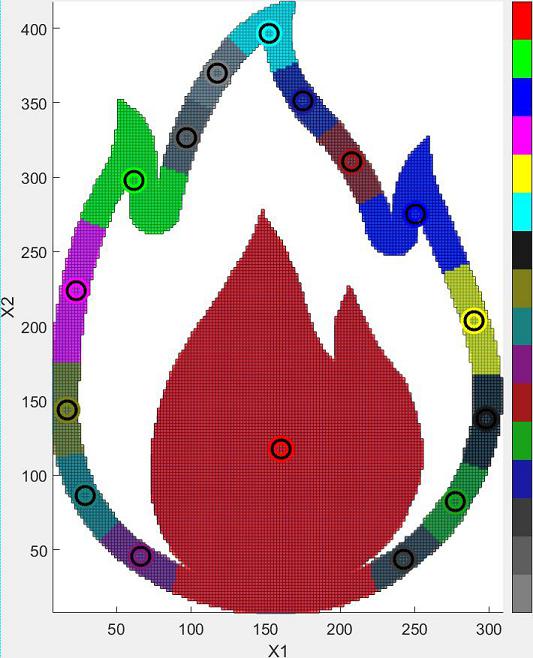} }
	\subfigure[] {
    	\label{fig:spectral9}
		\includegraphics[width=0.35\textwidth]{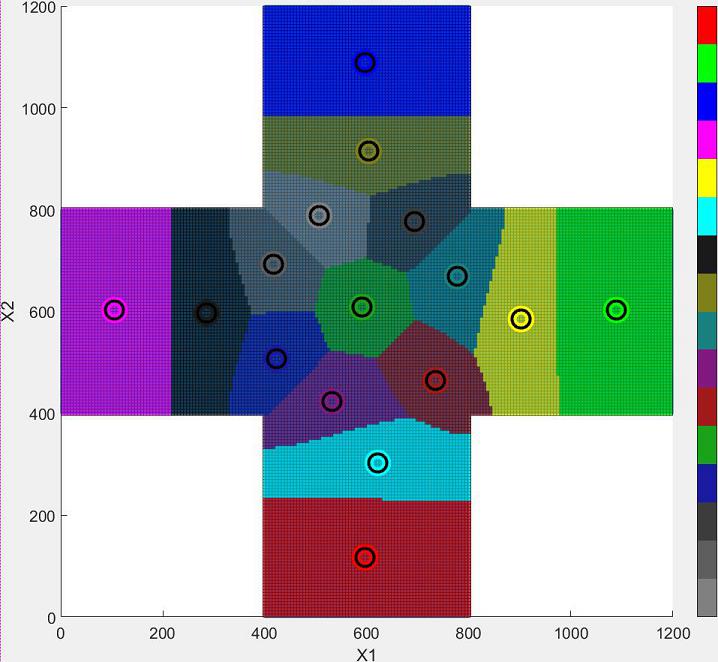} }
\qquad\hskip -0.4cm
	\subfigure[] {
    	\label{fig:spectral10}
		\includegraphics[width=0.35\textwidth]{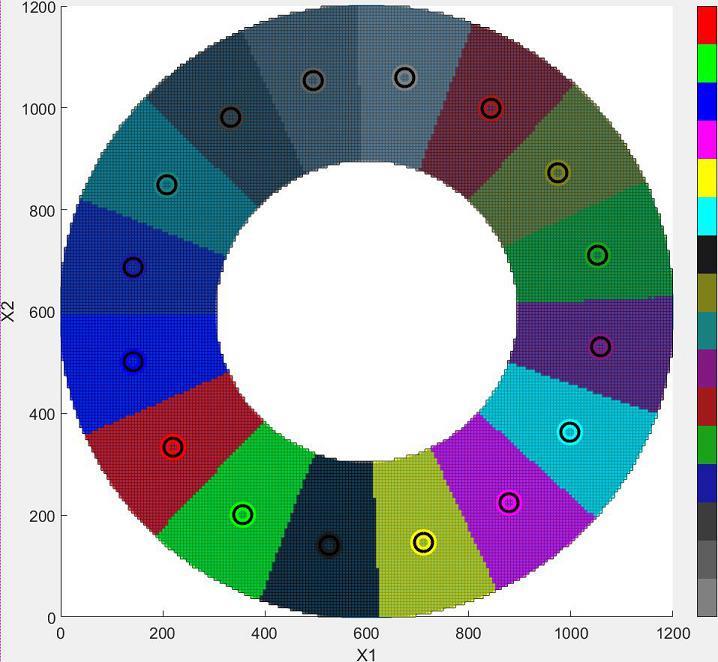} }
	\subfigure[] {
    	\label{fig:spectral11}
		\includegraphics[width=0.35\textwidth]{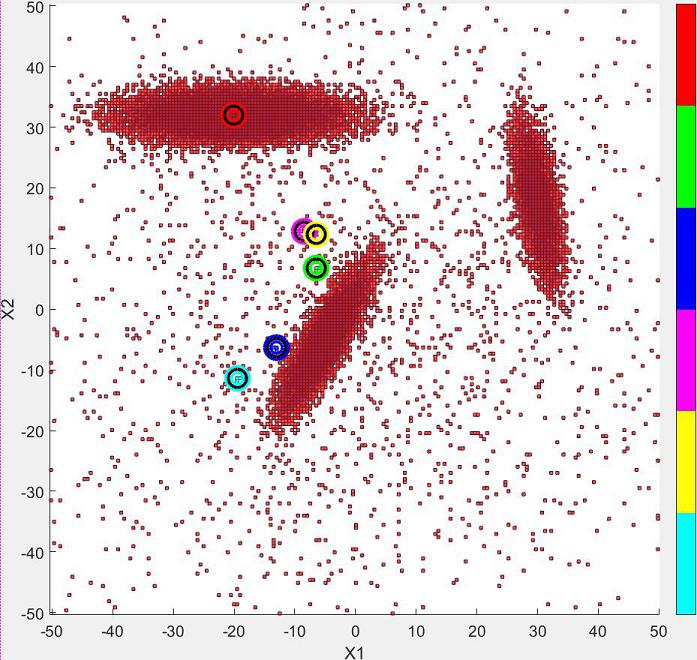} }
\qquad\hskip -0.4cm
	\subfigure[] {
    	\label{fig:spectral12}
		\includegraphics[width=0.35\textwidth]{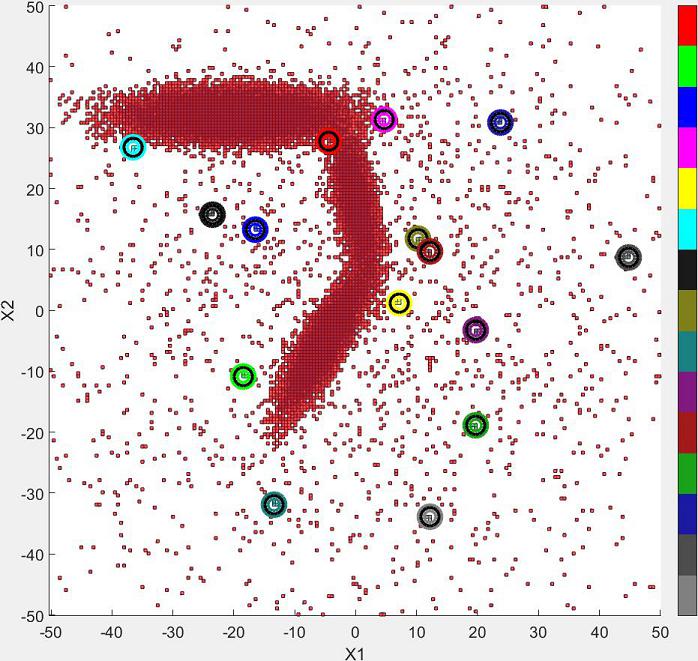} }
\vspace{-0.4cm}
	\caption{Test bank data sets for {\em L, Plus1, Concentric1, Data2D1, Data2D2} and {\em Flame1} are shown using spectral clustering with a Laplacian based on $NN1$ using eigenmodes two and three  collected in the eigenspace using K-means seeking 16 possible clusters.}
	\label{fig:testbank4}
\end{figure*}

\begin{figure*}[htbp]
	\centering
	\subfigure[] {
    	\label{fig:spectral13}
		\includegraphics[width=0.35\textwidth]{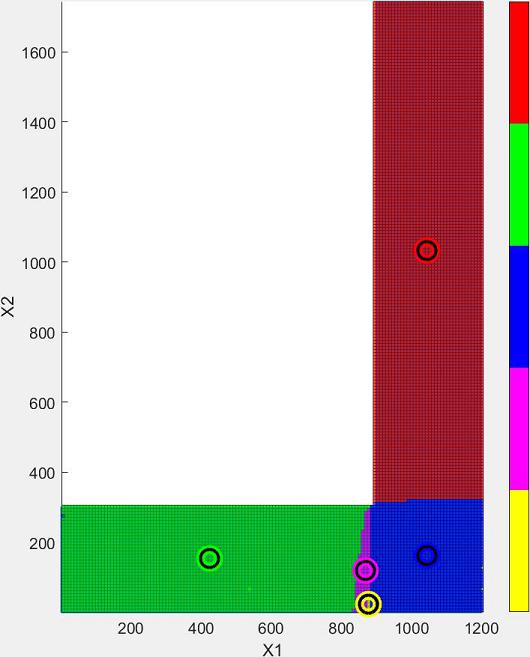} }
\qquad\hskip -0.4cm
	\subfigure[] {
    	\label{fig:spectral14}
		\includegraphics[width=0.35\textwidth]{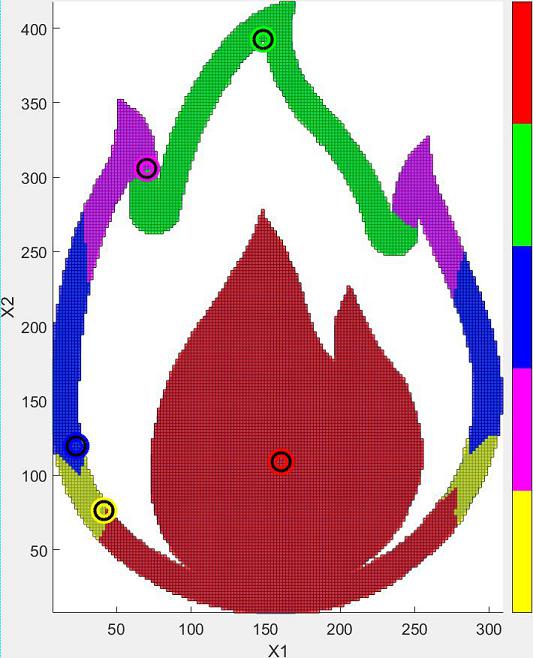} }
	\subfigure[] {
    	\label{fig:spectral15}
		\includegraphics[width=0.35\textwidth]{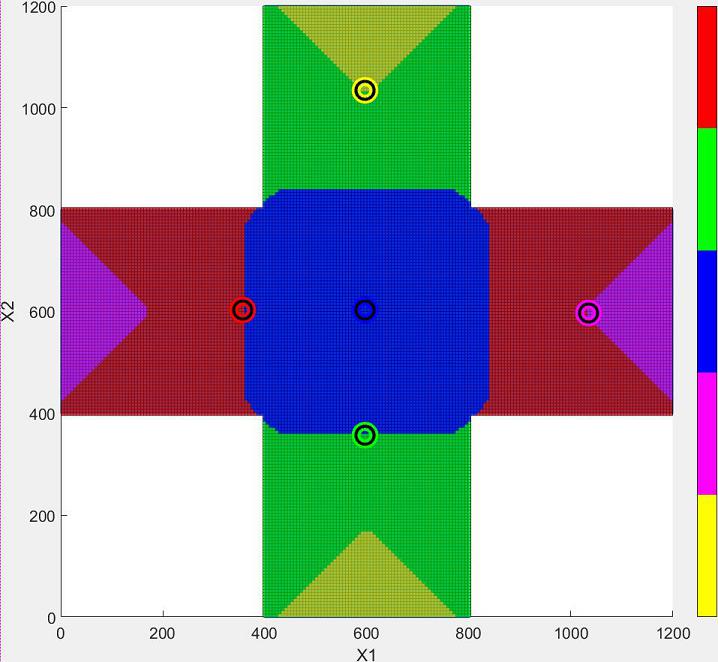} }
\qquad\hskip -0.4cm
	\subfigure[] {
    	\label{fig:spectral16}
		\includegraphics[width=0.35\textwidth]{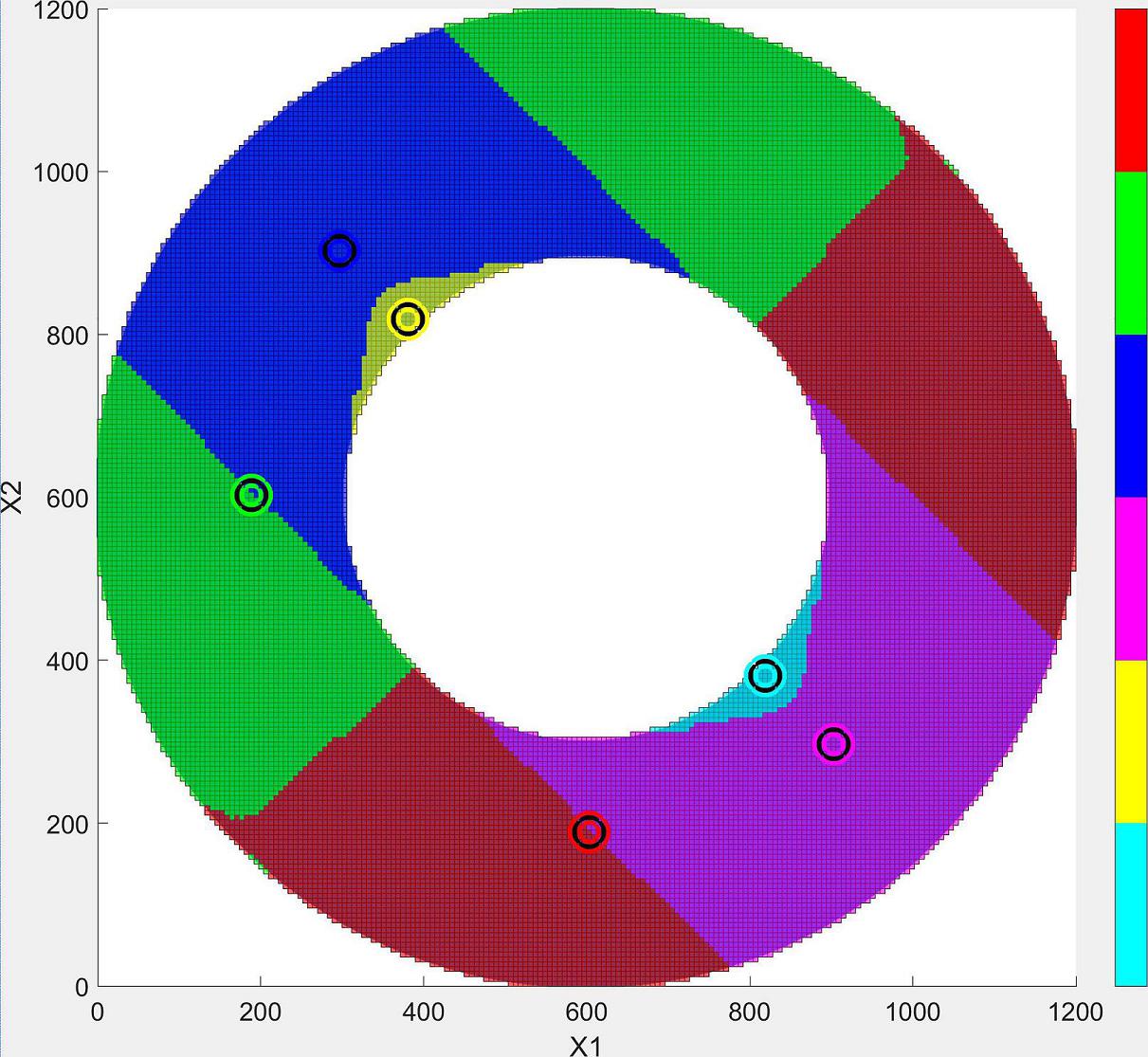} }
	\subfigure[] {
    	\label{fig:spectral17}
		\includegraphics[width=0.35\textwidth]{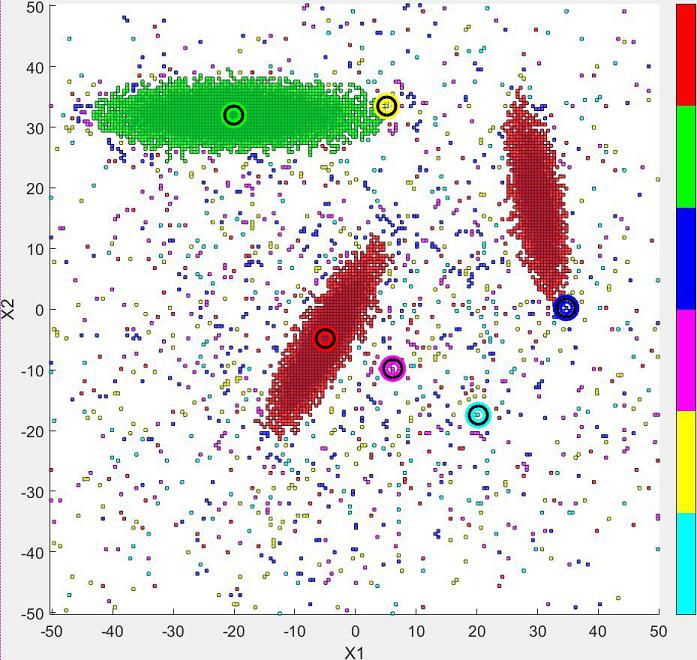} }
\qquad\hskip -0.4cm
	\subfigure[] {
    	\label{fig:spectral18}
		\includegraphics[width=0.35\textwidth]{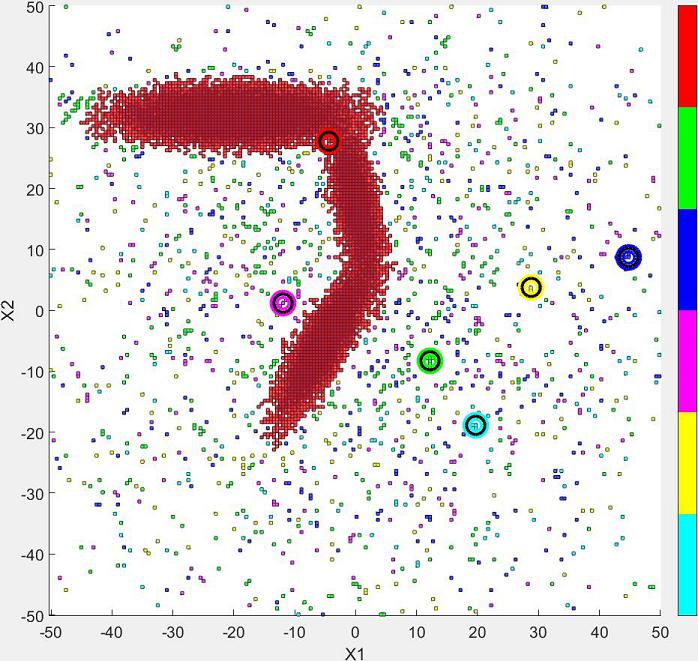} }
\vspace{-0.4cm}
	\caption{Test bank data sets for {\em L, Plus1, Concentric1, Data2D1, Data2D2} and {\em Flame1} are shown using spectral clustering with a Laplacian based on LOS using eigenmodes one and two collected in the eigenspace using a 2D histogram seeking 16 possible clusters.}
	\label{fig:testbank5}
\end{figure*}

A problem can occur when the partition domain is segregated by disconnected regions.  When solving for eigenvectors, through a similarity transformation the Laplacian matrix is effectively diagonalized so that the eigenvectors formed associate with the subblocks along the diagonal.  When the partition domain is fully connected, the subblocks become clusters {\em within} the connected region.  When the partitions are disconnected however, this process associates each subblock with a cluster, leaving the interior structure of the subblocks unassigned to clusters.  The effect is seen in Figs. \ref{fig:spectral6}, \ref{fig:spectral12}, \ref{fig:spectral18}, where the larger structure is assigned a single cluster due to the small disconnected clusters near its boundary.  This sensitivity to noise is the partition set is failing of spectral clustering, in that it requires a fully connected domain in order to resolve interior structure in the presence of ancillary distributions nearby.

Figures \ref{fig:spectral1}-\ref{fig:spectral6} show the results on select test cases for spectral clustering where the Laplacian is based on the ${\bf NN1}$ matrix and the first two eigenvectors were used for gathering within the eigenspace using K-means assuming 16 possible clusters ({\em SPEC-NN1-12}).

Figures \ref{fig:spectral7}-\ref{fig:spectral12} show the results using a Laplacian based on the ${\bf NN1}$ matrix with the second and third eigenvectors used for defining the eigenspace using K-means for gathering assuming 16 possible clusters ({\em SPEC-NN1-23}).

Figures \ref{fig:spectral13}-\ref{fig:spectral18} show the results using the LOS criteria to define neighbors on a graph that the Laplacian is created from.  The first two eigenvectors define the eigenspace  used for gathering with a 2D histogram with 4x4 bins ({\em SPEC-LOS-12}).

\begin{figure*}[htbp]
	\centering
	\subfigure[] {
    	\label{fig:spectral19}
		\includegraphics[width=0.35\textwidth]{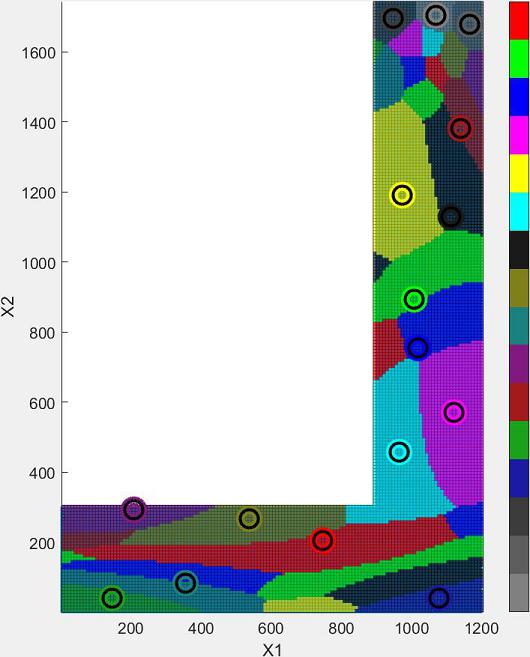} }
\qquad\hskip -0.4cm
	\subfigure[] {
    	\label{fig:spectral20}
		\includegraphics[width=0.35\textwidth]{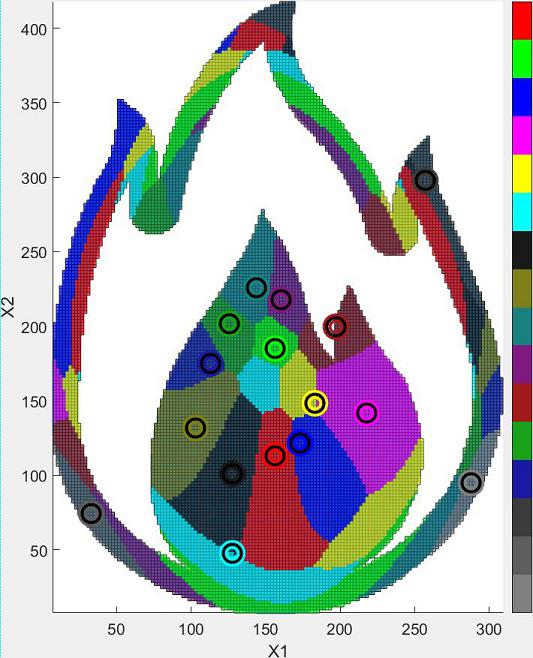} }
	\subfigure[] {
    	\label{fig:spectral21}
		\includegraphics[width=0.35\textwidth]{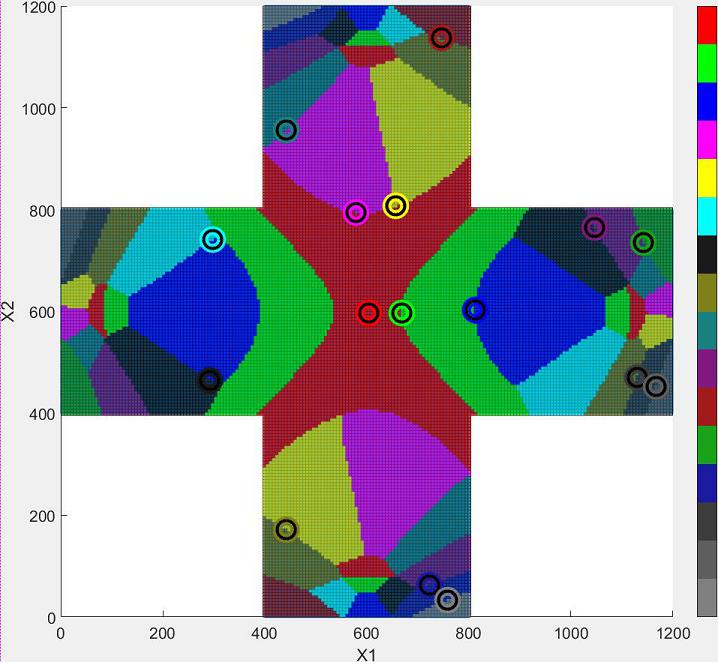} }
\qquad\hskip -0.4cm
	\subfigure[] {
    	\label{fig:spectral22}
		\includegraphics[width=0.35\textwidth]{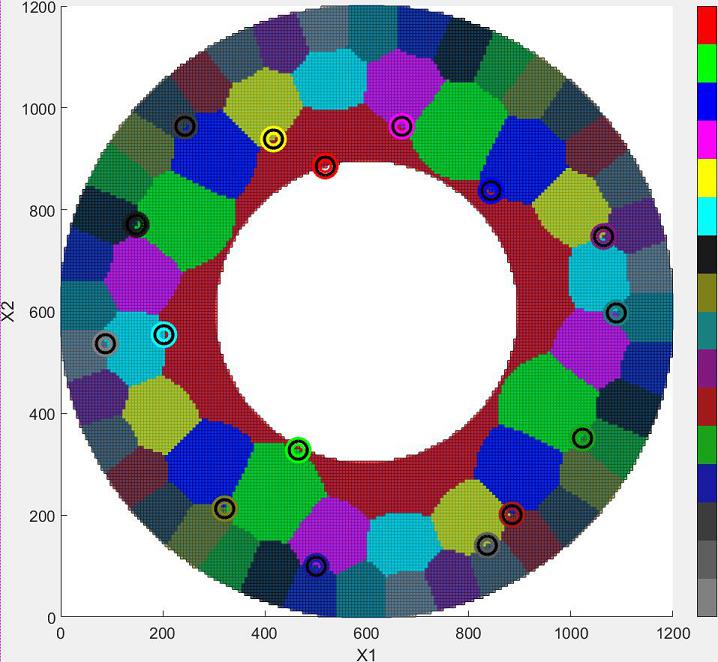} }
	\subfigure[] {
    	\label{fig:spectral23}
		\includegraphics[width=0.35\textwidth]{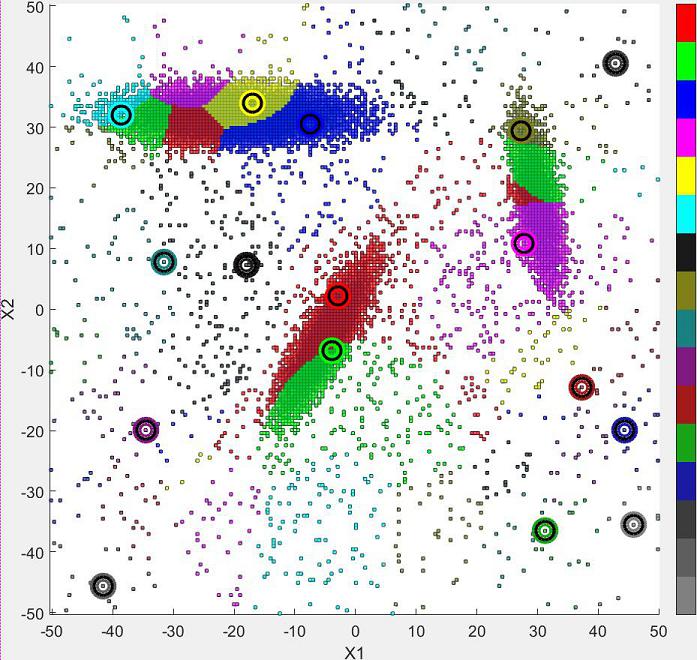} }
\qquad\hskip -0.4cm
	\subfigure[] {
    	\label{fig:spectral24}
		\includegraphics[width=0.35\textwidth]{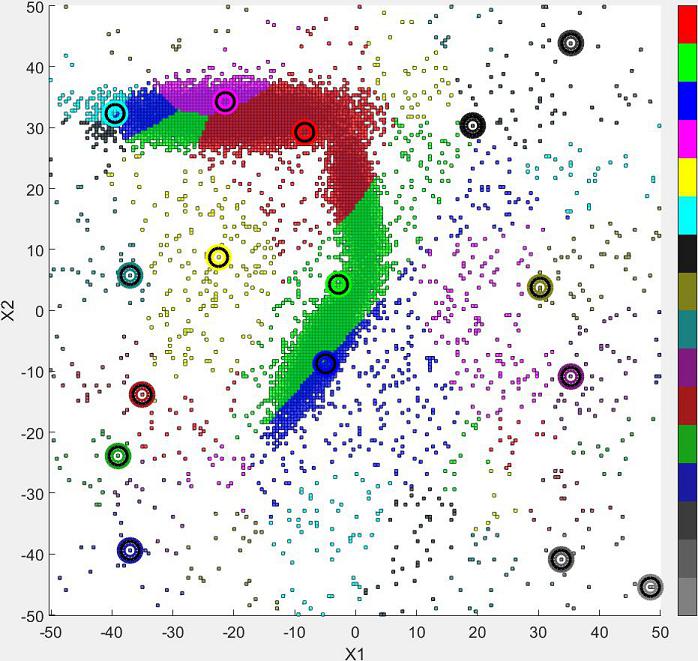} }
\vspace{-0.4cm}
	\caption{Test bank data sets for {\em L, Plus1, Concentric1, Data2D1, Data2D2} and {\em Flame1} are shown using spectral clustering with a Laplacian of a Guassian using eigenmodes two and three  collected in the eigenspace using K-means seeking 16 possible clusters.}
	\label{fig:testbank5}
\end{figure*}

Figures \ref{fig:spectral19}-\ref{fig:spectral24} show the results from using the Laplacian based on a Guassian, making it global in scope.  The first two eigenvectors were used for gathering and the eigenspace formed was gathered using K-medoids assuming 16 possible clusters ({\em SPEC-GAU-12}).

\subsection{Positional Clustering (LMH-POS)}
\label{sec:lmh-pos-2}
\begin{figure*}[htbp]
\caption{Positional clustering applied to the 2D data set (left) and the 3D data set (right). }
\label{fig:LMHpos01}
\centering
	\subfigure[] {
    	\label{fig:data2d2LMH}
		\includegraphics[width=0.45\textwidth]{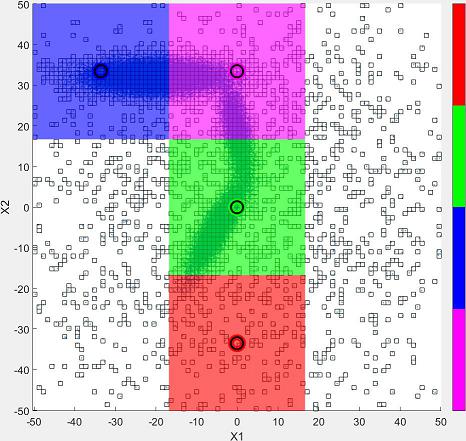} }
\qquad\hskip -0.3cm
	\subfigure[] {
    	\label{fig:data3d1LMH}
		\includegraphics[width=0.45\textwidth]{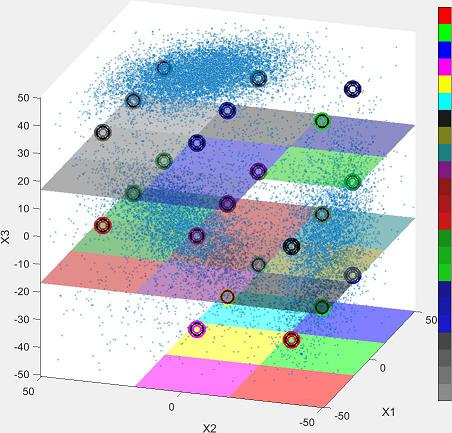} } \\
\end{figure*}

\cbk
 Figure \ref{fig:LMHpos01} shows positional clustering applied to two of the data test cases.  A low density threshold was applied so that only higher density partitions are considered for clustering.  In the two dimensional case, Fig. \ref{fig:data2d2LMH}, only four of the possible nine positional clusters are found, while in the three dimensional case, Fig. \ref{fig:data3d1LMH}, only 21 out of the possible 27 clusters are found.  As the number of dimensions grows, the number of possible clusters to be found increases as $3^{N_{_{D}}}$, yet in most cases, the data will likely fill only a fraction of the total possible set of clusters, making the LMHpos clustering a quick view of where data can be found within the data space.

\cbk

\begin{figure*}[t!p]
	\label{fig:robust01}
	\centering
	\subfigure {
	\label{fig:cutkmeans}
		\includegraphics[width=0.195\textwidth]{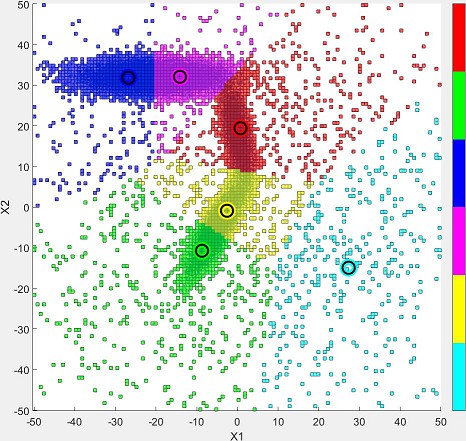} }
\hspace{-0.5cm}
	\subfigure{
	\label{fig:cutkmedoids}
		\includegraphics[width=0.195\textwidth]{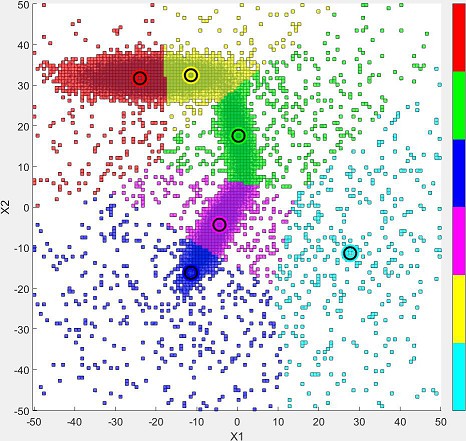} }
\hspace{-0.5cm}
	\subfigure {
	\label{fig:cutmaxglobal}
		\includegraphics[width=0.195\textwidth]{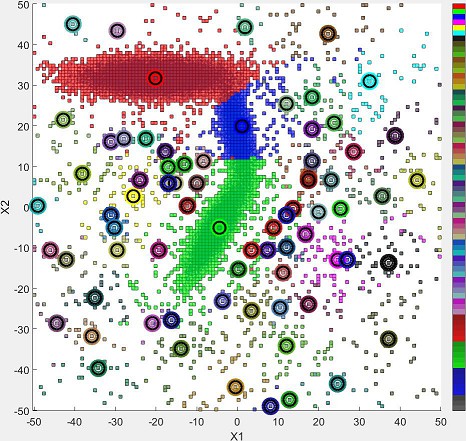} }
\hspace{-0.5cm}
	\subfigure {
	\label{fig:cutmaxpathl}
		\includegraphics[width=0.195\textwidth]{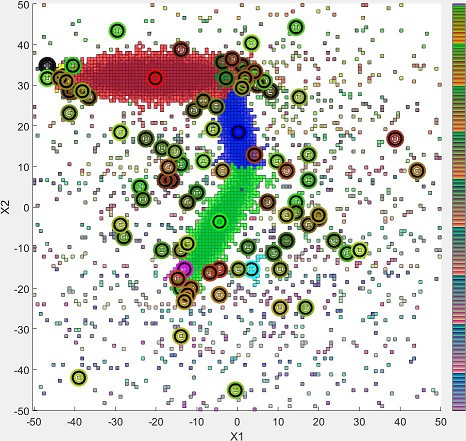} }
\hspace{-0.5cm}
	\subfigure {
	\label{fig:cutconn}
		\includegraphics[width=0.195\textwidth]{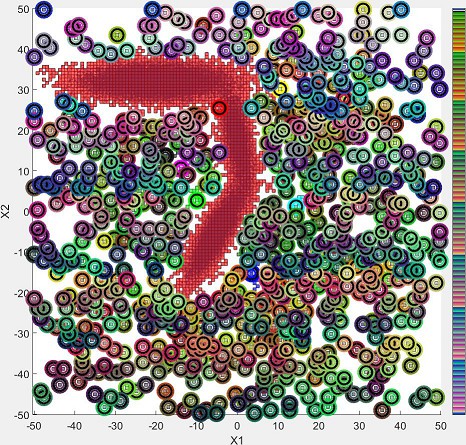} } \\
\qquad
\hspace{-0.9cm}
	\subfigure {
	\label{fig:cutmaxvis}
		\includegraphics[width=0.195\textwidth]{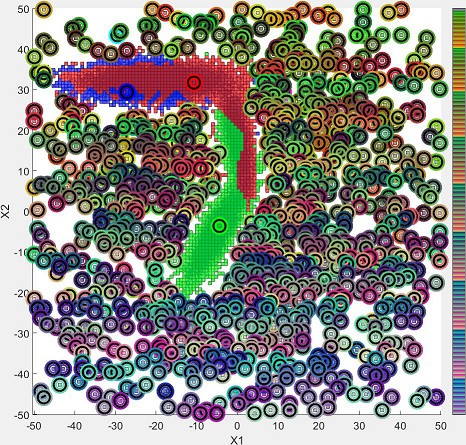} }
\hspace{-0.5cm}
	\subfigure {
	\label{fig:cutmaxmut}
		\includegraphics[width=0.195\textwidth]{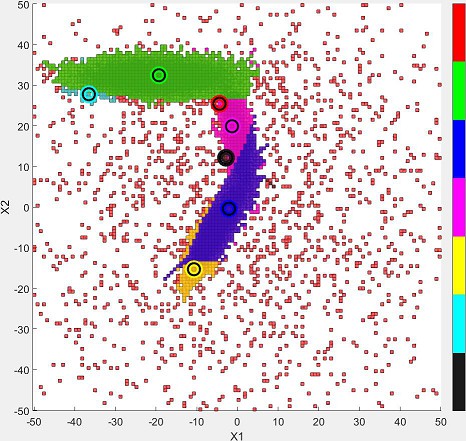} }
\hspace{-0.5cm}
	\subfigure {
	\label{fig:cutspec01}
		\includegraphics[width=0.195\textwidth]{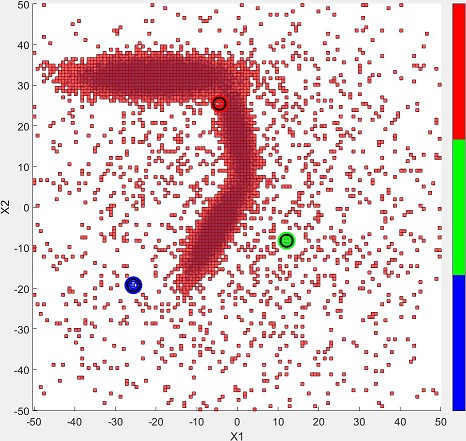} }
\hspace{-0.5cm}
	\subfigure {
	\label{fig:cutspec02}
		\includegraphics[width=0.195\textwidth]{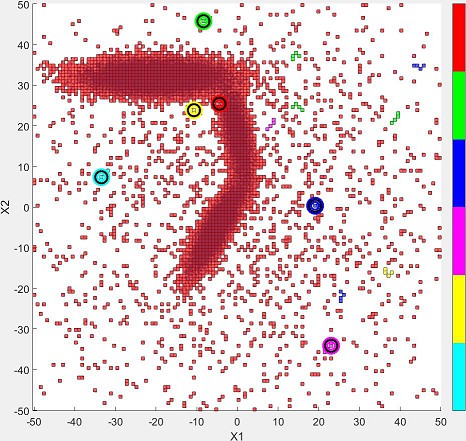} }
\hspace{-0.5cm}
	\subfigure {
	\label{fig:cutspec03}
		\includegraphics[width=0.195\textwidth]{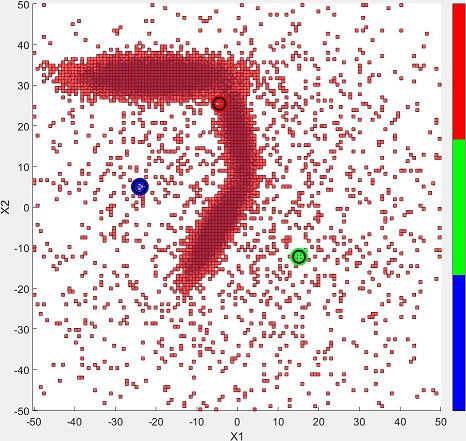} } \\
\qquad
\hspace{-0.9cm}
	\subfigure {
	\label{fig:cutspec04}
		\includegraphics[width=0.195\textwidth]{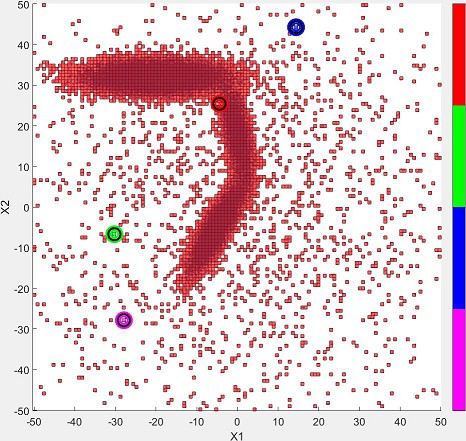} }
\hspace{-0.5cm}
	\subfigure {
	\label{fig:cutspec05}
		\includegraphics[width=0.195\textwidth]{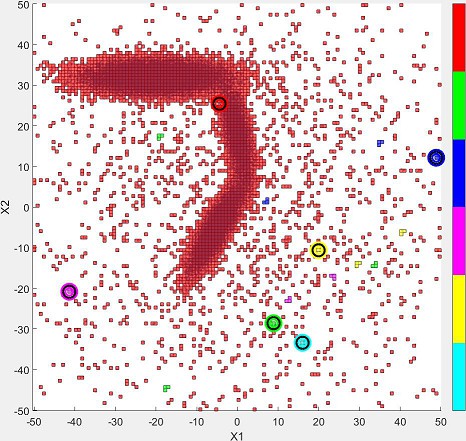} }
\hspace{-0.5cm}
	\subfigure {
	\label{fig:cutspec06}
		\includegraphics[width=0.195\textwidth]{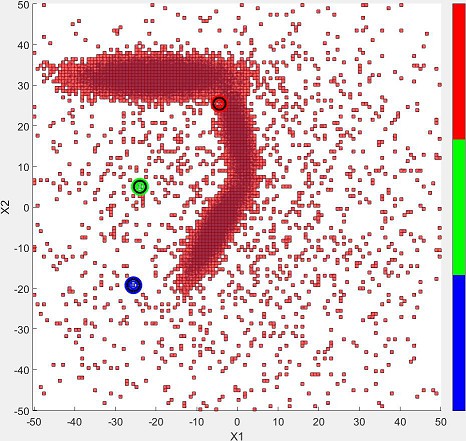} }
\hspace{-0.5cm}
	\subfigure {
	\label{fig:cutspec07}
		\includegraphics[width=0.195\textwidth]{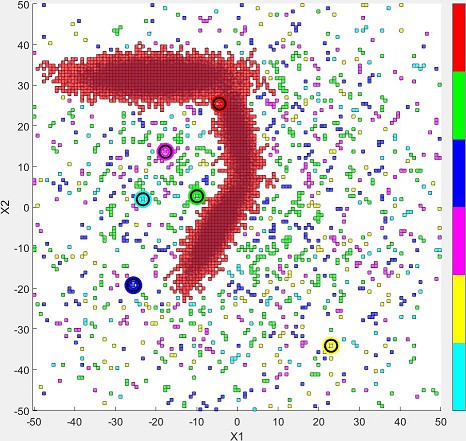} }
\hspace{-0.5cm}
	\subfigure {
	\label{fig:cutspec08}
		\includegraphics[width=0.195\textwidth]{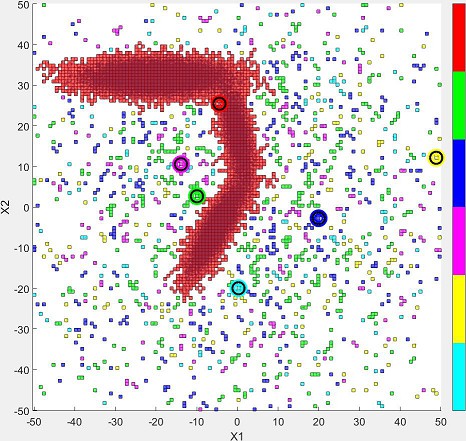} } \\
\qquad
\hspace{-0.9cm}
	\subfigure {
	\label{fig:cutspec09}
		\includegraphics[width=0.195\textwidth]{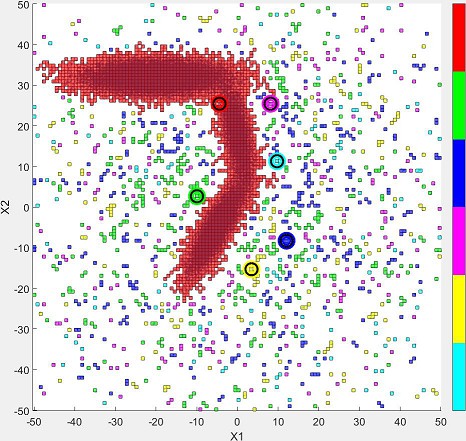} }
\hspace{-0.5cm}
	\subfigure {
	\label{fig:cutspec10}
		\includegraphics[width=0.195\textwidth]{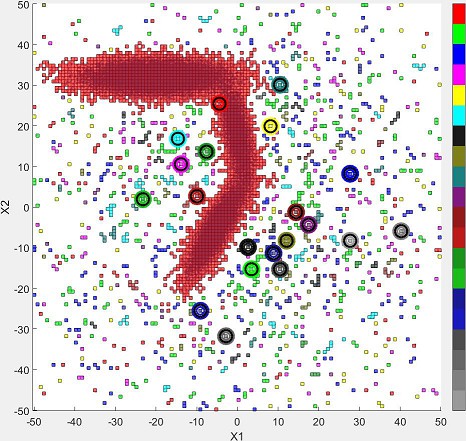} }
\hspace{-0.5cm}
	\subfigure {
	\label{fig:cutspec11}
		\includegraphics[width=0.195\textwidth]{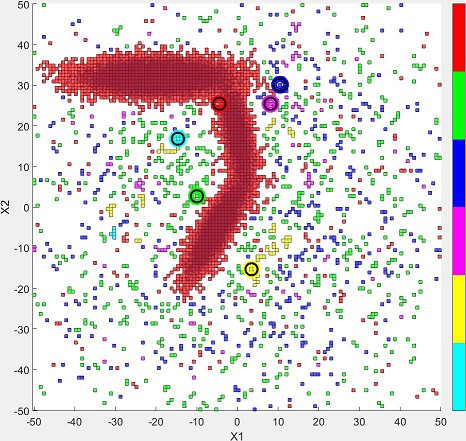} }
\hspace{-0.5cm}
	\subfigure {
	\label{fig:cutspec12}
		\includegraphics[width=0.195\textwidth]{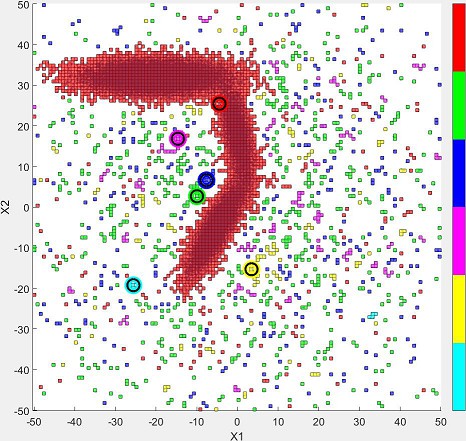} }
\hspace{-0.5cm}
	\subfigure {
	\label{fig:cutspec13}
		\includegraphics[width=0.195\textwidth]{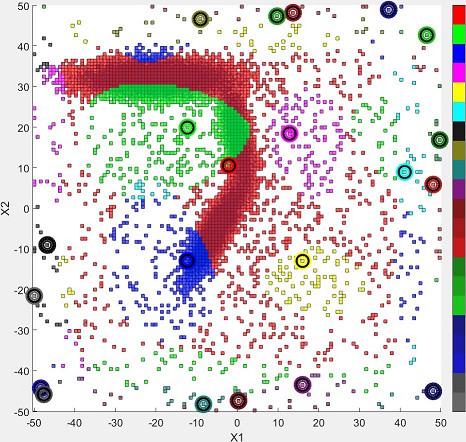} } \\
\qquad
\hspace{-0.9cm}
	\subfigure {
	\label{fig:cutspec14}
		\includegraphics[width=0.195\textwidth]{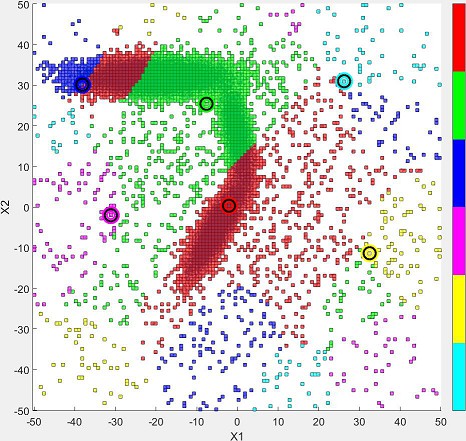} }
\hspace{-0.5cm}
	\subfigure {
	\label{fig:cutspec15}
		\includegraphics[width=0.195\textwidth]{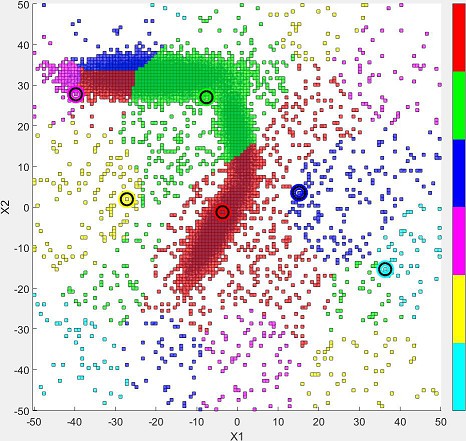} }
\hspace{-0.5cm}
	\subfigure {
	\label{fig:cutspec16}
		\includegraphics[width=0.195\textwidth]{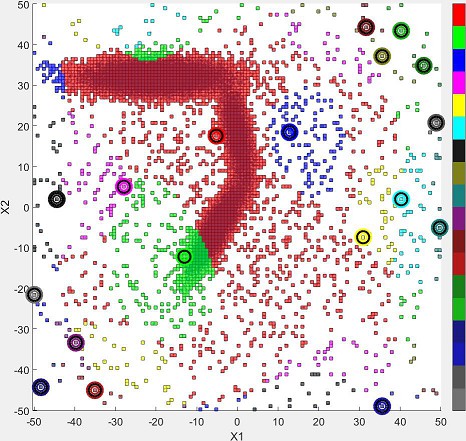} }
\hspace{-0.5cm}
	\subfigure {
	\label{fig:cutspec17}
		\includegraphics[width=0.195\textwidth]{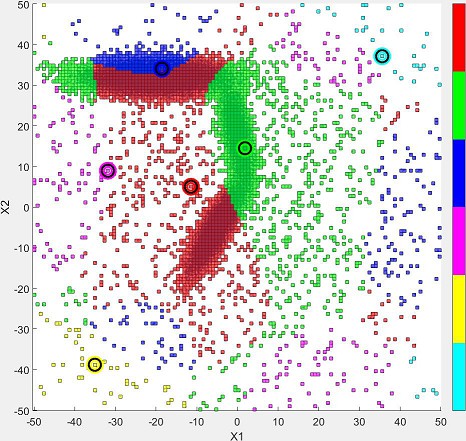} }
\hspace{-0.5cm}
	\subfigure {
	\label{fig:cutspec18}
		\includegraphics[width=0.195\textwidth]{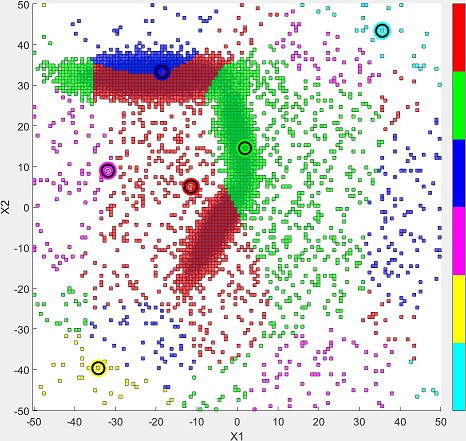} }
	\caption{Clustering techniques applied to the {\em Data2D2} set with no lower bound for partition population.  All techniques are shown excluding LMHpos clustering as it has differing partitions.  Clustering techniques are shown in the following order:  KMEANS, KMEDOIDS, MAXGLOBAL, CONN, MAXPATHL, LOS-MAXVIS, LOS-MUTUAL, SPECTRAL01, SPECTRAL01-18. }
\end{figure*}

\begin{figure*}[htbp]
    \label{fig:robust02}
	\centering
	\subfigure[] {
    	\label{fig:robust1}
		\includegraphics[width=0.35\textwidth]{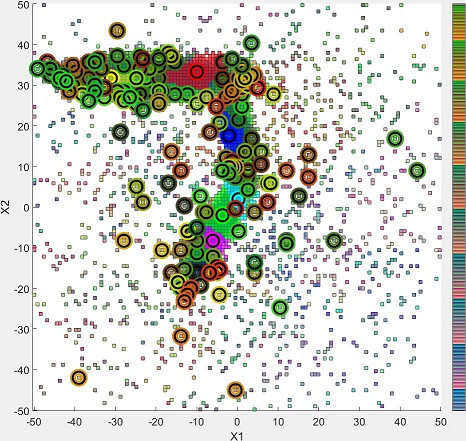} }
\qquad\hskip -0.4cm
	\subfigure[] {
    	\label{fig:robust2}
		\includegraphics[width=0.35\textwidth]{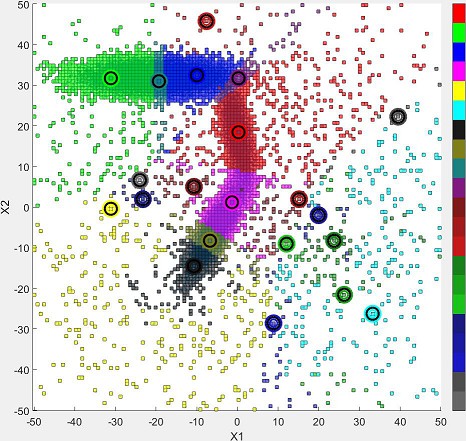} }
\qquad\hskip -0.4cm
	\subfigure[] {
    	\label{fig:robust3}
		\includegraphics[width=0.35\textwidth]{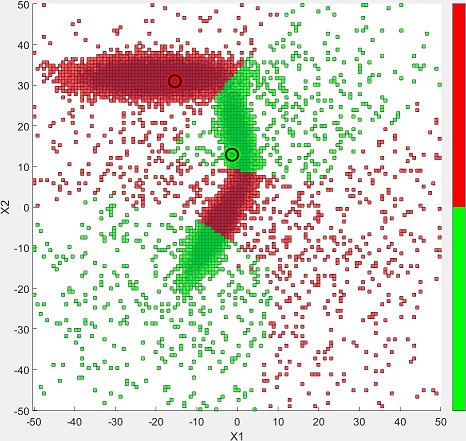} }
	\subfigure[] {
    	\label{fig:robust4}
		\includegraphics[width=0.35\textwidth]{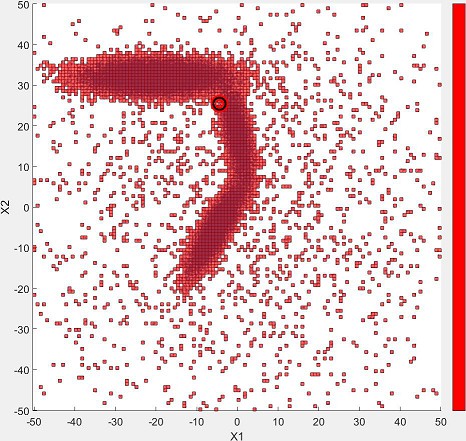} }
	\caption{Robust clustering results based on clustering techniques shown in Fig. \ref{fig:robust01}.  Panel \ref{fig:robust1} shows the clusters formed when any change across all clustering techniques occurs.  Panel \ref{fig:robust2}  shows clusters formed using a consensus with 75\% taken as the majority threshold.  Panel \ref{fig:robust3} shows clusters formed once all techniques (100\%) have encountered a change.  Panel \ref{fig:robust4} shows the cluster formed when requiring that all techniques register a change simultaneously. For K-means and K-medoids, a k-value of three was used. }
\end{figure*}


\section{Robust Clustering over Multiple Algorithms}
\label{sec:robust-1}

In this paper, multiple clustering algorithms have been presented and applied to several testcases.  Each technique has strengths as well as weaknesses which have been exposed through the cases presented.  When using multiple techniques, the possibility exists to leverage the information gathered from all techniques to arrive at a final cluster designation, based on the level of agreement or disagreement found between the algorithms \cite{strehl2003}.  This approach is comparable to ensemble modeling used in various fields \cite{hansen2002,tebaldi2007}.  This section proposes four possible robust ways to gather the cluster information.

In each approach taken, the cluster information for the partitions is represented by a matrix of cluster IDs, where each row represents a single cluster algorithm and each column is a partition.  The values along each row is then the cluster ID given to each partition.  The matrix formed is called the Cluster ID matrix and is $(N_C,N_P)$ in size.  In order find the agreement or disagreement between cluster IDs across many techniques, the columns are rearranged so that the cluster IDs are sequential starting from the first row and
maintaining the ordering as each row is subsequently reordered until all rows have been processed.  Table 3 illustrates this process for a sample of 40 partitions using six cluster algorithms.  The top matrix is  the initial partition cluster ID matrix unsorted.  The second matrix is the sorted cluster ID matrix described above.  Finally, the third matrix from the top shows the differences in cluster IDs {\em along each row}, where a one represents a change in cluster designation for that rows technique.   The process of assigning cluster IDs to partitions begins with the lowest numbered cluster IDs over all algorithms,  and proceeds in increasing cluster ID order.  In the table shown, this is equivalent to following the partitions from left to right across the page.

As examples of robust clustering, Fig. 14 as well as Tab. 3 are provided to illustrate the process.  The figures shown in Fig. 14 are all clustering techniques excluding the LMHpos algorithm for the {\em Data2D2} test case with no minimal population set for the partitions.  The LMHpos technique was excluded as its partition definitions do not align with the remaining 25 algorithms.  In cases where multiple techniques are compared using differing partition sizes, the robust technique is then applied {\em per datum}, using the same procedures, however, the sorting is performed over all data instead of partitions.

\subsection{Fractured Cluster ID}
\label{sec:fracture}
The {\bf Fractured} robust designation results by assigning each partition a new cluster ID starting from one and increasing the cluster ID each time {\em any} technique changes its ID.  This results in the largest set of clusters found.  This approach is the most sensitive to changes in the cluster designations.  Figure \ref{fig:robust1} shows the clusters formed, leading to the largest set of the robust techniques, also the most sensitive to changes in the partitions grouping.

\subsection{Majority Changed Cluster ID}
\label{sec:majority-1}
The {\bf Majority Changed} robust designation results by assigning each partition a new cluster ID starting from one and increasing the cluster ID each time the accumulated number of algorithms changing reaches a majority of the total number of algorithms.  Once the cluster ID has been changed, the accumulated sum of changes is reset to zero.  This results in a medium sized set of clusters found, where a significant number of algorithms found a change, however, not all algorithms are required to note the change in ID.  Figure \ref{fig:robust2}  shows the clusters formed using a consensus among the techniques with a 75\% threshold applied.  A simple majority places the threshold for consensus at 50\%, however, other values can be used to attain consensus.  Ideally, the best value would create the largest number of clusters with the highest average membership.

\subsection{All Changed Cluster ID}
\label{sec:majority-2}
The {\bf All Changed} robust designation results by assigning each partition a new cluster ID starting from one and increasing the cluster ID each time the accumulated number of algorithms changing reaches the total number of algorithms.  Once the cluster ID has been changed, the accumulated sum of changes is reset to zero.  This results in a small-medium sized set of clusters found, where every algorithm found a change, however, the changes may not have been at the same partition number, merely, that the total set of changes across all algorithms eventually required a change of ID.  Figure \ref{fig:robust3} shows the clusters formed by requiring that all techniques register a change in cluster definition before a partition is given a new identity.  This is equivalent to setting the consensus threshold to 100\%.   Once a clustering technique has changed a partitions cluster ID, any further changes from that technique are not registered until {\em all} techniques have shown a change as well.  This approach limits the number of final clusters formed.

\subsection{No Overlap Cluster ID}
\label{sec:no-overlap}
The {\bf No Overlap} robust designation results by assigning each partition a new cluster ID starting from one and increasing the cluster ID each time the total number of algorithms changes designation simultaneously.  This results in a smallest sized set of clusters found, where every algorithm found a change.  Figure \ref{fig:robust4} shows the cluster formed by requiring all techniques to simultaneously register a change in cluster identity.  Ideally, this would happen for each disconnected group of partitions, however, several techniques are ``global'' in scope and do not require a connection to exist to form clusters, leading - in this case - to a single large cluster.

\begin{table}[t!]
\label{tab:robust1}
\resizebox{\columnwidth}{!}{
\hspace{-0.6cm}\begin{tabular}{l|cccccccccccccccccccccccccccccccccccccccc}
\hline
        &                                         \multicolumn{39}{c}{ 40 Partition Cluster IDs} \\ \hline
$Alg_1$ & 5 & 7 & 2 & 4 & 7 & 1 & 2 & 4 & 1 & 1 & 2 & 3 & 8 & 6 & 4 & 1 & 5 & 5 & 4 & 1 & 1 & 4 & 8 & 8 & 9 & 6 & 9 & 4 & 1 & 2 & 4 & 4 & 4 & 1 & 7 & 4 & 3 & 3 & 4  \\
$Alg_2$ & 7 & 7 & 3 & 7 & 7 & 1 & 3 & 7 & 3 & 1 & 3 & 4 & 7 & 7 & 6 & 1 & 7 & 7 & 5 & 1 & 2 & 5 & 7 & 7 & 7 & 7 & 7 & 5 & 1 & 3 & 6 & 7 & 6 & 1 & 7 & 7 & 5 & 3 & 5  \\
$Alg_3$ & 6 & 6 & 2 & 6 & 7 & 1 & 3 & 6 & 2 & 1 & 3 & 3 & 8 & 6 & 5 & 1 & 6 & 6 & 5 & 1 & 1 & 5 & 8 & 8 & 8 & 6 & 8 & 5 & 1 & 2 & 5 & 6 & 6 & 1 & 6 & 6 & 3 & 3 & 5  \\
$Alg_4$ & 4 & 5 & 2 & 4 & 6 & 1 & 2 & 4 & 2 & 1 & 2 & 2 & 6 & 4 & 4 & 1 & 4 & 4 & 2 & 1 & 2 & 3 & 6 & 6 & 6 & 5 & 6 & 2 & 1 & 2 & 3 & 4 & 4 & 1 & 5 & 4 & 2 & 2 & 2  \\
$Alg_5$ & 5 & 6 & 1 & 4 & 6 & 1 & 1 & 5 & 1 & 1 & 1 & 2 & 6 & 6 & 4 & 1 & 6 & 5 & 3 & 1 & 1 & 4 & 6 & 6 & 6 & 6 & 6 & 4 & 1 & 1 & 4 & 4 & 4 & 1 & 6 & 4 & 2 & 1 & 4  \\
$Alg_6$ & 6 & 8 & 1 & 5 & 8 & 1 & 2 & 5 & 1 & 1 & 3 & 4 & 8 & 7 & 4 & 1 & 7 & 5 & 4 & 1 & 1 & 4 & 9 & 9 & 9 & 7 & 9 & 4 & 1 & 1 & 4 & 5 & 5 & 1 & 7 & 5 & 4 & 3 & 4  \\
\hline
\hline
        &                                         \multicolumn{39}{c}{ 40 Partition Cluster IDs - Resorted by Partitions in Ascending ID Order} \\ \hline
$Alg_1$ & 1 & 1 & 1 & 1 & 1 & 1 & 1 & 1 & 2 & 2 & 2 & 2 & 3 & 3 & 3 & 3 & 4 & 4 & 4 & 4 & 4 & 4 & 4 & 4 & 4 & 4 & 4 & 5 & 5 & 5 & 6 & 6 & 7 & 7 & 7 & 8 & 8 & 8 & 9  \\
$Alg_2$ & 1 & 1 & 1 & 1 & 1 & 1 & 2 & 3 & 3 & 3 & 3 & 3 & 3 & 4 & 5 & 5 & 5 & 5 & 5 & 5 & 6 & 6 & 6 & 7 & 7 & 7 & 7 & 7 & 7 & 7 & 7 & 7 & 7 & 7 & 7 & 7 & 7 & 7 & 7  \\
$Alg_3$ & 1 & 1 & 1 & 1 & 1 & 1 & 1 & 2 & 2 & 2 & 3 & 3 & 3 & 3 & 3 & 4 & 5 & 5 & 5 & 5 & 5 & 5 & 6 & 6 & 6 & 6 & 6 & 6 & 6 & 6 & 6 & 6 & 6 & 6 & 7 & 8 & 8 & 8 & 8  \\
$Alg_4$ & 1 & 1 & 1 & 1 & 1 & 1 & 2 & 2 & 2 & 2 & 2 & 2 & 2 & 2 & 2 & 2 & 2 & 2 & 2 & 3 & 3 & 4 & 4 & 4 & 4 & 4 & 4 & 4 & 4 & 4 & 4 & 5 & 5 & 5 & 6 & 6 & 6 & 6 & 6  \\
$Alg_5$ & 1 & 1 & 1 & 1 & 1 & 1 & 1 & 1 & 1 & 1 & 1 & 1 & 1 & 2 & 2 & 3 & 3 & 4 & 4 & 4 & 4 & 4 & 4 & 4 & 4 & 4 & 5 & 5 & 5 & 6 & 6 & 6 & 6 & 6 & 6 & 6 & 6 & 6 & 6  \\
$Alg_6$ & 1 & 1 & 1 & 1 & 1 & 1 & 1 & 1 & 1 & 1 & 2 & 3 & 3 & 4 & 4 & 4 & 4 & 4 & 4 & 4 & 4 & 4 & 5 & 5 & 5 & 5 & 5 & 5 & 6 & 7 & 7 & 7 & 7 & 8 & 8 & 8 & 9 & 9 & 9  \\
\hline
\hline
        &                                         \multicolumn{39}{c}{ 40 Partition Cluster Difference Flags (Logical) for Sorted IDs} \\ \hline
$Alg_1$ & 0 & 0 & 0 & 0 & 0 & 0 & 0 & 0 & 1 & 0 & 0 & 0 & 1 & 0 & 0 & 0 & 1 & 0 & 0 & 0 & 0 & 0 & 0 & 0 & 0 & 0 & 0 & 1 & 0 & 0 & 1 & 0 & 1 & 0 & 0 & 1 & 0 & 0 & 1  \\
$Alg_2$ & 0 & 0 & 0 & 0 & 0 & 0 & 1 & 1 & 0 & 0 & 0 & 0 & 0 & 1 & 1 & 0 & 0 & 0 & 0 & 0 & 1 & 0 & 0 & 1 & 0 & 0 & 0 & 0 & 0 & 0 & 0 & 0 & 0 & 0 & 0 & 0 & 0 & 0 & 0  \\
$Alg_3$ & 0 & 0 & 0 & 0 & 0 & 0 & 0 & 1 & 0 & 0 & 1 & 0 & 0 & 0 & 0 & 1 & 1 & 0 & 0 & 0 & 0 & 0 & 1 & 0 & 0 & 0 & 0 & 0 & 0 & 0 & 0 & 0 & 0 & 0 & 1 & 1 & 0 & 0 & 0  \\
$Alg_4$ & 0 & 0 & 0 & 0 & 0 & 0 & 1 & 0 & 0 & 0 & 0 & 0 & 0 & 0 & 0 & 0 & 0 & 0 & 0 & 1 & 0 & 1 & 0 & 0 & 0 & 0 & 0 & 0 & 0 & 0 & 0 & 1 & 0 & 0 & 1 & 0 & 0 & 0 & 0  \\
$Alg_5$ & 0 & 0 & 0 & 0 & 0 & 0 & 0 & 0 & 0 & 0 & 0 & 0 & 0 & 1 & 0 & 1 & 0 & 1 & 0 & 0 & 0 & 0 & 0 & 0 & 0 & 0 & 1 & 0 & 0 & 1 & 0 & 0 & 0 & 0 & 0 & 0 & 0 & 0 & 0  \\
$Alg_6$ & 0 & 0 & 0 & 0 & 0 & 0 & 0 & 0 & 0 & 0 & 1 & 1 & 0 & 1 & 0 & 0 & 0 & 0 & 0 & 0 & 0 & 0 & 1 & 0 & 0 & 0 & 0 & 0 & 1 & 1 & 0 & 0 & 0 & 1 & 0 & 0 & 1 & 0 & 0  \\
\hline
\hline
        &                                         \multicolumn{39}{c}{ 40 Partition Cluster Fractured IDs} \\ \hline
$Rob_1$ & 1 & 1 & 1 & 1 & 1 & 1 & 2 & 3 & 4 & 4 & 5 & 6 & 7 & 8 & 9 & 10 & 11 & 12 & 12 & 13 & 14 & 15 & 16 & 17 & 17 & 17 & 18 & 19 & 20 & 21 & 22 & 23 & 24 & 25 & 26 & 27 & 28 & 28 & 29  \\
\hline
\hline
        &                                         \multicolumn{39}{c}{ 40 Partition Cluster Majority Changed IDs} \\ \hline
$Rob_2$ & 1 & 1 & 1 & 1 & 1 & 1 & 1 & 2 & 2 & 2 & 3 & 3 & 3 & 4 & 4 & 5 & 5 & 6 & 6 & 6 & 6 & 6 & 7 & 7 & 7 & 7 & 7 & 8 & 8 & 8 & 9 & 9 & 9 & 10 & 10 & 11 & 11 & 11 & 11  \\
\hline
\hline
        &                                         \multicolumn{39}{c}{ 40 Partition Cluster All Changed IDs} \\ \hline
$Rob_3$ & 1 & 1 & 1 & 1 & 1 & 1 & 1 & 1 & 1 & 1 & 1 & 1 & 1 & 2 & 2 & 2 & 2 & 2 & 2 & 2 & 2 & 2 & 3 & 3 & 3 & 3 & 3 & 3 & 3 & 3 & 3 & 3 & 3 & 3 & 4 & 4 & 4 & 4 & 4  \\
\hline
\hline
        &                                         \multicolumn{39}{c}{ 40 Partition Cluster No Overlap IDs} \\ \hline
$Rob_4$ & 1 & 1 & 1 & 1 & 1 & 1 & 1 & 1 & 1 & 1 & 1 & 1 & 1 & 1 & 1 & 1 & 1 & 1 & 1 & 1 & 1 & 1 & 1 & 1 & 1 & 1 & 1 & 1 & 1 & 1 & 1 & 1 & 1 & 1 & 1 & 1 & 1 & 1 & 1  \\
\hline
\end{tabular}
}
\caption{A sample set of partitions that have had six differing cluster algorithms applied.  In each case, the cluster algorithm identified up to nine different clusters.
The set contains 40 partitions.  The top table represents the data initially unsorted.  Each row is a different cluster algorithm and each column is a partition where a cluster ID has been assigned.
The second table has sorted the each row while maintaining the assignments to each partition.  The third table from the top are the differences in cluster ID assignments from one column to the next.  The fourth table is the final cluster assignment given to the partitions when any one change occurs (a disagreement) between the cluster algorithms.  The fifth table requires a majority of the cluster algorithms to change (cumulatively) before a new cluster assignment is designated.  The sixth table only changes the cluster assignment once all cluster algorithms cumulatively have changed.  Finally, the last table requires that all algorithms change assignments simultaneously before a new cluster ID is designated (the clusters are disjoint - with no overlap). }
\end{table}

\subsection{Strategy with Robust Clustering}
Several of the clustering techniques used in this study require either a guess or fore-knowledge of the number of clusters sought, such as KMEANS and KMEDOIDS.  Further, the spectral methods applied k-means and k-medoids in order to identify clusters within the eigenspace formed from the eigenvectors chosen.  In these cases, an initial guess at the number of clusters sought is required, the {\em k-value}.   Robust clustering can provide a reasonable guess for the k-value, by first attaining consensus over all techniques that do {\em not} use a k-value, which are:  MAXGLOBAL, MAXPATHL, CONN, LOS-MAXVIS, LOS-MUTUAL and all spectral methods which use 2D histogram binning, although the 2D histogram binning itself requires a guess as to the number of bins to use; however, once the bins have been established, the 2D histograms simply find clusters which populate those bins, often finding fewer clusters than bins.   Using the ROBUST2 technique with a suitable choice in consensus threshold, $\Theta_{cons}$, a number of clusters will be found.  Taking the number of clusters found and setting k-value to this number then allows a reasonable guess to re-run the analysis utilizing the full complement of techniques.

This approach also accommodates using differing sets of variable choices as well as different binning across each dimension.  In this case, the partition definitions will not be the same from one set of analysis choices (variables, binning, techniques); however, each datum will still be assigned a cluster ID.  Robust clustering is then applied for each datum, leading to considerably longer sort times, however, the robust approach is still applicable, allowing various consensus clustering to performed.


\section{Conclusions}
\label{sec:conclusions}
A study using 26 clustering techniques has been performed over 12 test cases to illustrate both the strengths and weaknesses of clustering algorithms.  A robust form of clustering is achieved through consensus over all techniques, helping reduce clustering problems by finding consistent clustering definitions across many approaches.

\subsection{Overview - Multiple Clustering Techniques Employing Path Lengths}
\label{sec:strategy-2}
The approach taken by this study utilizes four main ideas to produce a robust clustering analysis:
\begin{itemize}
\item Reduce a large data set by binning the space into multi-dimensional partitions and create a serial index for each filled partition.
\item Algorithms use the path length between any connected partitions as well as traditional distance metrics (L1, L2, etc...).
\item Employ multiple clustering techniques to the set of partitions based on first nearest neighbors, distance weighted factors and geometrical properties of the set.
\item Establish a final cluster ID based on all the consensus of techniques employed as well as multiple resolutions and differing variable sets.
\end{itemize}
The combination of multiple clustering techniques, various distance metrics and traditional data reduction lead to a robust set of clusters defined.

From the arrays defined in Sec. \ref{sec:defs-3}, the following clustering techniques are employed:
\begin{itemize}
\item Clusters are sought by proximity using K-means and K-medoids.
\item Clusters are sought globally, finding local peaks based on the maxima of the weights.
\item Clusters are sought by path length, finding local peaks based on the maxima of the weights.
\item Clustering is calculated based on whether partitions are connected or not.
\item Clustering is calculated based on whether partitions are within LOS of each other using two gathering methods based on maximal visibility and highest mutual visibility.
\item Spectral clustering established from a $NN1$ Laplacian based on eigenmodes 1 and 2 or on eigenmodes 2 and 3 using either k-means, k-medoids or 2D histograms for gathering.
\item Spectral clustering established from a LOS Laplacian based on eigenmodes 1 and 2 or on eigenmodes 2 and 3 using either k-means, k-medoids or 2D histograms for gathering.
\item Spectral clustering established from a Laplacian of a Guassian based on eigenmodes 1 and 2 or on eigenmodes 2 and 3 using either k-means, k-medoids or 2D histograms for gathering.
\item Clusters are calculated based on a course binning (Low-Medium-High) of the absolute position of data within the data space.
\item A final cluster ID is assigned from one of four different robust techniques based on reaching consensus amongst the clustering algorithms.
\end{itemize}

\subsection{Analysis Choices}
By choosing to use the bin address space of filled bins only (partitions), the multidimensional nature of the study is reduced to working on a integer based rectilinear grid of bins, facilitating the choices in distance metrics used as well as their interpretation.  Keeping the distance metrics to either L1 or L2 norms and working within a global domain, using ${\bf \Delta R}$, or a local region, using ${\bf \Delta L}$, further simplifies the interpretation of the clusters identified.

The novel approach to data reduction used assists in reducing the computational cost associated with Big Data analysis.  The introduction of weighted partitions requires alterations to standard techniques as the proximity metrics used must be re-evaluated.  The computational advantage generally is orders of magnitude greater as most clustering problems are of the order $N^2$, where $N$ is reduced from the data set size to the partition set size.

Beyond standard clustering algorithms currently employed in the field, a path length approach as well as Line-Of-Sight conditions were added to this study.  Using path lengths provides greater sensitivity to the shape of a distribution, while using LOS criteria identifies convex hull regions within distributions.  Both additions were then further employed within standard techniques, enhancing their capability to find clusters.

Overall, the robust approach accomplishes its goals, where the second approach is the most appropriate in general.  The other three robust techniques help bound the result found in Robust-2.  Iterative applications of the robust approach further help to establish a reasonable k-value for those algorithms requiring an initial guess.

The inclusion of both path length based and LOS criteria enhance the ability to cluster data, accomplishing both the goals of data clustering by proximity as well as data identification by finding subdomains within a distribution that are either in the bulk of a group of data, or are in an overlapping region of two or more distributions.

\paragraph{\bf Acknowledgments.} SW acknowledges the support of ONR Grant No. N00014-01-1-0769. KM acknowledges the support by ONR grants: N00014-15-WX-01814, N00014-16-WX-01705 and N00014-17-WX-01705 as well as the Kinnear Fellowship from the USNA Foundation.

\clearpage


\appendix

\section{General Algorithms}
\label{sec:algorithms-general-1}

\subsection{Finding Local Maxima - FINDMAX}
\label{sec:local-max-1}

\begin{figure*}[t!p]
	\label{fig:maxima-2}
	\centering
	\subfigure[] {
	\label{fig:maxima-2-a}
		\includegraphics[width=0.40\linewidth]{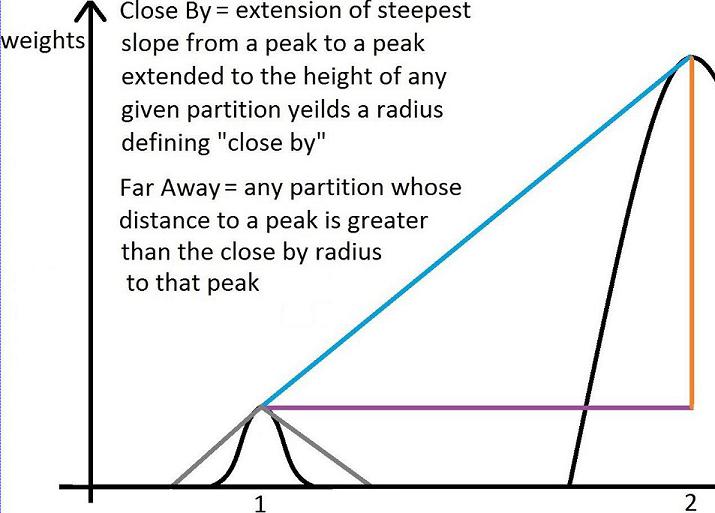} }
	\subfigure[] {
	\label{fig:maxima-2-b}
		\includegraphics[width=0.36\linewidth]{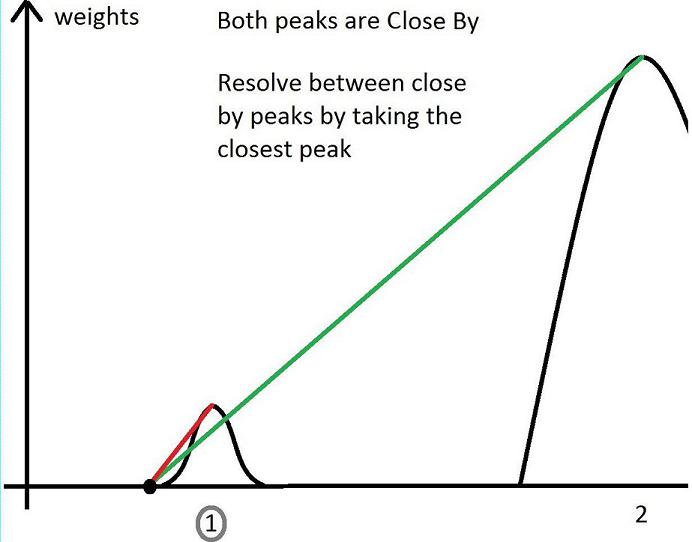} } \\
	\subfigure[] {
	\label{fig:maxima-2-c}
		\includegraphics[width=0.41\linewidth]{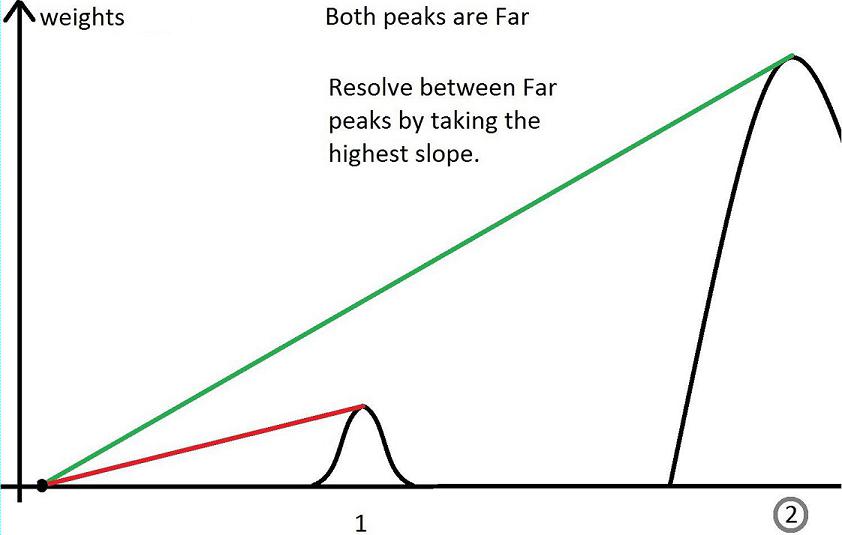} }
	\subfigure[] {
	\label{fig:maxima-2-d}
		\includegraphics[width=0.33\linewidth]{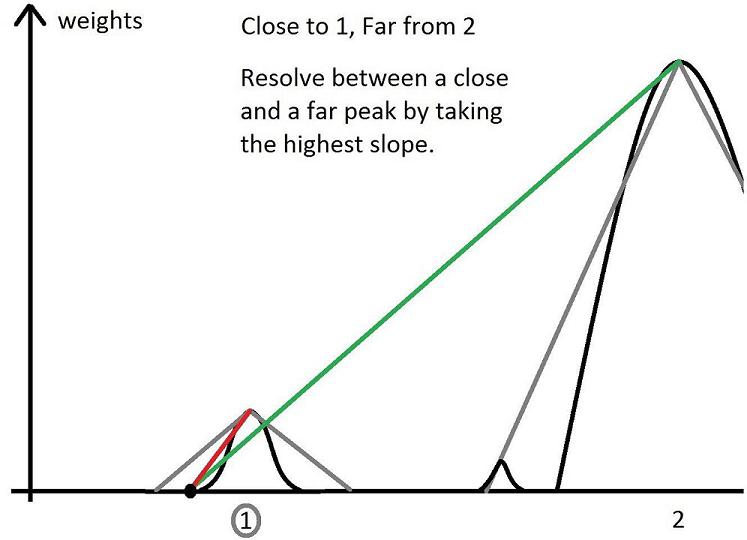} } \\
	\subfigure[] {
    	\label{fig:max-3}
		\includegraphics[width=0.79\linewidth]{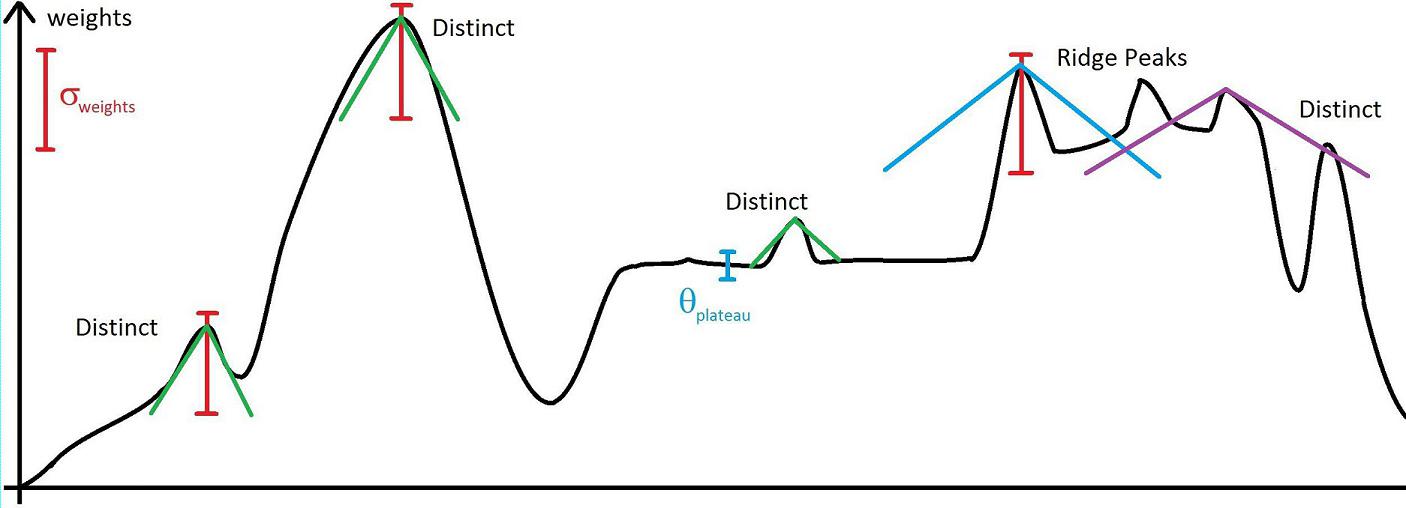} }\\
	\caption{Definition of the slope between peaks (shown in cyan).  Once a slope has been assigned to a peak, $s_q$, the closeness of a partition can be determined by calculating a radius for each peak, $R_{kq}$ compared to the distance to the peak, $\Delta R_{kq}$.  The slopes in the later panels are indicated by the grey lines. Determination of a peak, ridge or a plateau is based on the standard deviation of the weights across the domain, $\sigma(wgt_k)$.  The slopes are indicated by the lines stemming from each peak.  The radii are found by dividing the weight threshold by the slope, $R_q~=~\Theta_{peak}/s_q$.  Peaks are {\em distinct} if no other maxima are found within a radius of the peak, $R_q$.  A {\em ridge} is found when multiple peaks are connected within a range of weights and within the radius of each peak.  Finally, {\em plateaus} are the set of local maxima found within a narrow band of weights, $\Theta_{plat}$.}
\end{figure*}

Consider a contour map based on the weights of the partitions, similar to a multi-dimensional topographic map.  The peaks of the set of weighted partitions  fall into one of three categories:  distinct peaks, ridges and plateaus  Given that the bin resolution may be course over any one dimension, it is difficult to apply a critical value criteria to the weights of the partitions.  A simpler approach to finding the peaks is to first find all local maxima of the partitions by considering the first nearest neighbors ($NN1$) for a given partition.  If all of its neighbors have weights less than or equal to the center, then the center partition is a local maxima.  Among the local maxima, those that have partitions with neighbors of equal weight are set apart as plateau maxima, while the others are peak maxima.

When attempting to classify the various local maxima into categories, it is helpful to first calculate the standard deviation of the weights across the domain, $\sigma(wgt_k)$, where the index $k$ represents the partitions.  Once the standard deviation is calculated, a threshold parameter, $P_{peak}$ is set which together form a threshold, $\Theta_{peak}~=~P_{peak}~\sigma(wgt_k)$.   Similarly, a plateau parameter is also set, leading to the definition of plateaus, with the difference being that the band of weights is much smaller and in both directions, such that a ``plateau'' is defined as all of the connected partitions to a given local maxima within the threshold:
 \begin{equation}
 wgt_q-\Theta_{plat} < wgt_{k} \leq wgt_q + \Theta_{plat}
 \end{equation}
where $\Theta_{plat}~=~ P_{plat}~\sigma(wgt_k)$.  Finally, a check to ensure that all found maxima are connected to one another is performed.  All maxima found by this criteria are considered part of one plateau and given a single cluster index for the set.  A special plateau is formed from the set of all local maxima whose weight is close to zero (low population).  The same criteria given above is used to define this plateau, however, partitions found within this plateau may be released from this designation to allow these minimally populated partitions the chance to be associated with other higher maxima.  Once a partition has been classified as part of a plateau, it is no longer used in searches for further peaks, leaving the remaining peaks to be either distinct or part of a ridge.

\begin{figure*}[t!p]
	\centering
	\subfigure[] {
    	\label{fig:max-3}
		\includegraphics[width=0.79\linewidth]{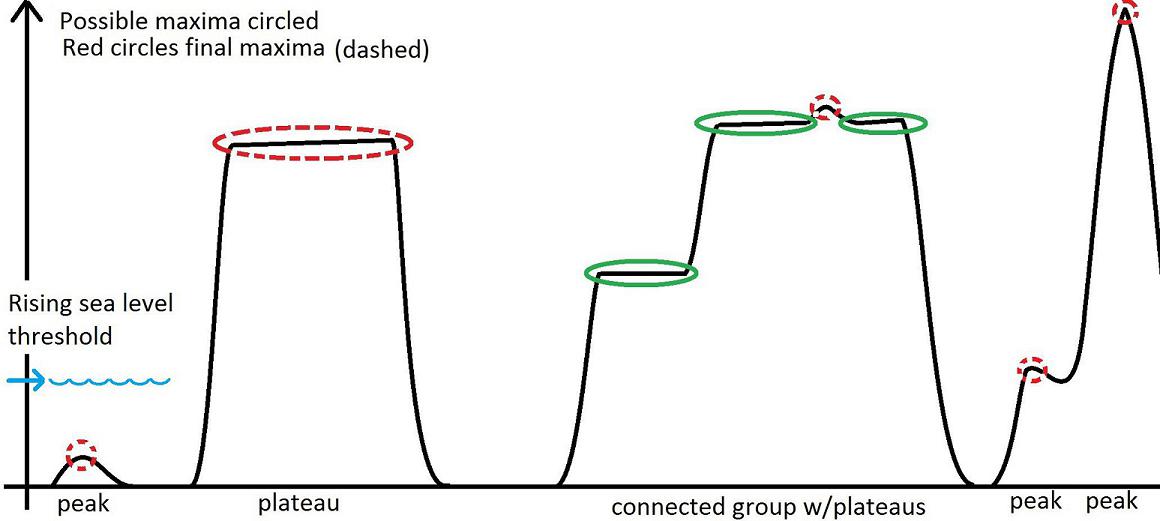} }\\
	\caption{Types of possible maxima, plateaus and peaks. While searching for the maxima within each group of partitions, first potential maxima are identified (shown circled).  By finding all connected partitions higher than a possible maxima, if any other maxima are connected to the lower maxima, the lowest is removed from consideration, leaving only successively higher and higher maxima until only one maxima is left per connected group of partitions.}
	\label{fig:maxima-3}
\end{figure*}

\begin{algorithm}
\caption{Find Maxima algorithm - Global and Path Length      \hspace{3.0cm}(uses FastConn)}
\label{alg:maxglobpathl}
\begin{algorithmic}[1]
\Procedure{FindMax}{$k,w_k,\Delta w_{k\ell},\Delta r_{k\ell},\Delta \ell_{k\ell}$}   \Comment {Requires a list of partitions, weights,~~~}
    \State ~\Comment {weight differences, distance differences}
    \State $\sigma_{_{P}} \gets stddev(w_k)$                 \Comment{Calculate standard deviation of weights}
    \State $\Theta_{plateau} \gets 0.02\sigma_{_{P}}$        \Comment{Threshold for weights to be within a {\em plateau}}
    \State $ik' \gets 1$                                     \Comment{index for possible maxima}

    \ForAll{$partitions,k$}                                    \Comment{Compare center to first nearest neighbors, $NN1$}
        \If{$w_{_{NN1,\ell}}<=w_k$}                          \Comment{If center  has the greatest weight (or tied)}
            \State {$possiblemax[ik',1]\gets k$}             \Comment{Add center to list of possible maxima}
            \State {$possiblemax[ik',2]\gets w_k$}           \Comment{Add center weight to list of possible maxima}
            \State $ik' \gets ik' + 1$                       \Comment{k' Index for {\em possiblemax}}
        \EndIf
    \EndFor     \Comment{See Fig.\ref{fig:maxima-3} where {\em possiblemax} are circled}.

    \ForAll{{\em possiblemax}}                               \Comment {Adds {\em plateaus} to {\em possiblemax}}
        \If{abs($\Delta w_{k'\ell})<\Theta_{plateau}$}       \Comment {Find other partitions weights within a}
            \State $possiblemax[ik',1] \gets \ell$           \Comment {narrow range of the current {\em possiblemax}}
            \State $possiblemax[ik',2] \gets w_\ell$
            \State $ik' \gets ik' + 1$                       \Comment {Add these partitions, $\ell$, to set of {\em possiblemax}}
        \EndIf
    \EndFor

    \State ~  \Comment{``Rising sea'' criteria to find one maxima per group of {\em possiblemax}~~~~~~~~~~~~~~~~~~~~~~~~~~~~~~}
    \State $possiblemax \gets sort(possiblemax, 2, ascend)$       \Comment {Reorder by ascending weights}
    \State $removemax \gets zeros$                                \Comment {Initialize removal array to zero}
    \State $ik' \gets 1$
    \ForAll{$possiblemax(ik')$}                                   \Comment {Find set of all partitions higher than}
        \State $partcheck \gets k(~w_\ell>possiblemax(ik',2)~)$   \Comment {~~~~~~~~current possiblemax, $w_\ell>w_{k'}~$}
        \State {$connpartcheck \gets FastConn(partcheck)$}        \Comment {Find all connected partitions to $k'$}
        \ForAll{$connpartcheck(\ell \neq k')$}
            \If{$connpartcheck \subset possiblemax(\ell)$}        \Comment {Set current {\em possiblemax}, k', for removal}
                \State {$removemax(ik') \gets 1$}                 \Comment {if there is a higher $possiblemax_\ell$ in the}
            \EndIf                                                \Comment { connected group of partitions}
        \EndFor
        \State $ik' \gets ik'+1$
    \EndFor
    \State {$possiblemax(removemax) \gets [empty]$}               \Comment {Remove all maxima from a group except highest}
    \State {$maximafinal \gets possiblemax$}
    \State \textbf{return} $maximafinal$                          \Comment{Returns list of partition maxima.}
\EndProcedure
\end{algorithmic}
\end{algorithm}

For each peak, a slope, $s_q$, is found which will be used to determine when another partition is close by or not.  The slope is used both to find ridges as well as determine clusters for
partitions to peaks.   Figure 16(a) illustrates the means of determining the slope for a given peak, labeled ``q''.  Taking the set of all local maxima found, for each maxima (q), a set of slopes can be found to all other local maxima, q'(including maxima found in plateaus).  The greatest slope in magnitude is taken as the slope assigned to each peak:
\begin{eqnarray}
s_{qq'} & = & \left|\frac{\Delta wgt_{qq'}}{\Delta R_{qq'}}\right|\\
s_q     & = & Max[~s_{qq'}~]
\end{eqnarray}
Although clustering can be restricted to just the definitions of peaks and plateaus, when a group of peaks are close together, they can act as a single entity called a {\em ridge}.  Clustering to a ridge is possible provided a distance can be found which determines when one peak is close to another.  For the purpose of determining whether a peak is distinct or part of a ridge, a radius is found for each peak by taking a weight difference and dividing by the slope.  Starting from the highest weighted partition, a band of partitions is found for all whose weights are between the given maxima weight down to the threshold:
\begin{equation}
(wgt_{min} \equiv wgt_q-\Theta_{peak}) < wgt_{k} \leq wgt_q
\end{equation}
with $q$ being the index of the current maxima and $k$ being the index of all connected partitions to $q$.  The set of partitions is further restricted by requiring that the distance between the maxima and the surrounding partitions be within the radius set by:
\begin{equation}
{\rm Radius~peak}~~~ \equiv~~ R_{q}  =  \frac{wgt_q-wgt_{min}}{s_q}
\end{equation}
Finally, a check to determine if the set of partitions is connected to the maxima is done.  A {\em distinct} peak is then any maxima which is a single maxima found within a connected group of partitions within a band of weight values which are close by to the peak, like the Matterhorn, Switzerland.   If other maxima are found within the connected set of partitions and within the radius, then the peak and others found form a ridge.  Subsequent checks on the other peaks are made to determine if the ridge continues to find further peaks.  When determining the radius for subsequent peaks in the ridge, the same lower value is used in the weight band, $wgt_{min}$.  In this manner, the full ridge is found by cascading the peak search to all other peaks found until the search within the set of ridge peaks is exhausted, at which point, a single cluster identification is given to the entire ridge, examples would include the Grand Tetons, USA or Valais, Switzerland.  Figure \ref{fig:maxima-3} illustrates how local maxima are classified into plateaus, distinct peaks or ridges.

Following the analogy to a height field, a criteria is established to determine which peaks are the most significant to their surrounding partitions.  By determining the set of partitions the most closely associated with a peak, a cluster is formed around that peak.  The criteria for determining which peak is the most appropriate is based on the maximal slope of any given peak to another peak - including the plateau local maxima.   When considering which peak a partition in the domain is associated with, the slope and distance to peak are the two parameters
used.  Figures 16(b)-16(e) show the scenarios that a partition may encounter.  In each case, the distance between a partition and a peak, $\Delta r_{kq}$, is
found to be either ``close by'' or ``far away'' based on a radius determined by the slope for each peak.    The radius is found by taking the partitions weight difference and dividing by the slope for the peak:
\begin{eqnarray}
{\rm Radius}~~~ \equiv~~ R_{kq} & = & \frac{wgt_q - wgt_k}{s_q} \\
{\rm Close~By}~~\equiv~~ \Delta R_{kq} & \leq & R_{kq} \\
{\rm Far~Away}~~\equiv~~ \Delta R_{kq} & > & R_{kq} \\
\end{eqnarray}

\begin{algorithm}
\caption{Connection Cluster algorithm  ~~~~~(Matlab notation)}
\label{alg:connclus1}
\begin{algorithmic}[1]
\Procedure{CONNCluster}{${\bf NN1}$}
    \State ${\bf NN1tmp} \gets {\bf NN1} + eye(N_P)$                       \Comment {Add the Identity matrix to NN1}
    \State $[p,q,r,s,cc,rr] = dmperm({\bf NN1tmp})$                  \Comment {Routine dbperm permutates the $NN1$ matrix ~}\\
                                                               \Comment {$p$, array of all partitions  ordered by connected clusters.}\\
                                                               \Comment {$r$, array of starting indices for each connected cluster.}
    \State $clustercnt \gets 0$
    \ForAll{$i \in r$ (except last entry)}
        \State $clustercnt \gets clustercnt+1$
        \State ${\bf CONN}(clustercnt,:) \gets p(r(i):r(i+1)-1)$
    \EndFor
    \State \textbf{return} ${\bf CONN}$                              \Comment {Returns matrix of clusters of connected partitions.}
\EndProcedure
\end{algorithmic}
\end{algorithm}

\subsection{Path Length Based Calculations}
\label{sec:pathl01}

This study employs clustering techniques utilizing the path length between two partitions.  When two partitions cannot form a path between them, the path length matrix entry is set to infinity. Dijkstra's algorithm (\citeyear{dijkstra}) and variations of it are used to calculate the shortest path length.  The first nearest neighbor matrix, ${\bf NN1}$, is a weighted adjacency matrix used to search for the path starting from one partition ($k$) continuing making NN1 steps until reaching a final partition ($\ell$).
\cbk

When partitions are connected it is useful to reorder the partitions so that their values in matrix form appear block-diagonal.  The algorithm of Dulmage and Mendelsohn permutes rows and columns until a block form is reached (\citeyear{dmperm}).  Algorithm \ref{alg:connclus1}  shows how partitions are reordered as well as assigned to clusters via connection.
\vspace{-0.1cm}
\begin{definition}[{\bf Path Type Values}]~\\
    $Stepwise      $\tabto{2.20cm}$\leftarrow$ A value defined between nearest neighbors summed as a path is traversed.  \\
    $Pathwise      $\tabto{2.20cm}$\leftarrow$ A value defined between the originating partition and each subsequent partition along a path which is summed as the path is traversed. \\
\label{def:paths1}
\end{definition}
The path length calculation uses the ${\bf NN1}$ matrix which gives the L2 distance from one partition to a neighboring partition.  When using the ${\bf NN1}$ matrix with Dijkstra's algorithm, the path length that is calculated sums the step-by-step distance when going from $[k,\ell]$, leading these calculations to be called {\em stepwise}.   Other path-based calculations needed in this study require distances to be summed from the $k^{th}$ partition to each subsequent partition along the path, dubbing these calculations {\em pathwise}.  The criteria used to establish a Line-Of-Sight (LOS) between two partitions relies on the summation of L1 distances to points along the path taken from $[k,\ell]$.   This distance has the property that when traversing a grid from $[k,\ell]$, the distance calculated is different than when returning from $[\ell,k]$.  The asymmetry of this measure proves useful in determining
the LOS condition.  Figure 18 illustrates how the distance is asymmetric with regard to the path taken.
\begin{algorithm}
\caption{Path Length Algorithm ~~~~~(Matlab notation)}
\label{alg:lah02}
\begin{algorithmic}[1]
\Procedure{PathLength}{${\bf NN1}$}                             \Comment{Compute a stepwise L2 path length.}
    \State $GraphPath \gets graph({\bf NN1})$                   \Comment{Form  graph from symmetric weighted adjacency matrix.}
    \State ${\bf PathL} \gets distances(GraphPath)$                   \Comment{Use Dijkstra's Algorithm for distance.}
    \State \textbf{return} ${\bf PathL} \equiv {\bf \Delta L}$        \Comment{Matrix of distances between partitions (stepwise).}
\EndProcedure
\end{algorithmic}
\end{algorithm}

\begin{algorithm}
\cbk
\caption{Path Length Algorithm (Asymmetric)~~~~~(Matlab notation)}
\label{alg:lah03}
\begin{algorithmic}[1]
\Procedure{PathLengthAsym}{${\bf NN1,AsymMatrix}$}\\
                                                                \Comment{Compute a pathwise, asymmetric path length.}
    \ForAll{$partitions,k$}                                     \Comment{Loop through each partition, row by row.}
        \State $AsymRow \gets {\bf AsymMatrix}(k,:)$                          \Comment{Copy the $k^{th}$ row to fill a square matrix}
        \State ${\bf AdjacencyAsym} \gets repmat(AsymRow,N_P,1) \circ logical({\bf NN1})$\\
                                                                        \Comment{Create an asymmetric Adjacency matrix.}
        \State $GraphPath \gets graph({\bf AdjacencyAsym})$ \\
                                                                \Comment{Forms a directed  graph from adjacency matrix.}
        \State ${\bf PathLAsym}(k,:) \gets shortestpath(GraphPath,k,:)$     \Comment{Shortest distance from $k$ to rest.}
    \EndFor
    \State \textbf{return} ${\bf PathLAsym}$                       \Comment{Matrix of pathwise distances between partitions.}
\EndProcedure
\end{algorithmic}
\cbk
\end{algorithm}

In path-based calculations, a graph is formed based on an adjacency matrix which will either be the ${\bf NN1}$, ${\bf SL1}$ or ${\bf SL1VAR}$ matrices.  The ${\bf NN1}$ matrix is symmetric and stepwise, which is used as a weighted adjacency matrix to form an undirected graph which is then used to calculate the path length based on L2 steps taken from $[k,\ell]$.   This application of the adjacency matrix to form a graph is a standard approach, outlined in algorithm \ref{alg:lah02}.

The ${\bf SL1}$ distances form an asymmetric and pathwise matrix, which requires an alteration to how Dijkstra's algorithm is applied to calculating the summed L1 path length.  Starting from the first partition, the starting partition is fixed ($k$) and the top row of the ${\bf L1}$ matrix is copied to fill a square matrix.  This matrix represents the L1 distances from $k$ to all other partitions.   To insure steps are taken by nearest neighbors only, the ${\bf NN1}$ matrix is first changed to logical values then multiplied element by element with the ${\bf L1_k}$ matrix just created for partition $k$ only.    By applying this matrix as the weighted adjacency matrix, a directed graph is formed, where each step along the path from $[k,\ell]$ is then summed using the ${\bf L1_k}$ value, but in each case, when taking a new step utilizing Dijkstra, the value used is always fixed to the distance from the initial partition, making this calculation pathwise rather than stepwise.  After all path distances have been calculated from the current partition to all others, the process is repeated for all other rows by fixing the next partition (row) and copying that row to fill a square matrix, at which point Dijkstra's algorithm is applied giving the pathwise summed L1 distances from the new partition to all others.  The final collection of summed L1 path distances form the matrix ${\bf SL1}$.

The ${\bf SL1VAR}$ matrix is asymmetric, pathwise and treated in a similar fashion as the ${\bf SL1}$ matrix in order to calculate the minimal variance of the summed L1 distance from the true summed L1 distance for a linear path.   Algorithm \ref{alg:lah03} outlines the procedure for treating asymmetric path calculations using Dijkstra.  The usage of ${\bf SL1}$ and ${\bf SL1VAR}$ in calculating the LOS criteria are discussed in App. \ref{sec:los-3}.
\cbk


\section{Line-of-Sight (LOS) Criteria }
\label{sec:los-3}
\cbk
A Line-Of-Sight algorithm determines whether two partitions can be connected by an unbroken line.  The ${\bf LOS}$  matrix produced is a logical matrix indicating no direct path (false) or direct path (true) exists for the $LOS_{k\ell}$ element.   The three LOS criteria will be discussed in this section, leading to the formation of the LOS matrix.
Each criteria places tighter restrictions on partitions to be LOS:
\begin{enumerate}
\item A path or set of paths must exist between $[k,\ell]$ equal to the true path length, $\Delta L_{T}$.  The set of all paths sharing this length form a parallelotope with partitions $k$ and $\ell$ placed at the far corners of the region,  called the $L2_{T}$ {\em convex hull}.  This criteria is the least restrictive and requires the least computational effort.
\item The summed L1 distance between $[k,\ell]$ along all paths  within the $L2_{T}$ convex hull must be less than or equal to the true summed L1 distance, $\Delta SL1_{T}$.  This is a strong restriction which removes all paths taken around a corner between $[k,\ell]$.   The computational time is greatest for this criteria.
\item The sum of the squared differences of the summed L1 distance and true summed L1 distance, $SL1VAR(k,\ell)$ must be less than the pathcount.  This is the most restrictive criteria demanding that a path form a line between $[k,\ell]$.  The computational time is generally less than the second criteria but can be as much.
\end{enumerate}

\begin{figure}[t!]
	\centering
 	\subfigure[] {
        \label{fig:interior01}
		\includegraphics[height=8.0cm]{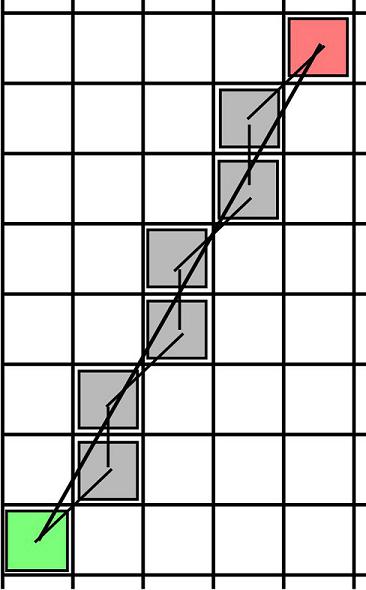} }
	\subfigure[] {
         \label{fig:interior02}
		\includegraphics[height=8.0cm]{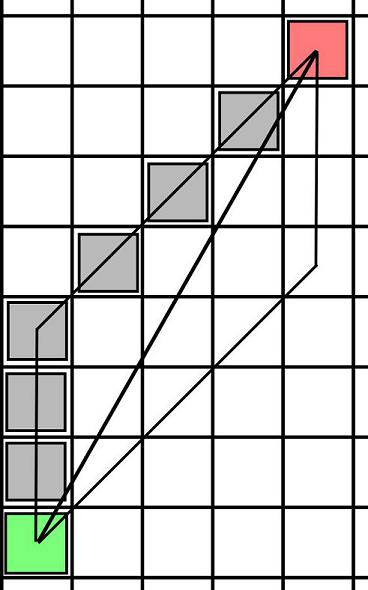} }
	\subfigure[] {
        \label{fig:interior03}
		\includegraphics[height=8.0cm]{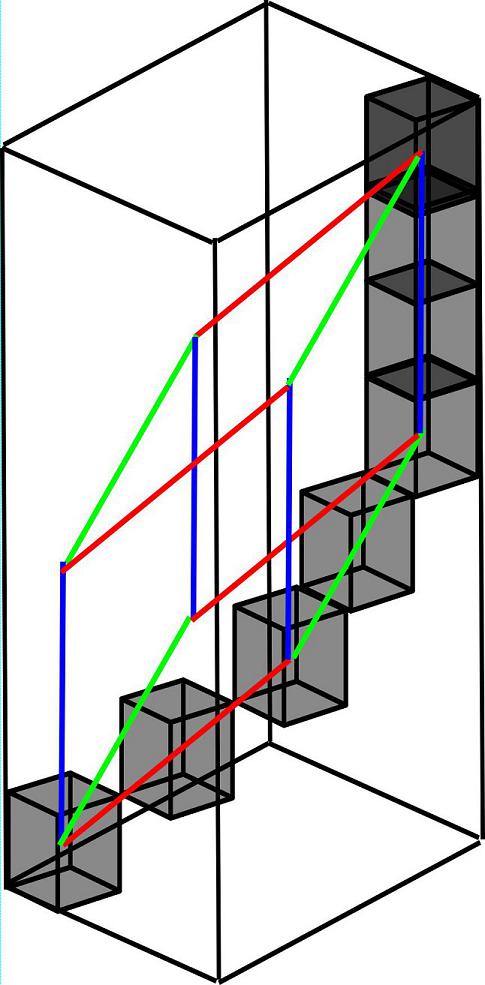} } \\
	\caption{Illustrating the True Path Length from the $k^{th}$ partition (green) to the $\ell^{th}$ (red).  The left panel shows the direct path and the nearest neighbor steps required to follow that path.   The middle panel shows that the same path length can be taken along the edge of a parallelogram, bounded by the rectangular region between $[k,\ell]$.  The right panel shows how to extend the idea into higher dimensions.}
\end{figure}

The first criteria determines whether two partitions are connected within a trapezoidal region, defined by the vector of differences of their individual bin addresses, ${\bf \Delta b}$.  In 2D, this forms a parallelogram while in $N_D$ this is a parallelotope.  If two partitions are connected within the parallelotope by some path, see Figs. \ref{fig:interior01}-\ref{fig:interior03}, they may be LOS, however, when a path is taken outside of this region, the two partitions are not LOS to each other.

\subsubsection{Path Length Calculations}
\label{sec:interior1}

All LOS criteria listed above require several calculations to be done based on two types of path lengths, L1 and L2 norms.  There are five path based calculations needed (in all cases between $[k,\ell]$):  the true L2 path length, ${\bf \Delta L_{T}}$, the true summed L1 path length, ${\bf \Delta SL1_{T}}$, the minimal summed L1 path length, ${\bf \Delta SL1_{min}}$, the maximal summed L1 path length, ${\bf \Delta SL1_{max}}$, and the summed L1 path difference, ${\bf \Delta SL1_{\Delta}}$.   Also essential to these calculations is the orthogonal convex hull formed between $[k,\ell]$, which defines the set of partitions that can be considered for a path that is LOS.  The true L2 path length is the shortest path length possible taking NN1 L2 steps, previously defined in Sec.\ref{sec:calcs-1}.  The true summed L1 path length has a similar definition, being the summed L1 distances taken along the straight line path.   The minimal summed L1 path is the path taken which minimizes the summed L1 distance and the maximal summed L1 path maximizes this length.  The L1 path difference is the difference between the maximal summed L1 and minimal summed L1 path distances.  In addition to the path based calculations, four vectors are required based on the orthogonal convex hull; the {\em corner rank vector}, ${\bf CR}$, which is simply $[1...N_D]$, which delineates the types of corner steps taken along a path, the {\em convex hull length} vector, ${\bf CVL}$, which is the difference of the bin address vectors of $[k,\ell]$, the reverse sorted convex hull length, ${\bf RSCVL}$, and the forward difference reverse sorted vector of the orthogonal convex hull length, ${\bf \Delta RSCVL}$.
\begin{eqnarray}
{\bf CVL}        & = & {\bf b}_{k} - {\bf b}_{\ell} \\
{\bf RSCVL}      & = & {\rm sort}\left({\bf CVL},{\rm ~'descend'}\right)                            \hspace{2.6cm}({\rm Matlab~notation})\\
{\bf CR}         & = & [1...N_D] \\
{\bf \Delta RSCVL} & = & {\rm diff}\left(\left[~{\bf RSCVL}~0\right]\right)                         \hspace{3.0cm}({\rm Matlab~notation})
\end{eqnarray}

Figures \ref{fig:interior01}-\ref{fig:interior03} illustrate how the path based calculations are made. The two partitions in this calculation form an orthogonal convex hull, with the $k^{th}$ partition in one corner and the $\ell^{th}$ partition in the farthest other corner.  The centers of each square (partition) are the bin indices, beginning with the lower left as (1,1) to the upper right at (5,8) for the 2D case; for the 3D case, the lower left bin is (1,1,1) and the upper right is (3,5,8).  The convex hull lengths, ${\bf CVL}$ are [4,7] and [2,4,7] respectively and the sorted and descending convex length differences, ${\bf \Delta RSCVL}$, are [3,4] and [3,2,2] respectively.  The reverse sorted convex hull length difference vector is the number of steps required to take at each successive corner rank, with a corner rank of one being a straight step, rank two a step in two dimensions, etc...

\begin{table}
\label{tab:l1}
\caption{Table of steps, coordinates and the summed L1 path distance for the two examples given by Figs. \ref{fig:interior02} and \ref{fig:interior03}.  The top three rows represent the number of steps taken along each dimension {\em from one partition to the next}.  The next three rows are the coordinates of the partitions along the paths taken, remembering that the path starts with the $k^{th}$ partition at the origin.  The final row is the summed L1 path distance.}
\resizebox{\columnwidth}{!}{
\hspace{-0.6cm}
\begin{tabular}{l|ccccccccccccccccccccccccccccccc}
\hline
dims  &           \multicolumn{31}{c}{Steps Taken Along Each Dimension - $k^{th}$ partition is the origin}           \\
& \multicolumn{7}{c}{2D Case (min path) } & & \multicolumn{7}{c}{2D Case (max path) } & & \multicolumn{7}{c}{3D Case (min path)} & & \multicolumn{7}{c}{3D Case (max path)}    \\ \hline
$x_1$ & 1 & 1 & 1 & 1 & 1 & 1 & 1 &   \hspace{0.3cm}   & 1 & 1 & 1 & 1 & 1 & 1 & 1 & \hspace{0.5cm} & 1 & 1 & 1 & 1 & 1 & 1 & 1 & \hspace{0.3cm} & 1 & 1 & 1 & 1 & 1 & 1 & 1   \\
$x_2$ & 0 & 0 & 0 & 1 & 1 & 1 & 1 &                    & 1 & 1 & 1 & 1 & 0 & 0 & 0 &                & 0 & 0 & 0 & 1 & 1 & 1 & 1 &                & 1 & 1 & 1 & 1 & 0 & 0 & 0   \\
$x_3$ &   &   &   &   &   &   &   &                    &   &   &   &   &   &   &   &                & 0 & 0 & 0 & 0 & 0 & 1 & 1 &                & 1 & 1 & 0 & 0 & 0 & 0 & 0   \\
\hline
\hline
      &           \multicolumn{31}{c}{Coordinates Of Path Along Each Dimension - $k^{th}$ partition is the origin}           \\
 & \multicolumn{7}{c}{2D Case (min path) } & & \multicolumn{7}{c}{2D Case (max path) } & & \multicolumn{7}{c}{3D Case (min path)} & & \multicolumn{7}{c}{3D Case (max path)}   \\ \hline
$x_1$ & 1 & 2 & 3 & 4 & 5 & 6 & 7 &                    & 1 & 2 & 3 & 4 & 5 & 6 & 7 &                & 1 & 2 & 3 & 4 & 5 & 6 & 7 &                & 1 & 2 & 3 & 4 & 5 & 6 & 7   \\
$x_2$ & 0 & 0 & 0 & 1 & 2 & 3 & 4 &                    & 1 & 2 & 3 & 4 & 4 & 4 & 4 &                & 0 & 0 & 0 & 1 & 2 & 3 & 4 &                & 1 & 2 & 3 & 4 & 4 & 4 & 4   \\
$x_3$ &   &   &   &   &   &   &   & & & & &  &   &   &  & \hspace{-4.5cm}$\Uparrow${\rm $x_2$~holds~at~4} & 0 & 0 & 0 & 0 & 0 & 1 & 2 &                & 1 & 2 & 2 & 2 & 2 & 2 & 2   \\
      &   &   &   &   &   &   &   & & & & &  &   &   &   & &  &  &  &  &  &  &  &                              &  &  &  &  &  &  &  \hspace{-5.7cm}$\Uparrow${\rm $x_3$~holds~at~2}  \\
      &   &   &   &   &   &   &   & & & & &  &   &   &   & &  &  &  &  &  &  &  &                              &  &  &  &  &  &  &  \hspace{-2.7cm}$\Uparrow${\rm $x_2$~holds~at~4}  \\
\hline\vspace{0.2cm}
${\bf SL1}$ & 1 & 3 & 6 &11 &18 &27 &38 &              & 2 & 6 &12 &20 &29 &39 &50 &                & 1 & 3 & 6 &11 &18 &28 &41 &                & 3 & 9 &17 &27 &38 &50 &63   \\
\hline
\hline
\end{tabular}
}
\end{table}

As figures \ref{fig:interior01},\ref{fig:interior02} show in 2D, the path length along the direct path between $[k,\ell]$ is the exact same length as if taken along the edge of a parallelogram bounded within the convex hull. In 3D, Fig.\ref{fig:interior03}, the same is true, however, two differing kinds of corner steps can be taken.  In higher dimensions, the set of all paths taken from $[k,\ell]$ that have the same L2 path length form a parallelotope.  The straight line path between $[k,\ell]$ shares the same path length as a path moving along the edge of the parallelotope, making the calculation for the true L2 path length simple by taking the dot product of the reverse sorted difference convex hull length vector with the square root of the corner rank vector:
\begin{equation}
\label{eqn:truepath}
{\bf \Delta L2}_{T} ~ = ~ {\bf \Delta RSCVL}\cdot\sqrt{{\bf CR}}
\end{equation}
The first LOS criteria demands that partitions considered in further LOS tests must have the path length equal to the true path length, ${\bf \Delta L}~=~{\bf \Delta L2_{T}}$, such that these partitions are located within the parallelotope, forming the more constrained  $L2_{T}$ {\em convex hull}.  The path length algorithm has been discussed in Sec.\ref{sec:pathl01}.

The summed L1 path length is a pathwise length described in Sec. \ref{sec:pathl01}, where the L1 norm is calculated for a set of partitions.  Using a modified Dijkstra's algorithm, the minimal path is found of the summed L1 norms taken along the path from $[k,\ell]$.  The minimal summed L1 path, ${\bf \Delta SL1}_{min}$, does not take steps directly from $k$ to $\ell$, rather, it takes steps which minimize the L1 norm by moving in a line taking no corners for as long as possible.  After exhausting all steps with no corners taken, the algorithm then finds the path with the most steps taken with the nest lowest rank corner, a 2D corner.  After exhausting this option, further corner steps are taken until the final partition, $\ell$, is reached.  Table \ref{tab:l1} outlines this process.  The columns in the table represent steps taken.  The rows represent the number of corners taken per step.  Initially, all steps are 1D (a line).  The number of possible steps that can be taken at each corner rank is given by the forward difference reverse sorted convex hull length vector, ${\bf \Delta RSCVL}$.  For the examples given, in the 2D case, the minimal summed L1 path will take three 1D steps followed by four steps in 2D.  For the 3D case, three steps will be 1D followed by two 2D steps then two 3D steps.  The table represents these steps by assigning a one or a zero for the number of dimensions used for the corner of that particular step.  The minimal summed L1 path takes the larger corners last in the sequence, which prevents the summed L1 value from becoming large.   In Fig. \ref{fig:interior02} the minimal path corresponds to the upper edge of the parallelogram.

By contrast, the maximal summed L1 path, ${\bf \Delta SL1}_{max}$, takes the highest rank corners first and ends with the lowest rank steps.  This path corresponds to the lower edge of the
parallelogram in Fig. \ref{fig:interior02}.  Table 4 shows the effect of taking the higher ranked corners first.  The number of steps at each rank is the reverse order of ${\bf \Delta RSCVL}$.   By taking the largest ranked corners first, once the steps are exhausted for these ranked corners, the L1 value for that dimension is held constant for the remainder of the summed L1 path length.  In the 2D case, four steps are taken at rank two, then all further steps have the value four as part of the L1 norm.  Finishing the steps in the example, the last three steps each have four added.   In the 3D case, two steps at rank three are taken, then two steps at rank two then three steps at rank one.  The middle three rows are the coordinates at each step. The last row shows the running sum along each dimension, which is added together to form the summed L1 path distance.

The straight line path between $[k,\ell]$ will have the {\em true} summed L1 path distance, ${\bf \Delta SL1_{T}}$, which is the average of the two extrema, ${\bf \Delta SL1}_{min},{\bf
\Delta SL1}_{max}$.  The difference between the two extrema is also necessary to calculate, ${\bf \Delta SL1}_{\Delta}$.   From table 4, the calculations for the extrema are easiest to see if calculated {\em along each dimension}.  In the minimal summed L1 case, this is simply the cumulative summation of ones for the convex hull length along each dimension. In the maximal summed L1 case, the calculation is similar, yet the highest coordinate reached along each dimension is now held for the remainder of the sum, indicated in the table.  In each calculation, the total number of steps is simply the path count, $P_C$, defined in Sec. \ref{sec:calcs-1}.
\begin{eqnarray}
{\bf \Delta SL1}_{min}     & = &  \sum_{i=1}^{N_D} \frac{1}{2} RSCVL_i \left(RSCVL_i +1\right)  \nonumber \\
                           & = &  \tfrac{1}{2}\left({\bf \Delta SL1} + {\bf \Delta L2}^2\right)   \\
{\bf \Delta SL1}_{max}     & = &  \sum_{i=1}^{N_D} \frac{1}{2} RSCVL_i \left(RSCVL_i +1\right) + \sum_{i=1}^{N_D} (P_C - RSCVL_i) RSCVL_i  \nonumber \\
                           & = &  \left(P_C + \tfrac{1}{2}\right) {\bf \Delta SL1} - \tfrac{1}{2}{\bf \Delta L2}^2 \\
{\bf \Delta SL1}_{T}       & = &  \tfrac{1}{2} \left(P_C + 1\right) {\bf SL1} \\
{\bf \Delta SL1}_{\Delta}  & = &  {\bf \Delta SL1}_{max} - {\bf \Delta SL1}_{max} ~~~~~=~~~~ P_C {\bf \Delta SL1} - {\bf \Delta L2}^2
\end{eqnarray}

When using Dijkstra's algorithm to find a path length, the minimal value at any given step along the path is always found, meaning that in the case of the summed L1 path distance, the algorithm will find the lowest value it can going from $[k,\ell]$.  If all paths within the $L2_{T}$ convex hull are available to traverse, then Dijkstra will find the path with ${\bf \Delta SL1}_{min}$.  When returning from $[\ell,k]$ Dijkstra's algorithm will find a new minimal pat, which is equivalent to the {\bf maximal} path when going from $[k,\ell]$.  The
asymmetry Dijkstra's algorithm creates is useful in finding the correct straight line path, illustrated in Fig. 17.   Due to this asymmetry, two values are considered when determining LOS, ${\bf \Delta SL1}$ for $[k,\ell]$ as well as the transpose, ${\bf \Delta SL1}^{\dag}$ for $[\ell,k]$.  The minimal and maximal paths in 2D form a plane which passes through the true summed L1 path.  In higher dimensions, the minimal and maximal paths are not constrained to a plane, however, they do follow two extreme edges of the $L2_{T}$ convex hull.  Fitting a line through each extreme path again gives a plane whose center line is the true summed L1 path.  The values of the all paths within the $L2_{T}$ convex hull are either below the true value (closer to the minimal path) or above the true value (closer to the maximal path) as shown in Fig. 17.

The second criteria to determine LOS compares the two values $\left({\bf \Delta Sl1}, {\bf \Delta SL1}^{\dag}\right)$ previously mentioned to determine if a path turns a corner or not.  When a limited number of paths are available between $[k,\ell]$, three possibilities can exist.  The first case is if the two partitions are around a corner from each other, where one of the two summed L1 distances will be forced to be greater than ${\bf \Delta SL1}_{T}$, which means that there is no LOS path from $[k,\ell]$.  The second case could be that two paths exist which give both summed L1 values below ${\bf \Delta SL1}_{T}$, however, a straight line path is not possible due to an empty bin along the straight path.  In this case, both paths encompass the empty bin, such that the criteria $\left({\bf \Delta SL1}, {\bf \Delta SL1}^{\dag}\right)~<~{\bf \Delta SL1}_{T}$ is a false positive, it fails LOS.  The third case is that one of the two summed L1 values is equal to ${\bf \Delta SL1}_{T}$, in which case, the two partitions are LOS to each other and no further criteria is required.

In order to resolve the second case, Dijkstra's algorithm is employed a third time, using the asymmetry from the two values given from $[k,\ell]$ and $[\ell,k]$, a new pathwise value is applied to Dijkstra's algorithm, the squared difference of the summed L1 path distance from the true summed L1 distance, called the summed L1 variance:
\begin{equation}
\label{sl1var}
{\bf SL1VAR} ~=~ \sum_{j=k}^{\ell} \mid {\bf \Delta SL1}_{kj} - {\bf \Delta SL1}_{T,kj} \mid^2
\end{equation}
The value of ${\bf SL1VAR}$ is similar to a variance in that it is minimized around the true value.  For this reason, using ${\bf SL1VAR}$ as the pathwise value applied to Dijkstra's algorithm will seek a minimal path from $[k,\ell]$ along the path closest to the straight path.  For each step along the {\em straight line path}, the difference between the summed L1 distance and the true distance should never be greater than one or one-half, such that the summation of steps along a straight line path should not be greater than the path count, $P_C$.  The third LOS criteria is that the summed L1 variance should be: ${\bf SL1VAR}~\leq~P_C^2$.  This criteria is too conservative as the $L2_{T}$ convex hull may be constrained such that of the paths available, all are closer to the straight path than this limit, in which case, an adaptive criteria for LOS is that ${\bf SL1VAR}~\leq~\tfrac{1}{4}{\bf \Delta SL1}_{\Delta}$ which is halfway between the minimal summed L1 distance and the true summed L1 distance.

With respect to Dijkstra's treatment of ${\bf SL1VAR}$, the value is both asymmetric and pathwise and requires caution as some of its elements will be zero.  In this case, the value being summed may be zero, which confuses its interpretation as an adjacency matrix, which typically has zero values for non-connected entries.  In order to avoid this confusion, a small non-zero value is applied to each entry in ${\bf SL1VAR}$ that is intended to be used as part of the path sought, keeping zero values for non-connected partitions.  Further, Dijkstra's algorithm treats multiple paths with exact same values by selecting one of the paths at random.  The small non-zero value applied can be used to force Dijkstra's algorithm to choose one path which might be slightly lower due to the artificial bias introduced by the small value.  To apply the small value, $\epsilon$, effectively, each entry in ${\bf SL1VAR}$ is summed with $(-1)^{L1}\epsilon$, which slightly increases and decreases values along the grid of partitions, which adjusts to the convex hull.

The three LOS criteria can be written as follows:
\begin{align}
&1)~{\bf \Delta L}~=~{\bf \Delta L2}_{T}   \hspace{7.8cm}{\rm fastest,~forms~the~}L2_{T}{\rm ~convex~hull} \nonumber\\
&2)~{\rm AND} \left( {\bf \Delta SL1} \leq {\bf \Delta L1}_{T}~,~{\bf \Delta SL1}^{\dag} \leq {\bf \Delta L1}_{T} \right) \hspace{2.85cm}{\rm slowest,~~~~eliminates~corner~paths} \nonumber\\
&3)~{\bf SL1VAR}~\leq~P_C^2 ~~~~~{\rm OR}~~~~~{\bf SL1VAR}~\leq~\tfrac{1}{4}{\bf \Delta SL1}_{\Delta}  \hspace{1.4cm} {\rm moderate,~~~~~~~~~~seeks~linear~path} \nonumber
\end{align}

Dijkstra's algorithm is used in each of the three LOS criteria as it is well-known and many optimized routines have been written for high computational efficiency.  The first criteria uses Dijkstra in the traditional symmetric, step-wise fashion, using a single adjacency matrix, running very quickly, typically 15-20 times faster than the pathwise application.  The second and third application of Dijkstra runs slower as each computation is pathwise requiring a new adjacency matrix to be generated as an input to the algorithm.  Each LOS criteria further restricts the number of partitions that are LOS to each other, suggesting that as time the user may choose the level of LOS applied by using all or fewer of the criteria based on the need and computational costs.  Finally, the third criteria may be tightened or relaxed based on the limit placed on the summed variance of the L1 distance.  As each LOS criteria becomes more restrictive, the number of partitions considered can be reduced at each step.  After criteria \#1, only those partitions within the $L2_{T}$ convex hull are considered for further paths.  After criteria \#2, only those partitions which survived the corner condition are then considered for the criteria \#3.  For this reason, the first criteria runs the fastest, being stepwise, while the second criteria requires the most computational effort, and the third criteria may take as long as the second, but generally will be faster as fewer partitions are considered for LOS.

\cbk

\begin{figure*}[t!p]
	\centering
\label{fig:l1asym}
	\label{fig:losl1-5}
		\includegraphics[width=0.80\textwidth]{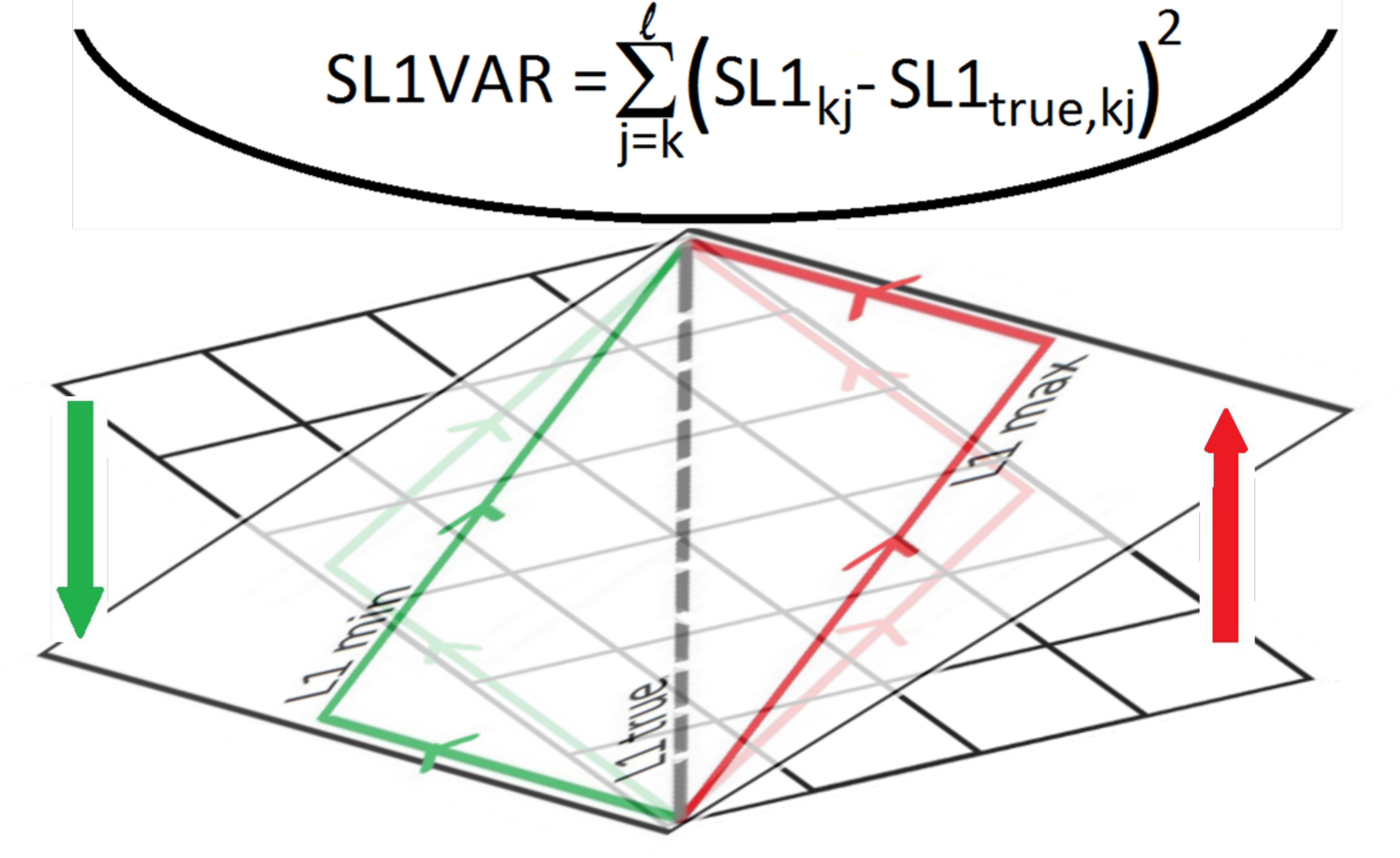}
	\caption{Figure illustrates the two paths of ${\bf \Delta SL1}_{min}$ and ${\bf \Delta SL1}_{max}$, which in 2D is a plane.  The average between them gives the straight line ``true'' value, ${\bf \Delta SL1}_{T}$.  When all paths within the $L2$ convex hull are available, Dijkstra's algorithm will seek the ${\bf \Delta SL1}_{min}$ path when going from $[k,\ell]$ and will seek the ${\bf \Delta SL1}_{max}$ path going from $[\ell,k]$.  This asymmetric behavior is exploited in order to find the straight line path by applying Dijkstra on a third pass, where the pathwise value applied is the squared difference between the ${\bf \Delta SL1}$ and ${\bf \Delta SL1}_{T}$, as is shown above the plane as a parabolic bowl.}
\end{figure*}

\bibliography{clustering_2017}

\end{document}